\pdfoutput=1
\documentclass[a4paper,fleqn,usenatbib,useAMS]{mnras}

\usepackage{subcaption} 
\usepackage{graphicx}	
\usepackage{amsmath}	
\usepackage{booktabs} 
\usepackage{tikz}
\usetikzlibrary{calc}

\DeclareMathOperator{\Lya}{\mathrm{Ly}\alpha}
\DeclareMathOperator{\LyAndreu}{A_{\mathrm{Ly}\alpha}}
\DeclareMathOperator{\LyBusca}{B_{\mathrm{Ly}\alpha}}
\DeclareMathOperator{\DKL}{\mathcal{D}_{KL}}
\DeclareMathOperator{\Like}{\mathcal{L}}
\DeclareMathOperator{\Prior}{\pi}
\DeclareMathOperator{\Bayes}{\mathcal{B}}
\DeclareMathOperator{\PO}{\mathcal{P}}
\DeclareMathOperator{\Evidence}{\mathcal{Z}}
\DeclareMathOperator{\Model}{\mathcal{M}}
\DeclareMathOperator{\Data}{\mathcal{D}}
\DeclareMathOperator{\Prob}{\mathrm{Pr}}
\DeclareMathOperator{\param}{\theta}

\DeclareMathOperator{\nats}{\mathrm{nats}}

\newcommand{\codeF}[1]{\textsc{#1}}

\newcommand{\movablecross}[2]{%
  \draw[->] (#1) -- ++ (0:#2);
  \draw[->] (#1) -- ++ (90:#2);
  \draw[->] (#1) -- ++ (180:#2);
  \draw[->] (#1) -- ++ (270:#2);
  \fill[green!40!black] (#1) circle (2pt);
}

\newcommand{\movablevert}[2]{%
  \draw[->] (#1) -- ++ (90:#2); 
  \draw[->] (#1) -- ++ (270:#2);
  \fill[green!40!black] (#1) circle (2pt);
}

\usepackage[T1]{fontenc}
\usepackage{ae,aecompl}

\usepackage{txfonts}

\title[Constraining the dark energy equation of state]{Constraining the dark energy equation of state using Bayes theorem and the Kullback--Leibler divergence}

\author[S. Hee et al.]{S. Hee,$^{1,2}$\thanks{Contact e-mail: \href{mailto:sh767@cam.ac.uk}{sh767@cam.ac.uk}}
  J.A. V\'azquez,$^{3}$
  W.J. Handley,$^{1,2}$
  M.P. Hobson$^{1}$ and
  A.N. Lasenby$^{1,2}$
\\
$^{1}$ Astrophysics Group, Battcock Centre, Cavendish Laboratory, JJ Thomson Avenue, Cambridge CB3~0HE, UK \\
$^{2}$ Kavli Institute for Cosmology Cambridge, Madingley Road, Cambridge, CB3~0HA, UK \\
$^{3}$ Brookhaven National Laboratory, 2 Center Road, Upton, NY 11973, USA}

\date{Last updated 2016 Jul 1; in original form 2016 July 1}

\pubyear{2016}

\begin{document}
\label{firstpage}
\pagerange{\pageref{firstpage}--\pageref{lastpage}}
\maketitle

\date{Received 1 July 2016}
\pubyear{2016}

\begin{abstract} 
\noindent
Data-driven model-independent reconstructions of the dark energy equation of state $w(z)$ are presented using Planck 2015 era CMB, BAO, SNIa and Lyman-$\alpha$ data. These reconstructions identify the $w(z)$ behaviour supported by the data and show a bifurcation of the equation of state posterior in the range $1.5{<}z{<}3$. 
Although the concordance $\Lambda$CDM model is consistent with the data at all redshifts in one of the bifurcated spaces, in the other a supernegative equation of state (also known as `phantom dark energy') is identified within the $1.5 \sigma$ confidence intervals of the posterior distribution. 
To identify the power of different datasets in constraining the dark energy equation of state, we use a novel formulation of the Kullback--Leibler divergence. 
This formalism quantifies the information the data add when moving from priors to posteriors for each possible dataset combination. 
The SNIa and BAO datasets are shown to provide much more constraining power in comparison to the Lyman-$\alpha$ datasets. Further, SNIa and BAO constrain most strongly around redshift range $0.1-0.5$, whilst the Lyman-$\alpha$ data constrains weakly over a broader range. 
We do not attribute the supernegative favouring to any particular dataset, and note that the $\Lambda$CDM model was favoured at more than $2$ log-units in Bayes factors over all the models tested despite the weakly preferred $w(z)$ structure in the data.
\end{abstract}




 
\section{Introduction}
\label{sec:intro}

The nature of dark energy (DE) remains a significant outstanding problem in cosmology. The $\Lambda$CDM model considers a constant equation of state (EoS) parameter $w{=}{-}1$ motivated by vacuum energy. The most frequent generalisation of the $\Lambda$CDM dark energy EoS is to allow an alteration of the time-independent EoS parameter so that $w\ne{-}1$ (hereafter referred to as $w$CDM). Allowing $w$ to vary in time $w=w(z)$ results in quintessence DE models. Many quintessence models~\citep{Ratra1988,Caldwell1998,Tsujikawa2013}, including phantom DE~\citep{Caldwell2002,Sahni2004}, as well as modified GR theories~\citep{Sahni2004} make predictions for the behaviour of $w(z)$ which may be tested against cosmological datasets~\citep{PlanckCollaboration2015_DE}. Time-dependent behaviour can also be investigated by choosing equations that are simple or mathematically appealing, to test as a DE model. These phenomenological models include the CPL~\citep{Chevallier:2000qy, Linder:2002et}, JPB~\citep{Jassal2004} and FNT~\citep{Felice2012} models. Lastly, free-form approaches attempt to avoid any commitment to particular equations and instead aim to allow the observational data to define any time-dependent features in $w(z)$~\citep{Huterer2003, Zunckel2007, Zhao2008, Serra2009, Lazkoz2012, Vazquez2012}. Other free-form reconstruction methods include gaussian processes~\citep{Holsclaw2010a, Holsclaw2010b, Seikel2012}. We refer the reader to an older review by \cite{Sahni2006} which describes the general reconstruction process and new results by \cite{PlanckCollaboration2015_DE} for further reading on dark energy constraints.

In this paper, we use Bayes factors combined with a `nodal' free-form method, which reconstructs a function using a spline between nodes whose amplitude and position can vary, as first proposed by~\cite{Vazquez2012c}, to investigate the constraints on $w(z)$. This approach has subsequently been used by~\cite{Vazquez2012, Aslanyan2014, PlanckCollaboration2015_inflation, Hee2015} and has the benefit of remaining general and focussing on the cosmological datasets rather than a specific model. The first aim of this paper is to investigate potential deviations from the $\Lambda$CDM constant dark energy equation of state using Bayesian model selection. The second aim is to analyse the constraining power on $w(z)$ of the datasets using the Kullback-Leibler divergence ($\DKL$). Observational data are improving in quality with many upcoming missions promising to increase our ability to understand DE models. Assessing the datasets in the manner this paper proposes provides a robust, quantitative measure of DE information that may easily be compared with past or future missions.

The paper is structured as follows: We first identify the datasets and computational techniques in Section~\ref{sec:method}. An analysis of $w(z)$ constraints from Planck satellite era cosmological datasets is presented in Section~\ref{sec:wz} and the analysis of these additional datasets using the $\DKL$ approach is presented in Section~\ref{sec:dkl}. We conclude in Section~\ref{sec:conclusions}, considering the findings in relation to $\Lambda$CDM and constraints on $w(z)$ and comment on the efficacy of the techniques used for quantifying dataset constraining power and information content.

\section{Datasets and Computation}
\label{sec:method}

We update the work of~\cite{Vazquez2012} and~\cite{Hee2015}, where time dependent behaviour in $w(z)$ within a CDM universe is identified using a sequence of nodal reconstructions weighted by their Bayes factors.
In addition we use the Kullback-Leibler divergence to analyse information content, expanding on similar work by~\cite{Trotta2008, Bridges2009}.

\subsection{Bayes theorem and model selection}
In order to reconstruct the $w(z)$ plane we perform Bayesian parameter estimation and model comparison~\citep{Bayes1763,MacKay2003,Sivia2006} on cosmological models to be defined shortly.

Bayesian parameter estimation is the process of determining the posterior probability distribution of a set of parameters $\param$ for a given model $\Model$ via Bayes theorem:
\begin{equation}
  \Prob(\param | \Data, \Model) = \frac{\Prob(\Data | \param , \Model) \Prob(\param | \Model)}{\Prob(\Data | \Model)} \equiv \frac{\Like\Prior}{\Evidence}.
  \label{eqn:bayes_theorem}
\end{equation}
This requires a prior on the model parameters, $\Prob(\param | \Model) = \Prior(\param)$, and a means to calculate the likelihood, $\Prob(\Data | \param , \Model) = \Like(\param)$. The evidence $\Evidence$ (or marginal likelihood) may be computed from the priors and likelihoods via:
\begin{equation}
  \Evidence \equiv \Prob(\Data |\Model) = \int \Prob(\Data | \param , \Model) \Prob(\param | \Model) \: d\param \equiv \int \Like\Prior.
  \label{eqn:evidence}
\end{equation}
Our priors are defined in Table~\ref{tab:priors}, whilst the likelihood codes are defined in the references in the dataset section below. 

Bayesian model comparison uses Bayes theorem to make inferences about how probable a model is in light of the data: 
\begin{equation}
  P(\Model | \Data) = \frac{\Prob(\Data | \Model) \Prob(\Model)}{\Prob(\Data)}.
  \label{eqn:bayes_theorem_models}
\end{equation}
Taking log-ratios of this equation for two different models yields the posterior odds ratio:
\begin{equation}
  \exp(\PO_{ij}) \equiv \frac{P(\Model_j | \Data)}{P(\Model_i | \Data)} = \frac{P(\Data |\Model_j)}{P(\Data |\Model_i)} \frac{P(\Model_j)}{P(\Model_i)}.
  \label{eqn:posterior_odds_ratios}
\end{equation}
Thus, the critical data-dependent quantity is the Bayes factor:
\begin{equation}
  \Bayes_{ij} = \ln( \Evidence_j/\Evidence_i ),
  \label{eqn:bayes_factor}
\end{equation}
where $\Prob(\Data | \Model_i) = \Evidence_i$ is the evidence of model $i$. We quantify a model's favouring using the~\cite{Jeffreys1961} guideline defined in Table~\ref{tab:jeffreys}. We use Bayes factors and posterior odds ratios interchangeably as we assume uniform model priors ${P(\Model_j)} = {P(\Model_i)}$ throughout.

\begin{table}
  \centering
  \begin{tabular}{l l l} 
    \toprule \toprule
    Parameter                   &   Prior Range         &   Prior Type \\
    \midrule
    $\Omega_{b} h^2$            &   $[0.019, 0.025]$    &   uniform   \\[0.3\normalbaselineskip]
    $\Omega_{c} h^2$            &   $[0.095, 0.145]$    &   uniform   \\[0.3\normalbaselineskip]
    $100\theta_{MC}$            &   $[1.03, 1.05]$      &   uniform   \\[0.3\normalbaselineskip]
    $\tau$                      &   $[0.01, 0.4]$       &   uniform   \\[0.3\normalbaselineskip]
    $n_s$                       &   $[0.885, 1.04]$     &   uniform   \\[0.3\normalbaselineskip]
    $\ln(10^{10}A_s)$           &   $[2.5, 3.7]$        &   uniform   \\[0.3\normalbaselineskip]
    \midrule
    $y_{\scriptscriptstyle \rm cal}$                                            &   $[0.9, 1.1]$    &   uniform   \\[0.3\normalbaselineskip]
    $\alpha_{\scriptscriptstyle JLA}$                                           &   $[0.01, 2.00]$  &   uniform   \\[0.3\normalbaselineskip]
    $\beta_{\scriptscriptstyle JLA}$                                            &   $[0.9, 4.6]$    &   uniform   \\[0.3\normalbaselineskip]
    $A^{\scriptscriptstyle CIB}_{\scriptscriptstyle 217}$                       &   $[0, 200]$      &   uniform   \\[0.3\normalbaselineskip]
    $\xi^{\scriptscriptstyle tSZ-CIB}$                                          &   $[0, 1]$        &   uniform   \\[0.3\normalbaselineskip]
    $A^{\scriptscriptstyle tSZ}_{\scriptscriptstyle 143}$                       &   $[0, 10]$       &   uniform   \\[0.3\normalbaselineskip]
    $A^{\scriptscriptstyle PS}_{\scriptscriptstyle 100}$                        &   $[0, 400]$      &   uniform   \\[0.3\normalbaselineskip]
    $A^{\scriptscriptstyle PS}_{\scriptscriptstyle 143}$                        &   $[0, 400]$      &   uniform   \\[0.3\normalbaselineskip]
    $A^{\scriptscriptstyle PS}_{\scriptscriptstyle 143 \times 217}$             &   $[0, 400]$      &   uniform   \\[0.3\normalbaselineskip]
    $A^{\scriptscriptstyle PS}_{\scriptscriptstyle 217}$                        &   $[0, 400]$      &   uniform   \\[0.3\normalbaselineskip]
    $A^{\scriptscriptstyle kSZ}$                                                &   $[0, 10]$       &   uniform   \\[0.3\normalbaselineskip]
    $A^{\scriptscriptstyle {\rm dust}TT}_{\scriptscriptstyle 100}$              &   $[0, 50]$       &   uniform   \\[0.3\normalbaselineskip]
    $A^{\scriptscriptstyle {\rm dust}TT}_{\scriptscriptstyle 143}$              &   $[0, 50]$       &   uniform   \\[0.3\normalbaselineskip]
    $A^{\scriptscriptstyle {\rm dust}TT}_{\scriptscriptstyle 143 \times 217}$   &   $[0, 100]$      &   uniform   \\[0.3\normalbaselineskip]
    $A^{\scriptscriptstyle {\rm dust}TT}_{\scriptscriptstyle 217}$              &   $[0, 400]$      &   uniform   \\[0.3\normalbaselineskip]
    $c_{\scriptscriptstyle 100}$                                                &   $[0, 30]$       &   uniform   \\[0.3\normalbaselineskip]
    $c_{\scriptscriptstyle 217}$                                                &   $[0, 30]$       &   uniform   \\[0.3\normalbaselineskip]
    \midrule
    $w(z_i)|_{i=1\dots5}$           &   $[{-}2, {-}0.01]$               &   uniform   \\[0.3\normalbaselineskip]
    $z_i|_{i=2\dots4}$              &   $[0.01, 3.0]$                   &   sorted-uniform   \\[0.3\normalbaselineskip]
    \bottomrule \bottomrule
  \end{tabular}
  \caption{The 31 priors that define the parameter space. The top set of parameters are the CDM parameters, the middle ones show the nuisance parameters associated with the datasets, and the bottom set are the parameters introduced by the free-form dark energy model extensions. \protect\cite{PlanckCollaboration2015_likelihoods} has more details about the CDM and nuisance parameters, whilst the dark energy extension parameters are defined in the text.}
\label{tab:priors}
\end{table}

\begin{table}
  \begin{center}
    \begin{tabular}{ll}
      \toprule 
      Posterior odds ratio          & Favouring of $\Model_j$ over $\Model_i$ \\
      \midrule
      $0.0 \le \PO_{i j} \le 1.0$   & None \\
      $1.0 \le \PO_{i j} \le 2.5$   & Slight\\
      $2.5 \le \PO_{i j} \le 5.0$   & Significant\\
      $5.0 \le \PO_{i j}        $   & Decisive\\
      \bottomrule
    \end{tabular}
  \end{center}
  \caption{Jeffreys guideline for interpreting posterior odds ratios. As $\PO_{j i} {=} {-}\PO_{i j}$, negative values imply model favouring is reversed.}
\label{tab:jeffreys}
\end{table}

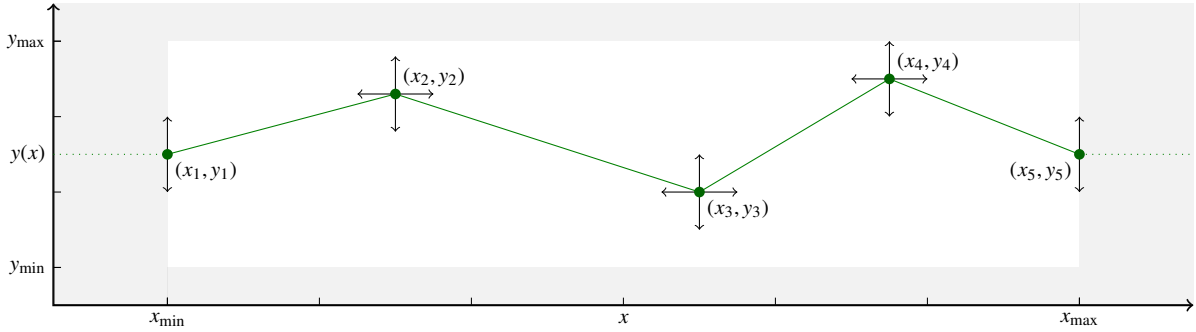
\begin{figure*}
  \centering
  \begin{tikzpicture}

  \def\xmin{0.0}  
  \def\xmax{12.0}  
  \def\ymin{0.0}  
  \def\ymax{3.0}  
  \def\dx{1.5}    
  \def\dy{0.5}    
  \def\dxtick{0.1}   
  \def\dytick{0.1}   
  \def\xnodeA{\xmin}  
  \def\xnodeB{3}      
  \def\xnodeC{7}      
  \def\xnodeD{9.5}    
  \def\xnodeE{\xmax}  
  \def\ynodeA{1.5}      
  \def\ynodeB{2.3}      
  \def\ynodeC{1}      
  \def\ynodeD{2.5}      
  \def\ynodeE{1.5}    
  \def\crosslen{0.5}

  \pgfmathsetmacro\xlabel{\xmax / 2}
  \pgfmathsetmacro\ylabel{\ymax / 2}
  \draw [<->,thick] 
  (\xmin-\dx, \ymax+\dy) 
  |-
  (\xmin-\dx, \ylabel) node (yaxis) [left] {$y(x)$}
  |- 
  (\xlabel, \ymin-\dy) node (xaxis) [below] {$x$}
  |- 
  (\xmax+\dx, \ymin-\dy);


  \pgfmathtruncatemacro\xincrement{\xmin+2}
  \pgfmathtruncatemacro\yincrement{\ymin+1}

  \foreach \x in {\xmin,\xincrement,..., \xmax} {%
    \draw ($(\x, \ymin -\dy)$) -- ($(\x, \ymin - \dy + \dxtick)$);
  }

  \foreach \y in {\ymin,\yincrement,..., \ymax} {%
    \draw ($(\xmin - \dx, \y)$) -- ($(\xmin - \dx + \dytick, \y)$);
  }

  \draw[fill=gray, draw=none, opacity=0.1] (\xmin-\dx, \ymin-\dy) rectangle (\xmin, \ymax);
  \draw[fill=gray, draw=none, opacity=0.1] (\xmin, \ymin-\dy) rectangle (\xmax+\dx, \ymin);
  \draw[fill=gray, draw=none, opacity=0.1] (\xmin-\dx, \ymax) rectangle (\xmax, \ymax+\dy);
  \draw[fill=gray, draw=none, opacity=0.1] (\xmax, \ymin) rectangle (\xmax+\dx, \ymax+\dy);

  \coordinate (dashleft) at (\xnodeA-\dx, \ynodeA);
  \coordinate (A) at (\xnodeA, \ynodeA);
  \coordinate (B) at (\xnodeB, \ynodeB);
  \coordinate (C) at (\xnodeC, \ynodeC);
  \coordinate (D) at (\xnodeD, \ynodeD);
  \coordinate (E) at (\xnodeE, \ynodeE);
  \coordinate (dashright) at (\xnodeE+\dx, \ynodeE);

  \draw [green!50!black] (A) -- (B) -- (C) -- (D) -- (E);
  \draw [green!50!black, dotted] (dashleft) -- (A);
  \draw [green!50!black, dotted] (E) -- (dashright);
  \draw (\xmin, \ymin-\dy+\dytick) -- (\xmin, \ymin-\dy) node[below] {$x_{\min}$};
  \draw (\xmax, \ymin-\dy+\dytick) -- (\xmax, \ymin-\dy) node[below] {$x_{\max}$};
  \draw (\xmin-\dx+\dxtick, \ymin) -- (\xmin-\dx, \ymin) node[left] {$y_{\min}$};
  \draw (\xmin-\dx+\dxtick, \ymax) -- (\xmin-\dx, \ymax) node[left] {$y_{\max}$};

  \movablevert{A}{\crosslen};
  \movablecross{B}{\crosslen};
  \movablecross{C}{\crosslen};
  \movablecross{D}{\crosslen};
  \movablevert{E}{\crosslen};

  \draw (A) node[below right] {$(x_{1},y_{1})$};
  \draw (B) node[above right] {$(x_{2},y_{2})$};
  \draw (C) node[below right] {$(x_{3},y_{3})$};
  \draw (D) node[above right] {$(x_{4},y_{4})$};
  \draw (E) node[below left]  {$(x_{5},y_{5})$};

\end{tikzpicture}
  \caption{%
    Piecewise linear interpolation function. We place a $n$ internal nodes $(x_i,y_i)$ in the rectangle bounded by $(x_{\min},y_{\min})$ and $(x_{\max},y_{\max})$, where the positions $x_i$ and amplitudes $y_i$ are model parameters to be varied. At $x_{\min}$ and $x_{\max}$ fixed-position nodes are placed with varying amplitude only, such that for the model defined by $n$ internal nodes there are $2+2n$ parameters. Linear interpolation between the nodes $(x_i,y_i)$ is used to construct $y$ at all points, with $y(x)$ set constant outside the range $[x_{\min}, x_{\max}]$.
    }
\label{fig:NodalMethod}
\end{figure*}

\subsection{Datasets}

In order to investigate possible time-dependent behaviour in the dark energy equation of state we use likelihood codes from Planck CMB measurements, baryonic acoustic oscillations (BAO), type-Ia supernovae (SNIa) and Lyman-$\alpha$ BAO data ($\Lya$). Specifically, for the CMB data we use the low-$l$ TEB and high-$l$ TT likelihoods from the Planck satellite 2015 data release~\citep{PlanckCollaboration2015_overview, PlanckCollaboration2015_likelihoods, PlanckCollaboration2015_parameters}, which we will refer to as $Planck$. For the BAO data we use the BOSS data release 11 likelihoods~\citep{Anderson2014}, or $BAO$. For the SNIa data we use the JLA supernovae catalogue likelihoods~\citep{Betoule2014}, $JLA$ for short. For the $\Lya$ data we use the likelihood codes described by~\cite{Font-Ribera2014} ($\LyAndreu$; BOSS auto-correlation) and~\cite{Delubac2015} ($\LyBusca$; BOSS cross-correlation). For a good summary of the BAO data see~\cite{Aubourg2015}. Using the above notation, the whole dataset combination can be referred to as $Planck+BAO+JLA+\LyAndreu+\LyBusca$.

\subsection{Computational tools}

To carry out Bayesian inference we use \codeF{CosmoMC}~\citep{Lewis:2002ah} containing the Boltzmann \codeF{CAMB} code \citep{Lewis:1999bs,Howlett2012}. We substitute the default Metropolis-Hastings sampler with the \codeF{PolyChord} nested sampling plug-in \citep{Handley2015,Handley2015b}, an effective nested sampling implementation \citep{Sivia2006, Skilling2004,Skilling2006} for evidence calculations and parameter estimation with proven efficacy using Planck era data \citep{PlanckCollaboration2015_inflation}. Aside from the $\Lya$ datasets, all datasets used are default \codeF{CosmoMC} options. To facilitate deviations from the standard $\Lambda$CDM equation of state parameter $w {=} {-}1$ we implement the Parameterized Post-Friedmann framework (PPF) modification to \codeF{CAMB} \citep{Fang2008}, which has sound speed equal to $c$ and no scalar anisotropic stress. The free-form reconstruction we use is the nodal reconstruction as proposed by~\cite{Vazquez2012} and successfully used in several cosmological applications to date~\citep{Vazquez2012c,Vazquez2012,Aslanyan2014,PlanckCollaboration2015_inflation,Hee2015}.


\subsection{Nodal reconstruction}
\label{sec:method_nodal}

We model a one-dimensional function $y(x)$ using a piecewise linear interpolation between a set of $n$ nodes (Figure~\ref{fig:NodalMethod}), where the positions of the nodes are model parameters to be varied. Alternative interpolation schemes may be used, for example, the cubic spline studied by~\cite{Vazquez2012c}, although we do not consider these here since the continuity requirements of the interpolation functions and its derivatives limit its ability to model sharply changing functions $y(x)$.

A model is defined by how many nodes are used in reconstructing $y(x)$. We use Bayes factors to compare models with increasing numbers of nodes, which quantify how many nodes are needed to fit the data. 

Further, as each posterior sample defines a function in $y(x)$, we can calculate the posterior probability of $y$ in normalised slices of constant $x$, $\Prob(y|x,\Data,\Model)$, to obtain the plane reconstruction of a model. We plot these as a function of $\sigma$ confidence intervals to show the statistical significance of deviations from the maximal $y$ at each $x$. One can plot $\Prob (y | x, n_{\star})$, where $n_{\star}$ denotes the number of nodes in the most favoured model. In order to identify the nature of constraints from various models, one should also plot $\Prob (y|x)$ averaged over all models weighted by their posterior odds ratios~\citep{Parkinson2013,PlanckCollaboration2015_inflation,Hee2015}.

A key strength of this reconstruction procedure is its free-form nature, which can capture any shape of function in the $y(x)$ plane by adding arbitrarily large numbers of nodes. The Bayes factor penalises over-complex models, identifying how much complexity the data is able to support. Model selection techniques can thus be used to solve questions on the constraining power of the data in cosmological applications \citep{Vazquez2012c, Vazquez2012, Aslanyan2014, PlanckCollaboration2015_inflation, Hee2015}. 

\begin{table}
  \begin{center}
    \begin{tabular}{ll}
      \toprule 
      Model name          & Description \\
      \midrule
      $\Lambda$CDM & $w=-1$\\
            $w$CDM & $w$ constant in $z$, but allowed to vary\\
            $t$CDM & tilted spectrum: two fixed-position nodes at $z=0,3$\\
            $1$CDM & One internal node\\
            $2$CDM & Two internal nodes\\
            $3$CDM & Three internal nodes\\
      \bottomrule
    \end{tabular}
  \end{center}
  \caption{The six models we consider. Priors on each $w$ parameter are uniform on the range $[{-}2, 0]$, and were chosen to be conservative~\citep{Vazquez2012}. Priors on each $z$ parameter are uniform on $[0, 3]$ and sorted, such that for more than one internal node we have $z_i {<} z_{i{+}1}$ (i.e.\ sorted uniform priors).}
\label{tab:models}
\end{table}

We apply this reconstruction to $w(z)$.
The models we consider, along with their priors are detailed in Table~\ref{tab:models}.
 The previous work using WMAP satellite era data by~\cite{Vazquez2012} found that $\Lambda$CDM was favoured, whilst $2$CDM had the second largest evidence, pointing to structure in $w(z)$ that could not be captured by a constant equation of state parameter $w$CDM, or even the $1$ internal node model. Subsequent work with Planck 2013 era data by~\cite{Hee2015} showed that $\Lambda$CDM was again favoured, and that each model of increasing complexity was more disfavoured than the last. We now investigate this more fully with Planck 2015 era datasets, the addition of $\Lya$ data and further dataset analysis tools.

\subsection{Kullback-Leibler divergence and dataset analysis}
\label{sec:method_DKL}

We expand on the model selection complexity analysis through the use of the {\em Kullback-Leibler\/} (KL) divergence. The KL divergence of $P$ from $Q$ is defined as
\begin{equation}
    \DKL(P||Q) \equiv \int_{-\infty}^{\infty} p(x) \ln \left[ \frac{p(x)}{q(x)} \right] dx = \int \ln \left[ \frac{dP}{dQ} \right] dP,
    \label{eqn:KL_divergence}
\end{equation}
where $p(x)$ and $q(x)$ are the probability density functions of probability distributions $P$ and $Q$. 
Evaluating the KL-divergence~\eqref{eqn:KL_divergence} of a posterior distribution from its prior provides a measure of information gained from the data~\citep{Kullback1951, Trotta2008, Bridges2009, Seehars2014, Seehars2016, Grandis2016}. 

We wish to restrict our analysis to the constraining power of the datasets on $w(z)$, and not the other cosmological and nuisance parameters as a whole. 
First, we can calculate the KL divergence of the marginalised posterior $\Prob(w|z)$ from the marginalised prior $\Prior(w|z)$ for $w$ at each $z$ to obtain a function:
\begin{equation}
  \DKL(z) = \int \Prob(w|z) \ln \left[ \frac{\Prob(w|z)}{\Prior(w|z)} \right] dw 
  \label{eqn:DKL_z}
\end{equation}
 which defines the gain in information on $w$ at each $z$.
Second, we calculate the $\DKL$ for the whole plane by using the function $\Prob(w, z)$ and its prior, which can be written as 
\begin{equation}
  \DKL = \int \DKL(z) \Prob(z) dz
  \label{eqn:DKL}
\end{equation}
where $\Prob(z)$ is flat (as $z$ is not constrained by the analysis given that every posterior sample for a nodal reconstruction passes through every point in $z$). Note that it is also possible to integrate over $da$ or $d\log(a)$ to compress higher redshifts, however $dz$ is more natural here given how we have defined our reconstruction.
Together the two values allow us to analyse the gain in information due to different datasets using $\DKL$ as well as to understand where each dataset provides the greatest gains in information using $\DKL(z)$. We obtain the posterior plane reconstructions from \codeF{PolyChord} and the prior distributions based on $\Prior(z_i)$ and $\Prior(w_i)$ together with the physical restrictions imposed by \codeF{CosmoMC}. 

Typically, a gain in information can occur for two reasons: either due to an increase in parameter constraints, or due to a shift in the position of the peak from prior to posterior~\citep{Trotta2008,Seehars2014,Seehars2016, Grandis2016}. It is not yet possible to differentiate between the two cases for non-Gaussian distributions. In order to identify the constraining power of the data, we supplement our analysis by calculating the $\DKL$ and $\DKL(z)$ when moving from a completely flat prior on $w(z)$ to the posterior. As there is no peak to shift from for a flat posterior, this measure only identifies how tightly constrained the posterior is, due both to the priors and data. In cases where the \codeF{CosmoMC} prior divergences are larger than those from the flat prior we can deduce that a significant shift is present.

\section{Results: dark energy equation of state reconstruction}
\label{sec:wz}

\begin{figure*}
  \centering
  $\Bayes_{\Lambda \, w} = -2.34 \pm 0.29$ \\
  \includegraphics[width=0.33\textwidth, height=0.21\textwidth]{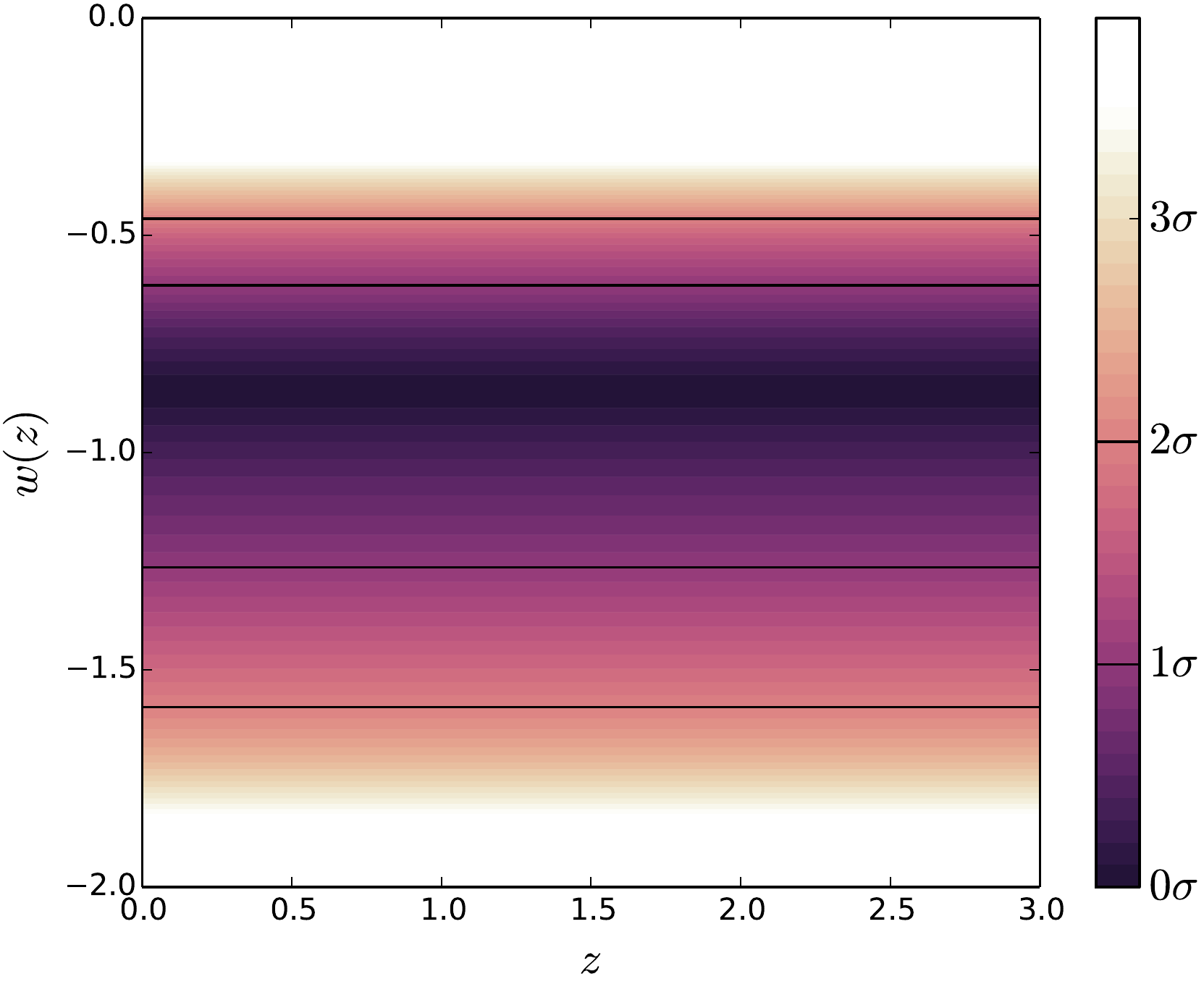}
  \includegraphics[width=0.33\textwidth, height=0.21\textwidth]{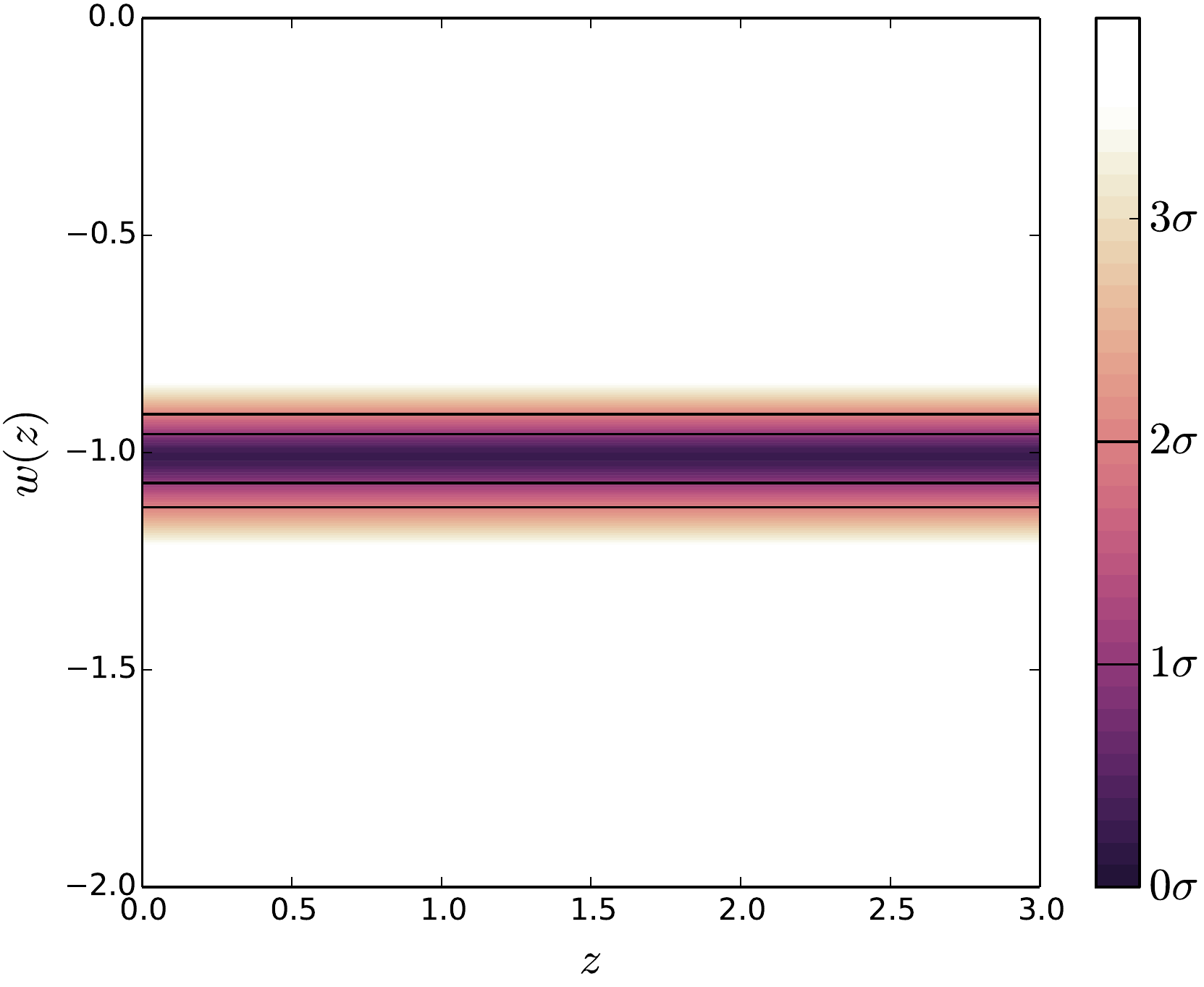}
  \includegraphics[width=0.33\textwidth, height=0.21\textwidth]{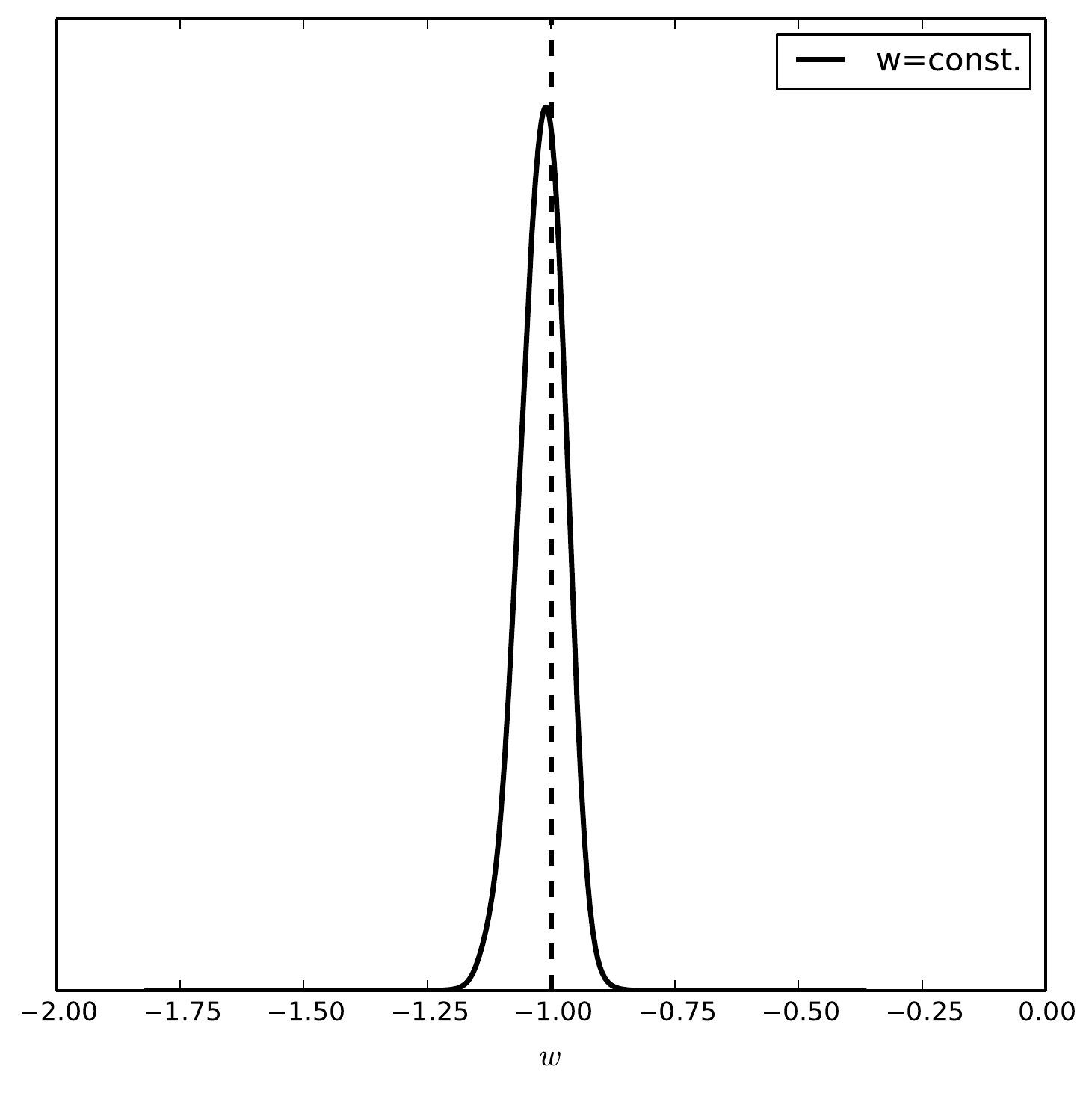}

  $\Bayes_{\Lambda \, t} = -2.68 \pm 0.29$ \\
  \includegraphics[width=0.33\textwidth, height=0.21\textwidth]{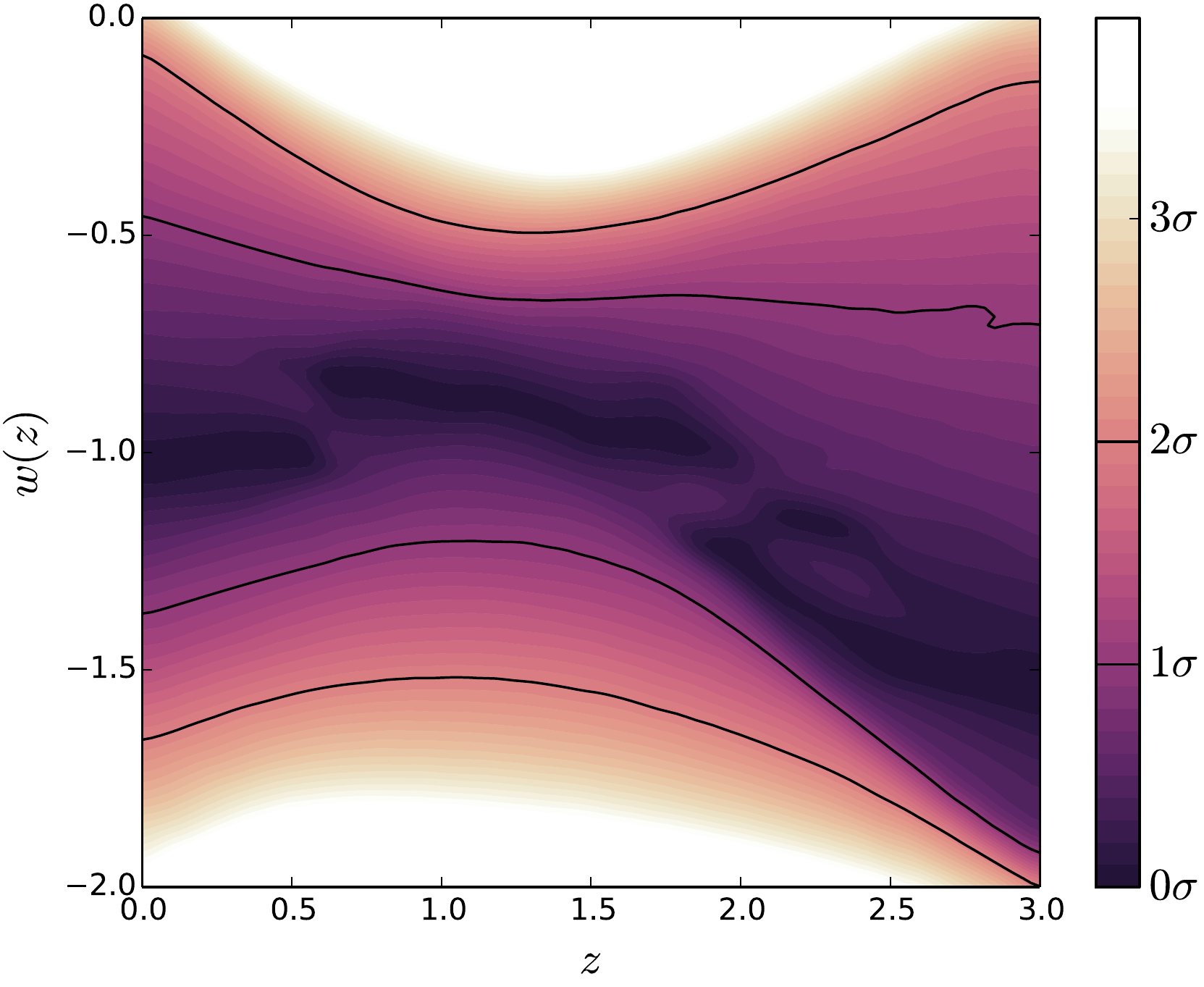}
  \includegraphics[width=0.33\textwidth, height=0.21\textwidth]{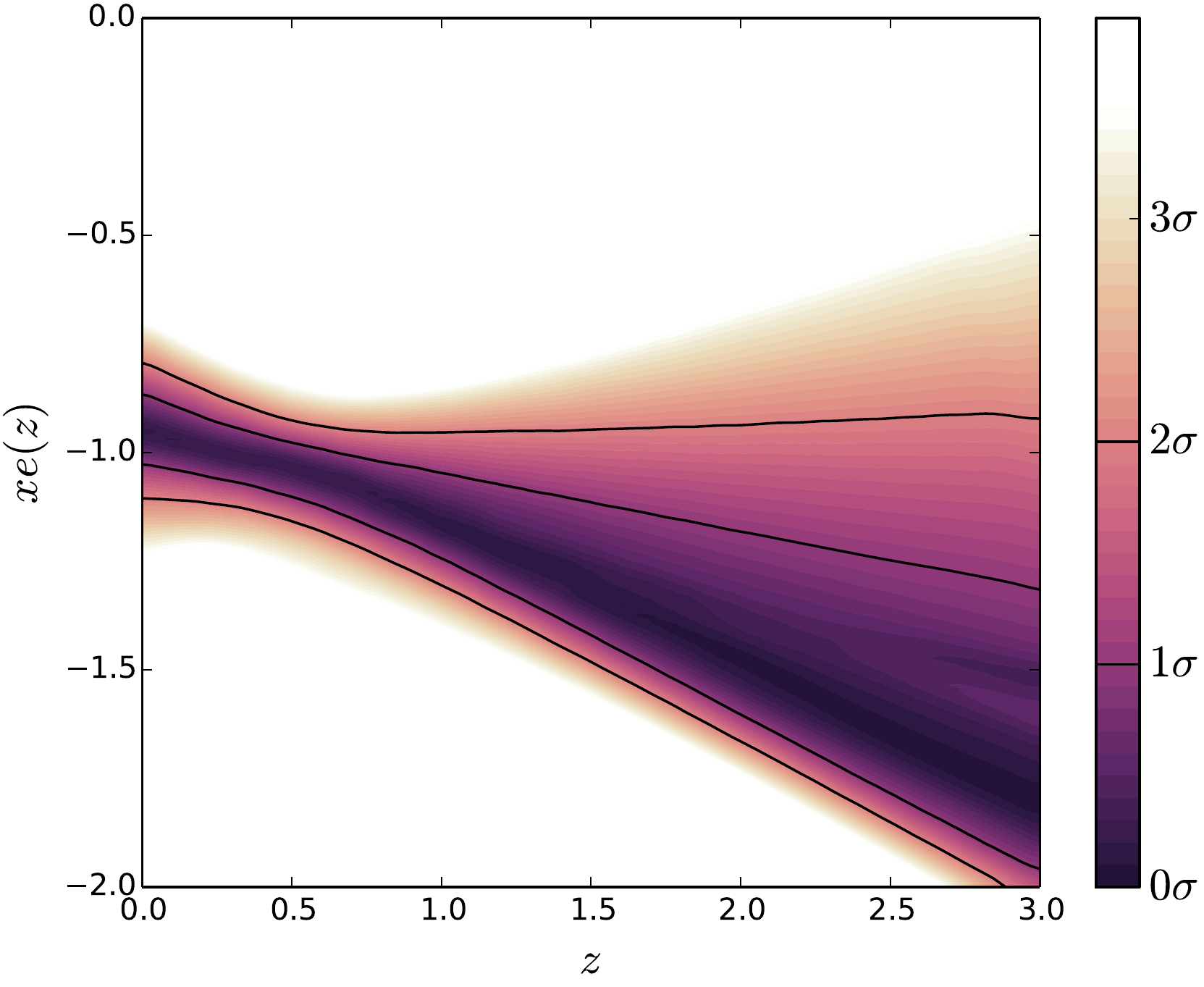}
  \includegraphics[width=0.33\textwidth, height=0.21\textwidth]{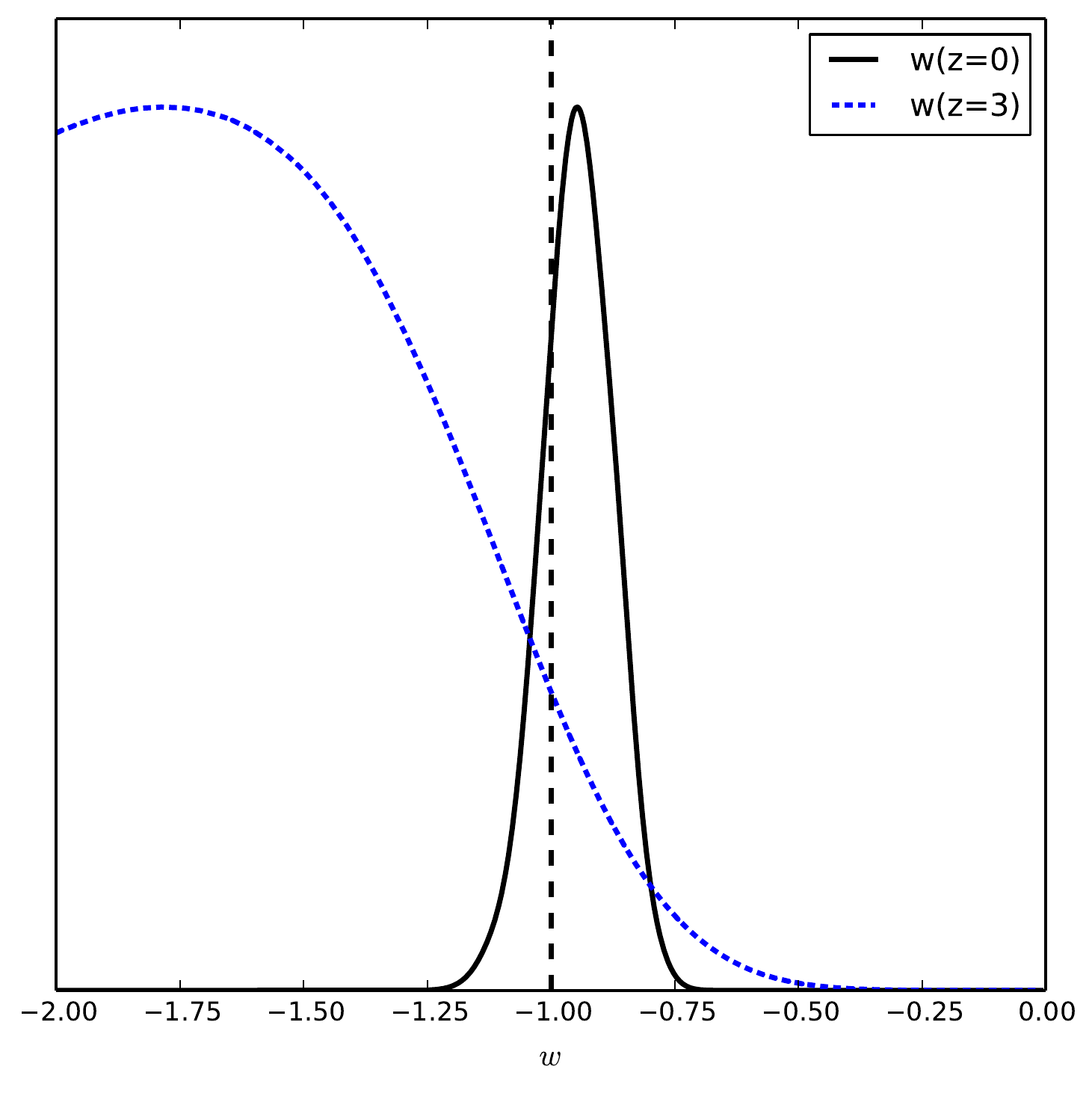}

  $\Bayes_{\Lambda \, 1} = -2.68 \pm 0.29$ \\
  \begin{minipage}{0.33\textwidth}
    \includegraphics[width=1.00\textwidth, height=0.63\textwidth]{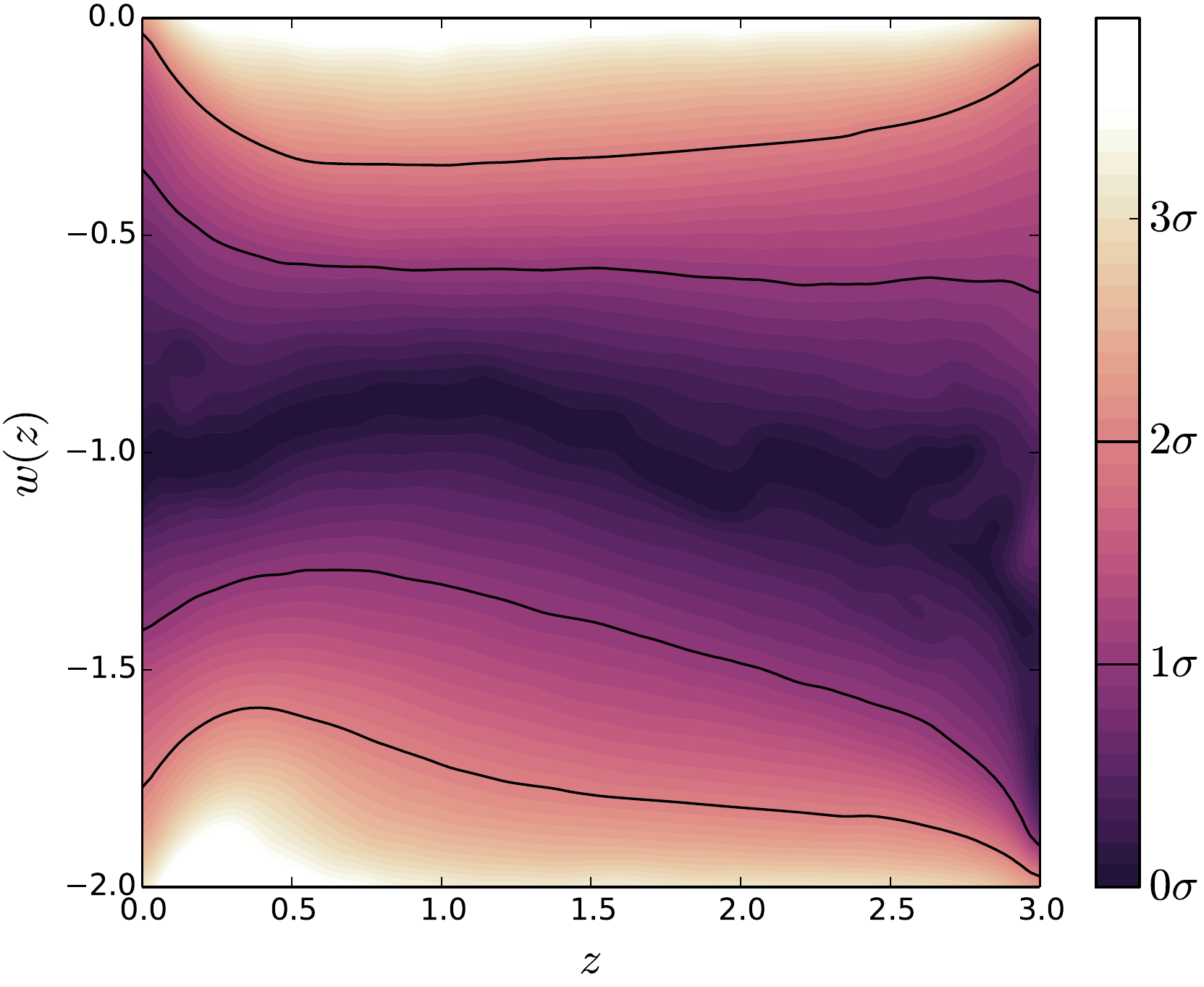}
  \end{minipage}
  \begin{minipage}{0.33\textwidth}
    \includegraphics[width=1.00\textwidth, height=0.63\textwidth]{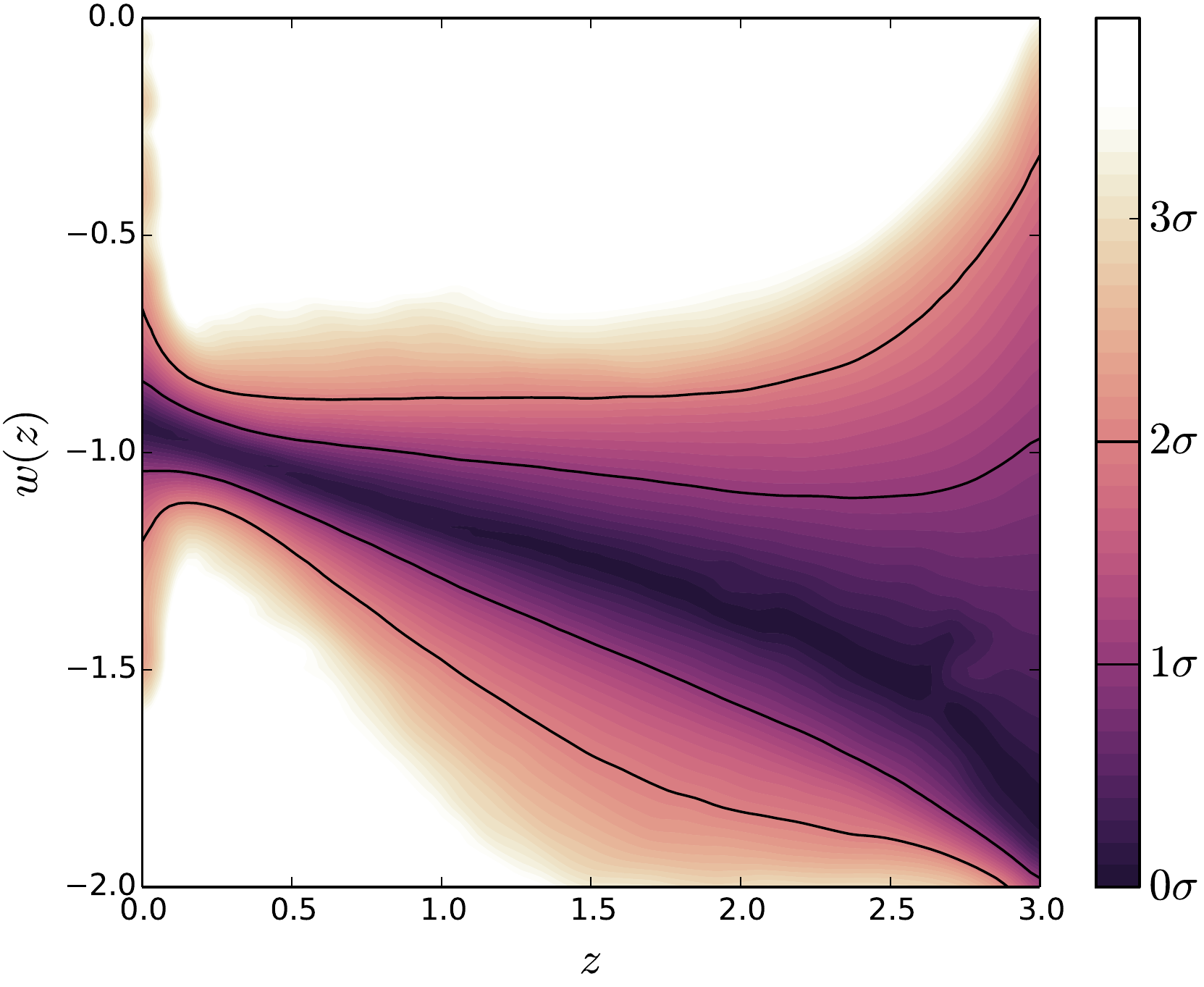}
  \end{minipage}
  \begin{minipage}{0.33\textwidth}
    \includegraphics[width=1.00\textwidth, height=0.32\textwidth]{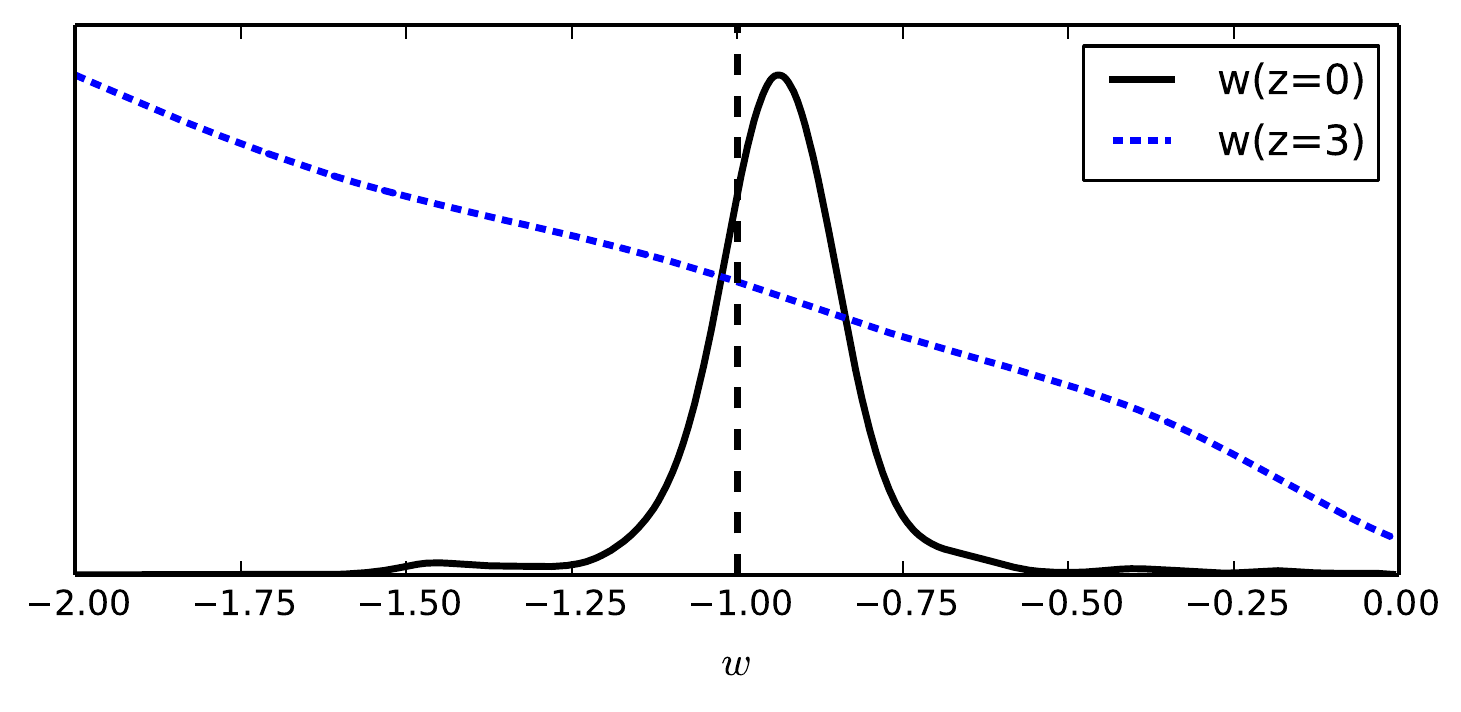}
    \includegraphics[width=1.00\textwidth, height=0.32\textwidth]{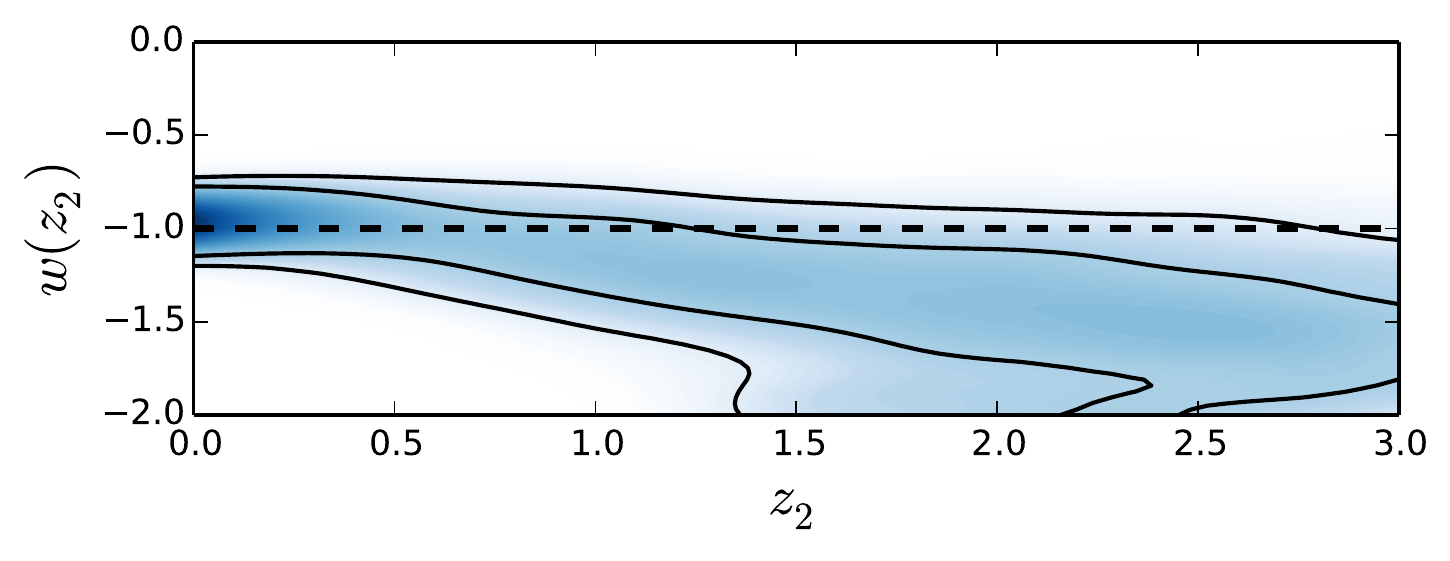}
  \end{minipage}

  $\Bayes_{\Lambda \, 2} = -2.82 \pm 0.29$ \\
  \begin{minipage}{0.33\textwidth}
    \includegraphics[width=1.00\textwidth, height=0.63\textwidth]{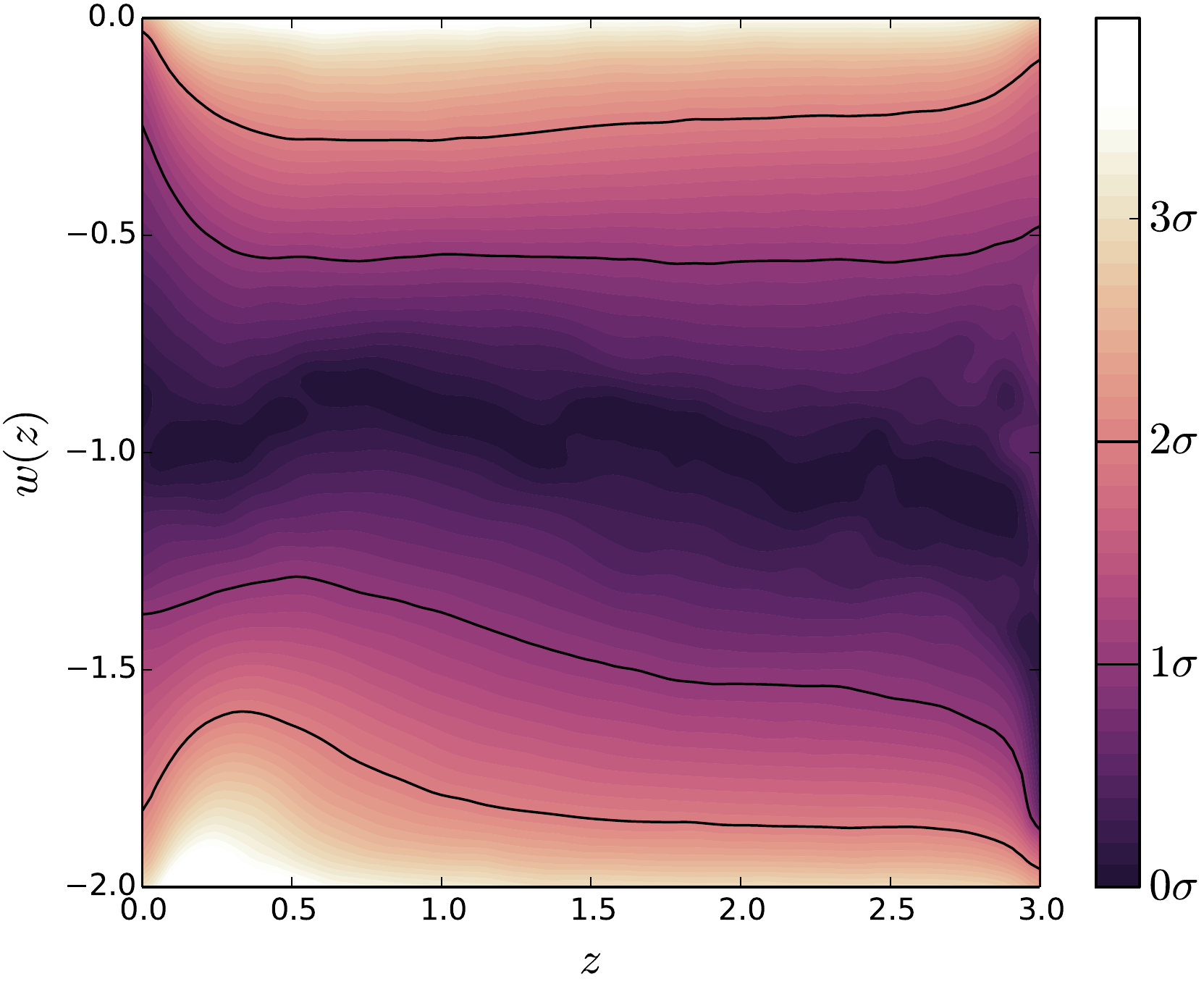}
  \end{minipage}
  \begin{minipage}{0.33\textwidth}
    \includegraphics[width=1.00\textwidth, height=0.63\textwidth]{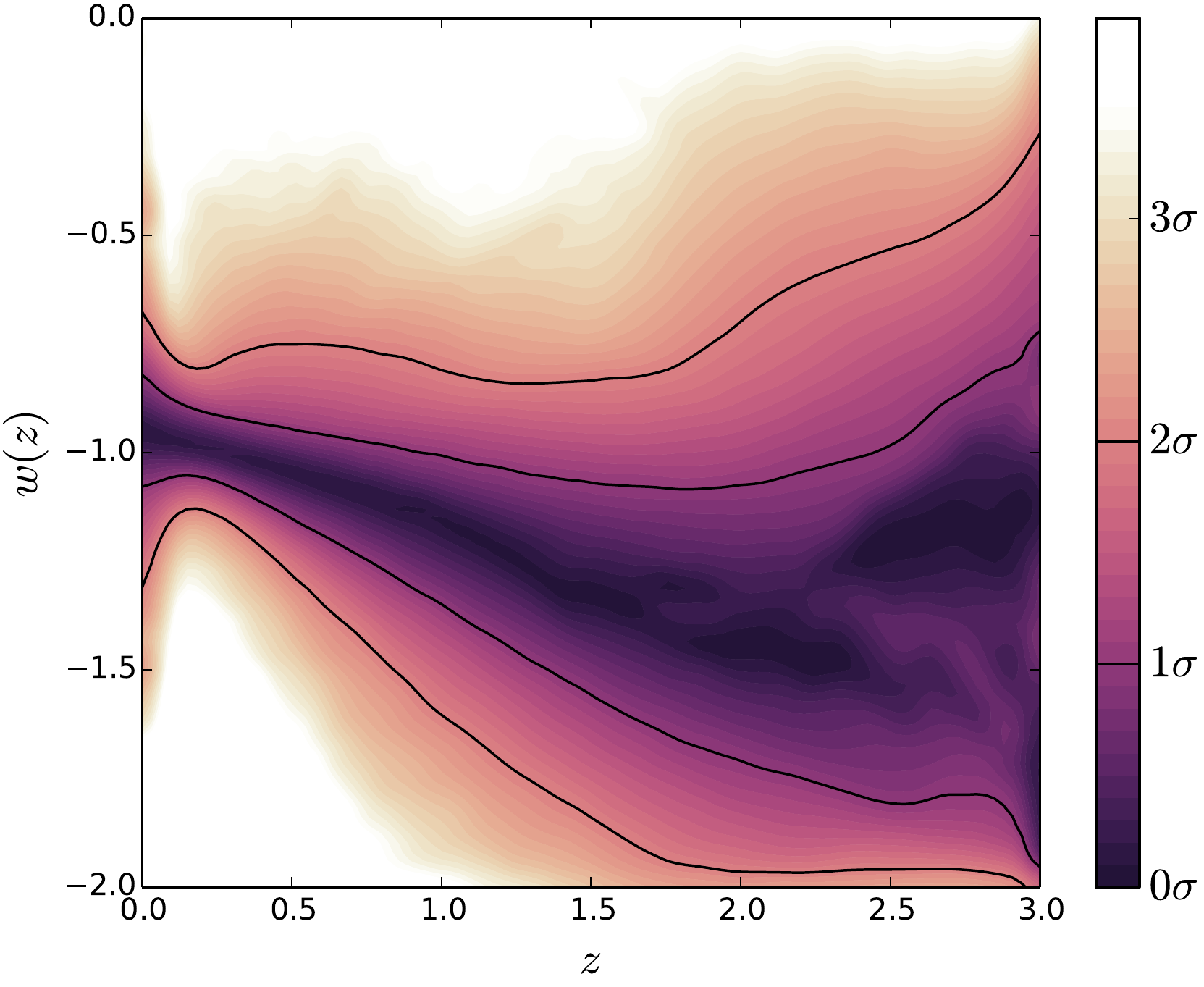}
  \end{minipage}
  \begin{minipage}{0.33\textwidth}
    \includegraphics[width=1.00\textwidth, height=0.32\textwidth]{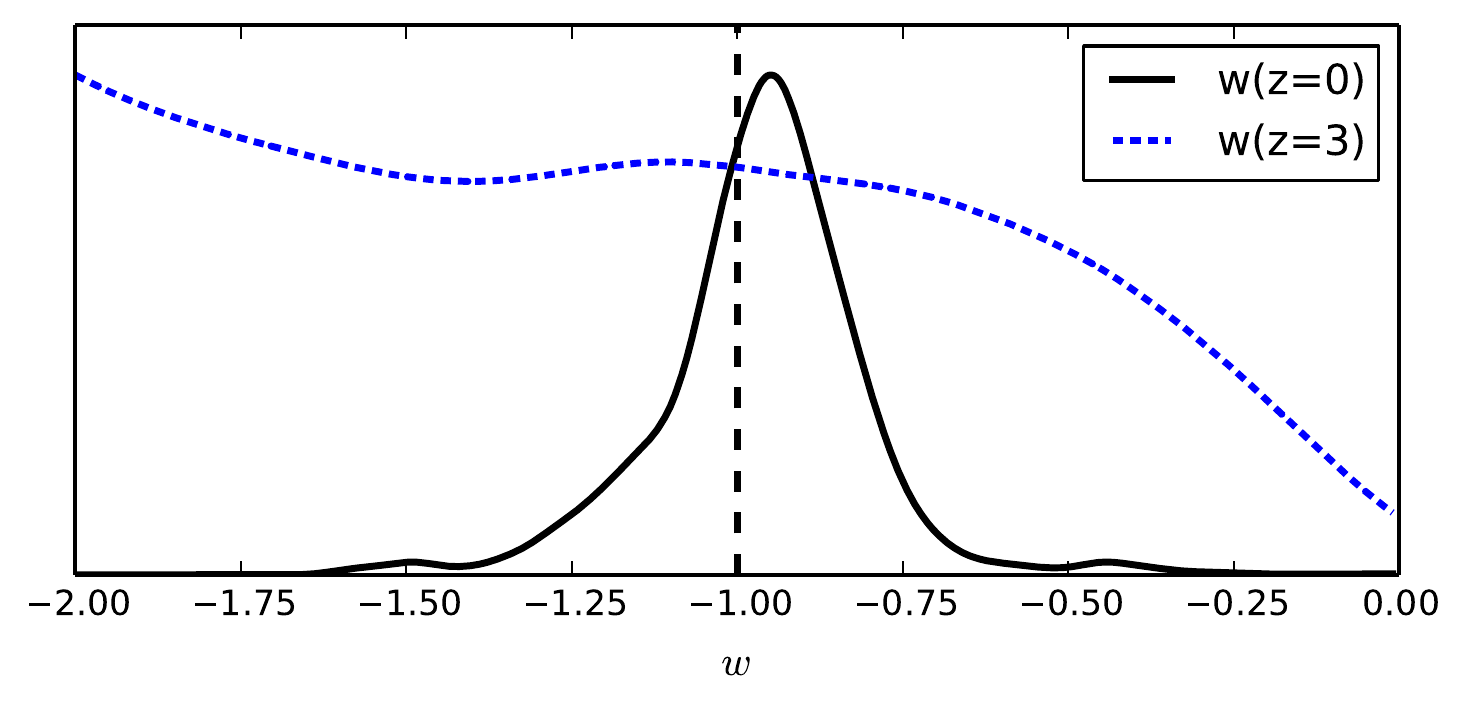}
    \includegraphics[width=0.49\textwidth, height=0.32\textwidth]{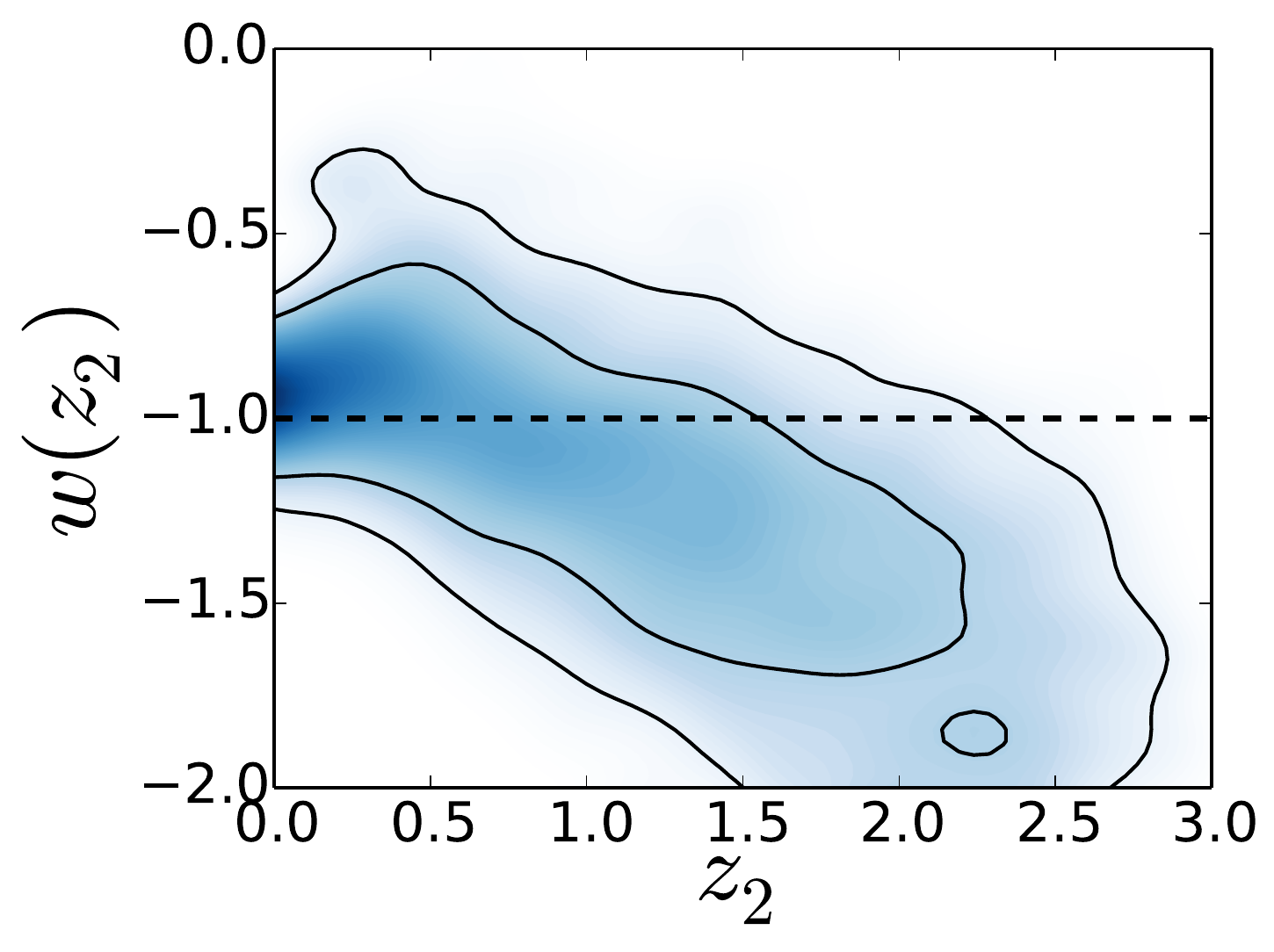}
    \includegraphics[width=0.49\textwidth, height=0.32\textwidth]{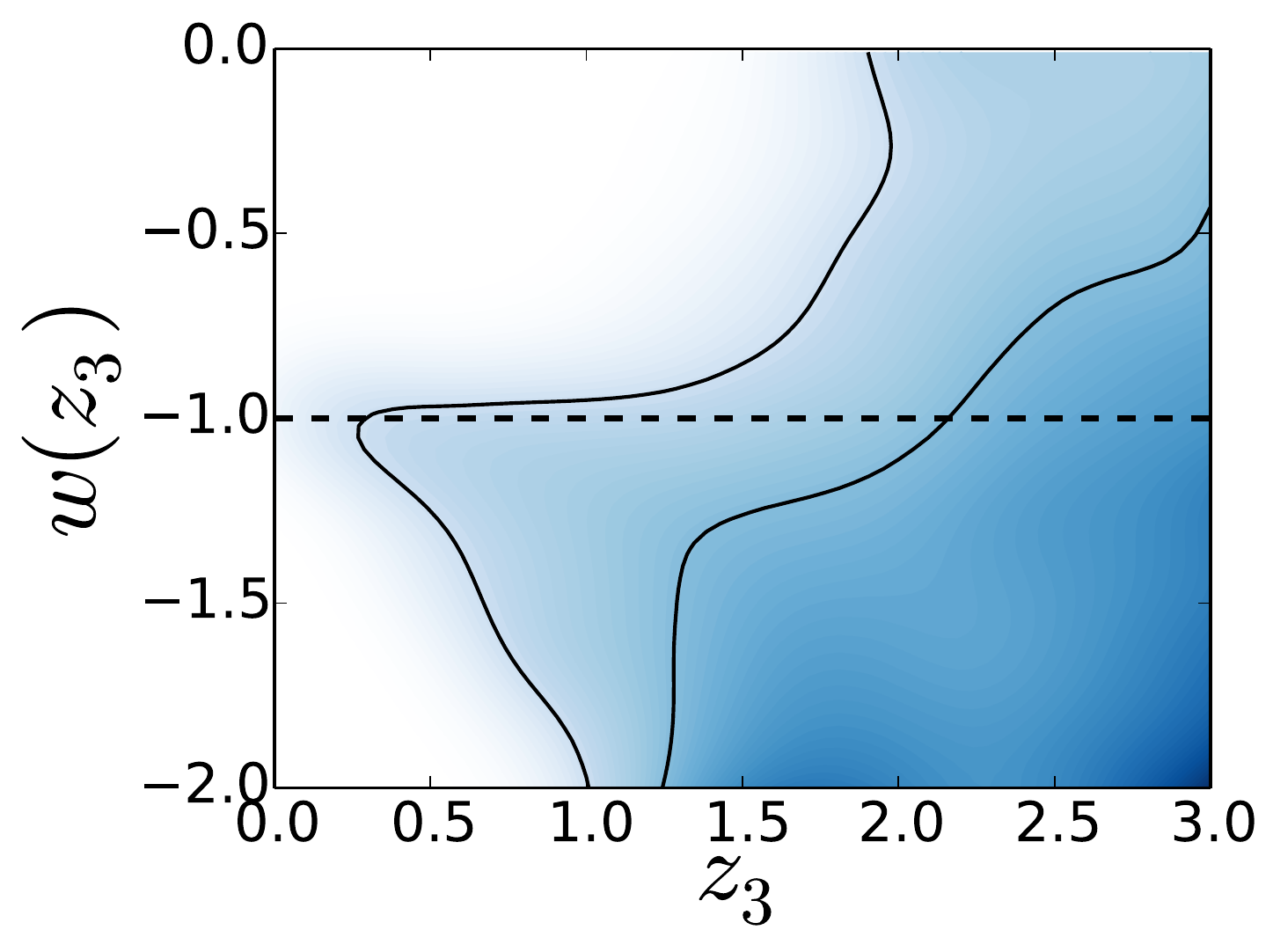}
  \end{minipage}

  $\Bayes_{\Lambda \, 3} = -3.36 \pm 0.29$ \\
  \begin{minipage}{0.33\textwidth}
    \includegraphics[width=1.00\textwidth, height=0.63\textwidth]{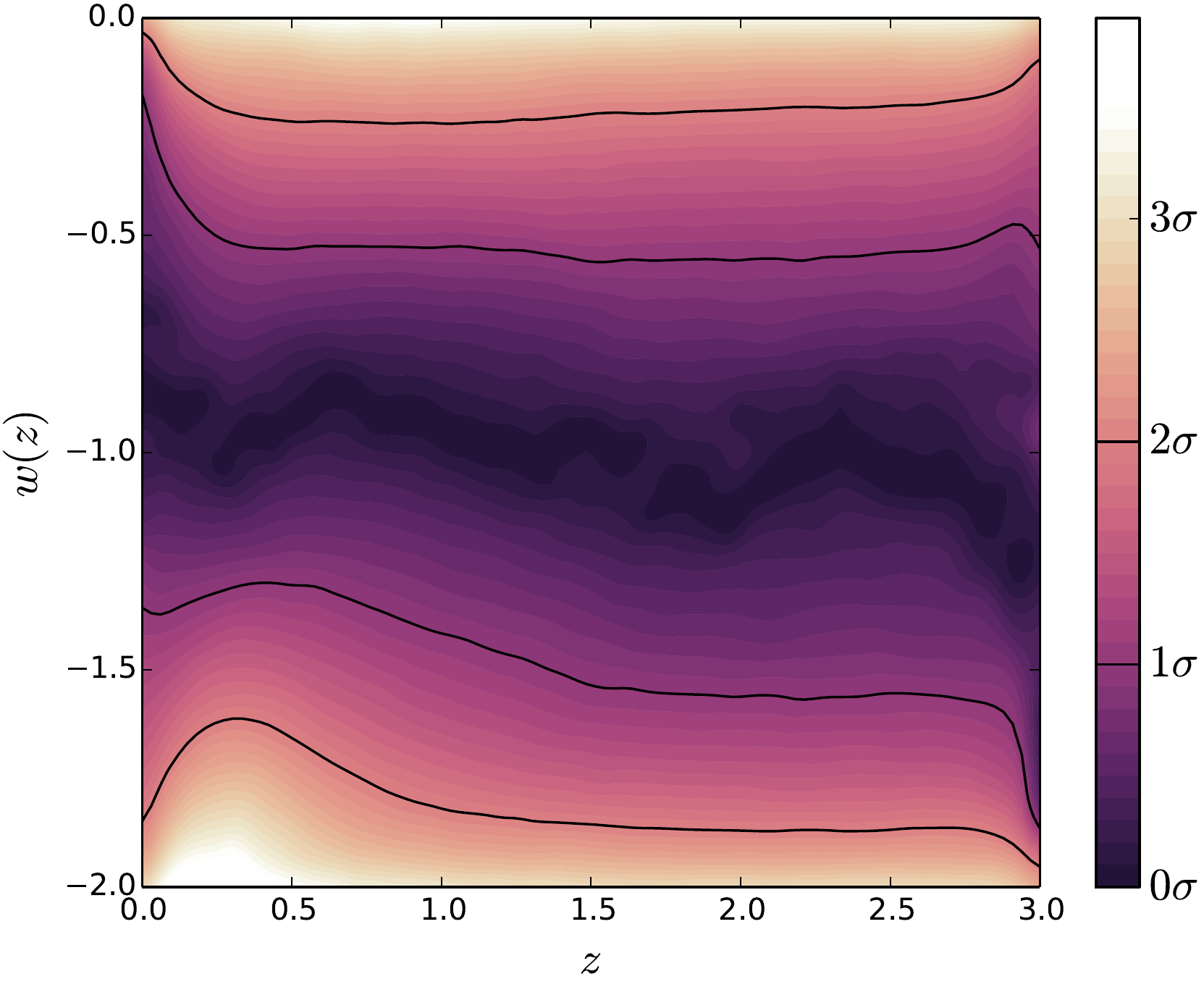}
  \end{minipage}
  \begin{minipage}{0.33\textwidth}
    \includegraphics[width=1.00\textwidth, height=0.63\textwidth]{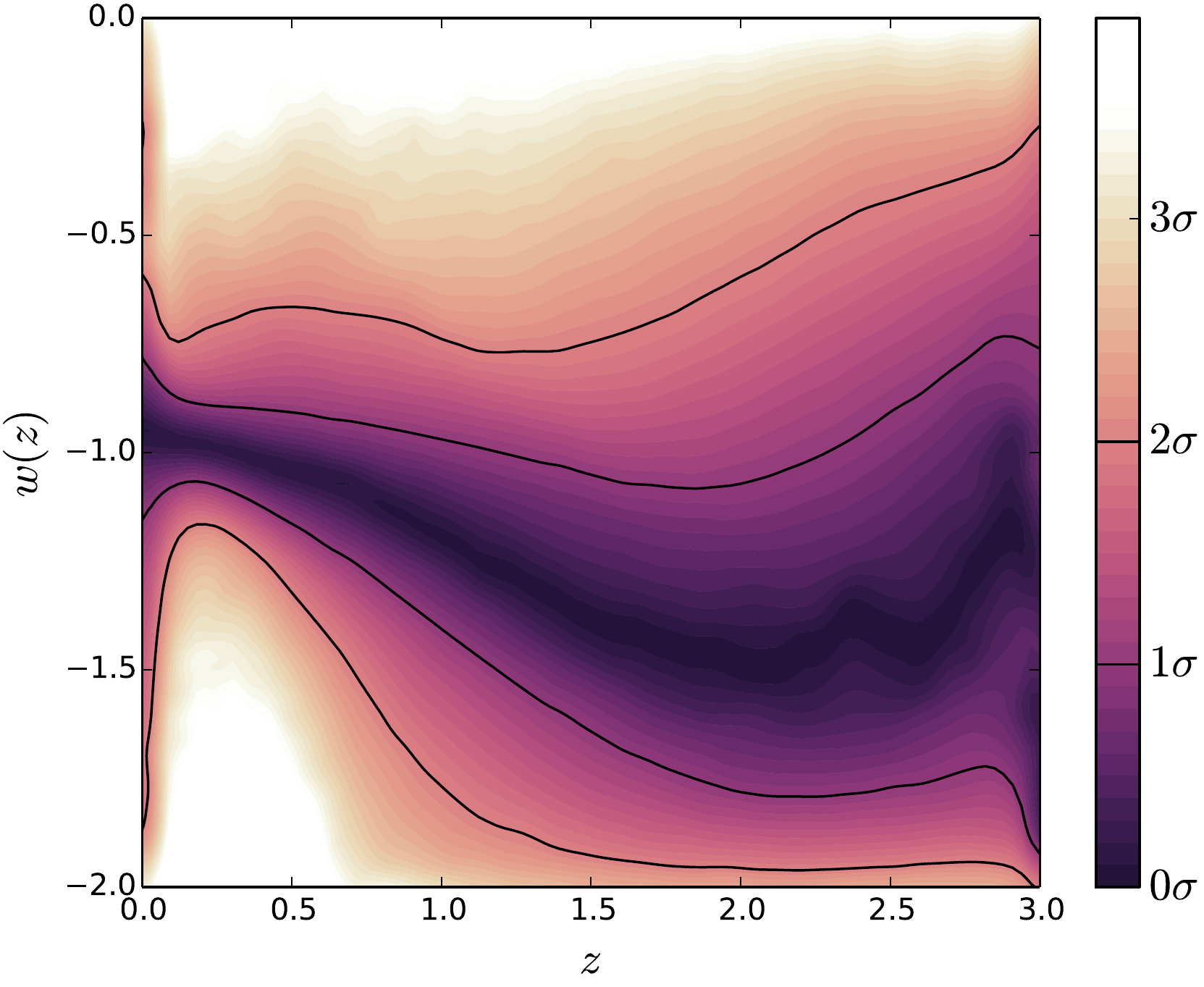}
  \end{minipage}
  \begin{minipage}{0.33\textwidth}
    \includegraphics[width=1.00\textwidth, height=0.32\textwidth]{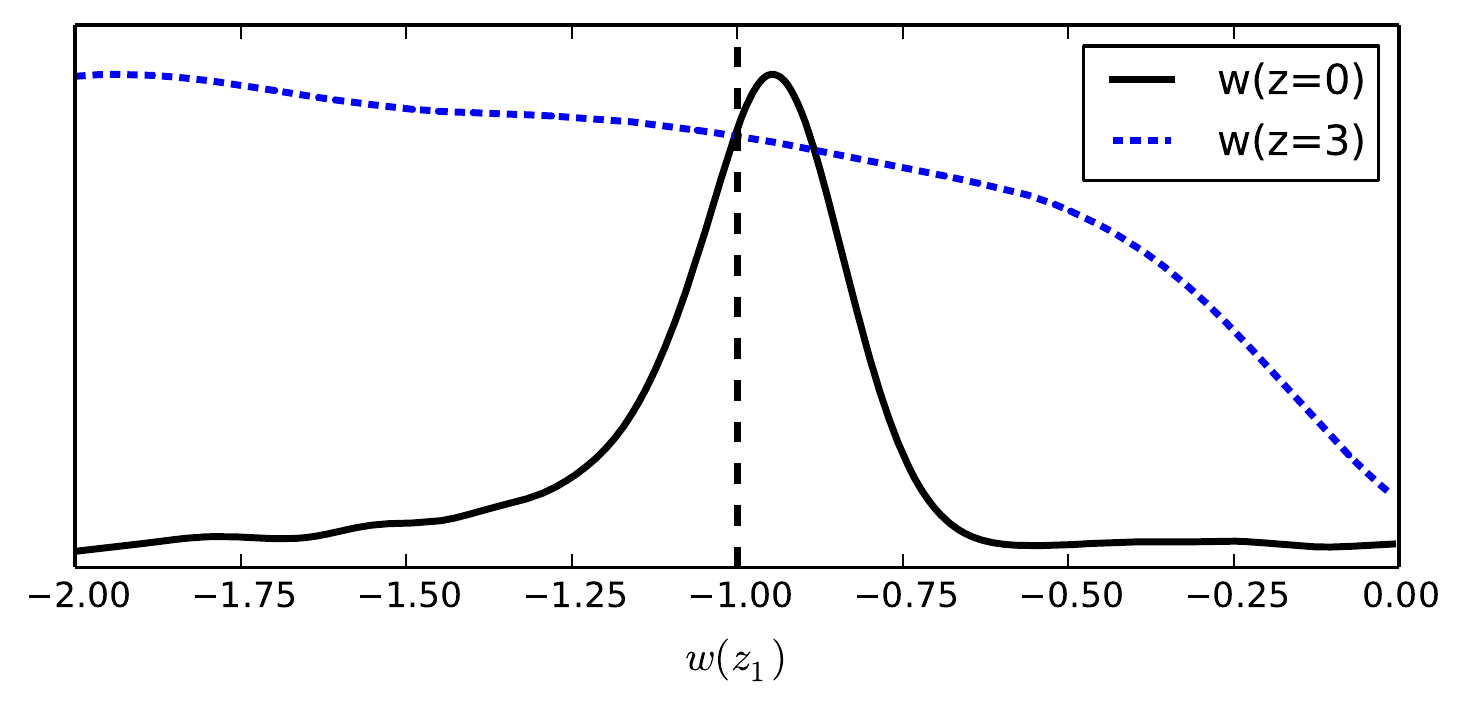}
    \includegraphics[width=0.32\textwidth, height=0.32\textwidth]{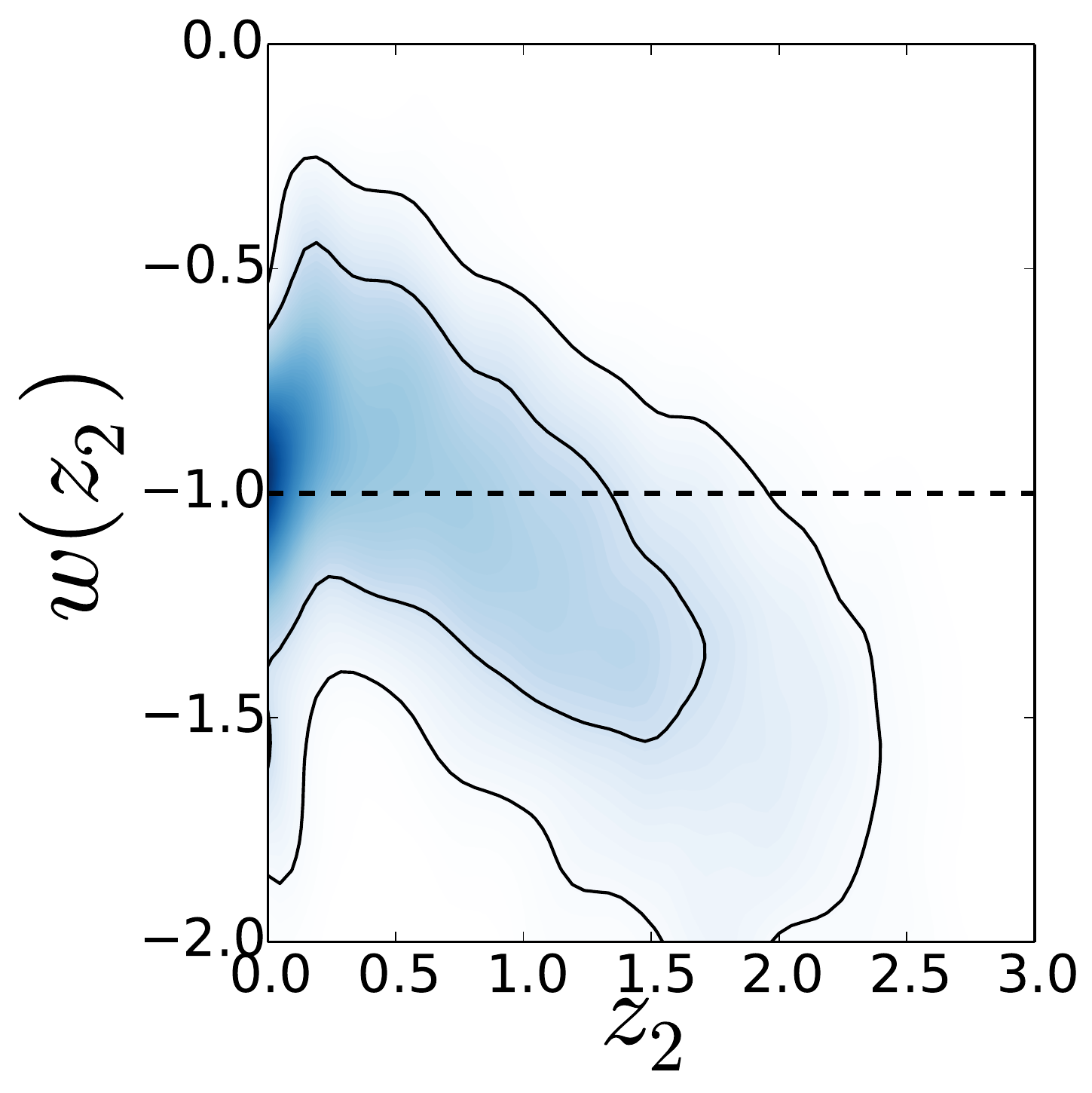}
    \includegraphics[width=0.32\textwidth, height=0.32\textwidth]{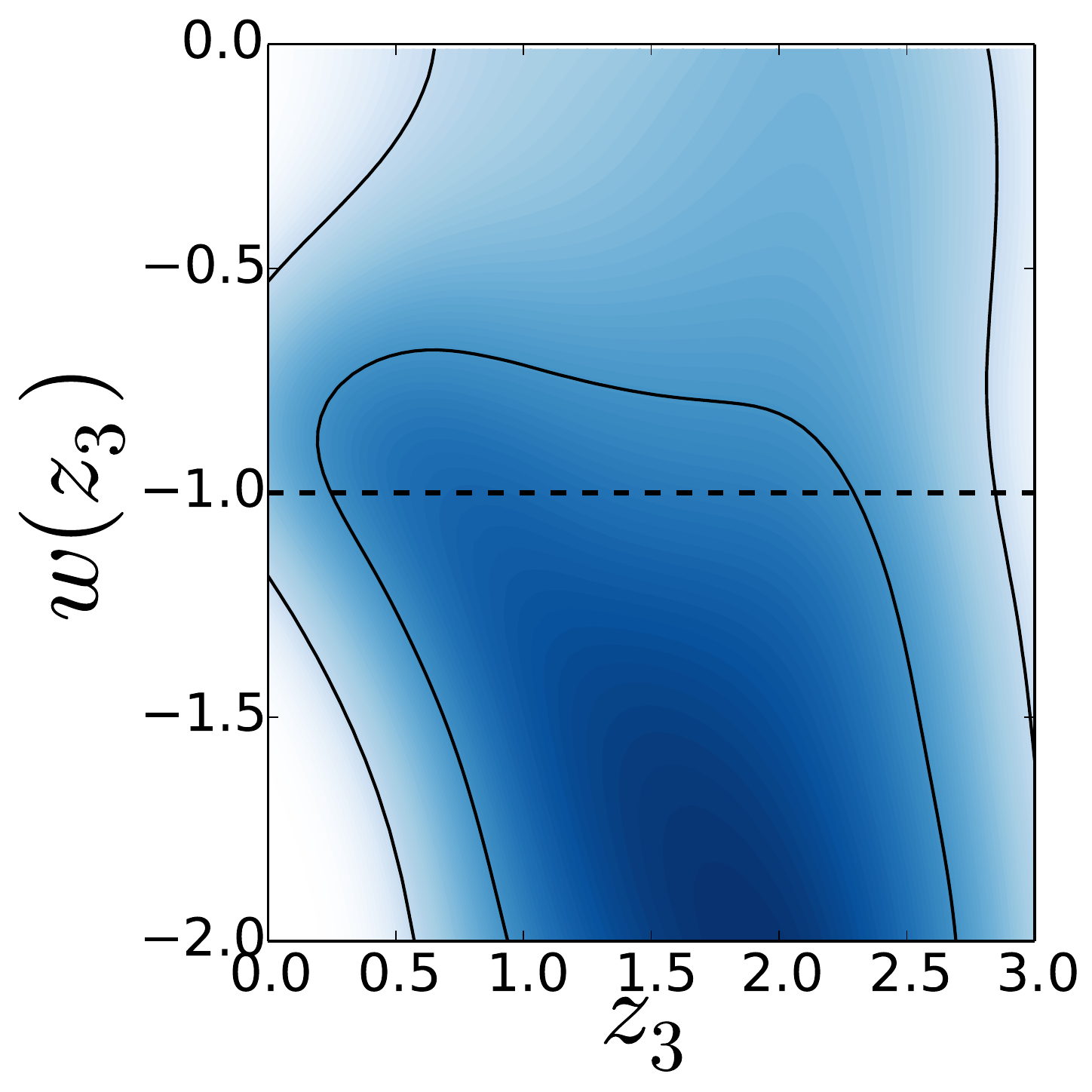}
    \includegraphics[width=0.32\textwidth, height=0.32\textwidth]{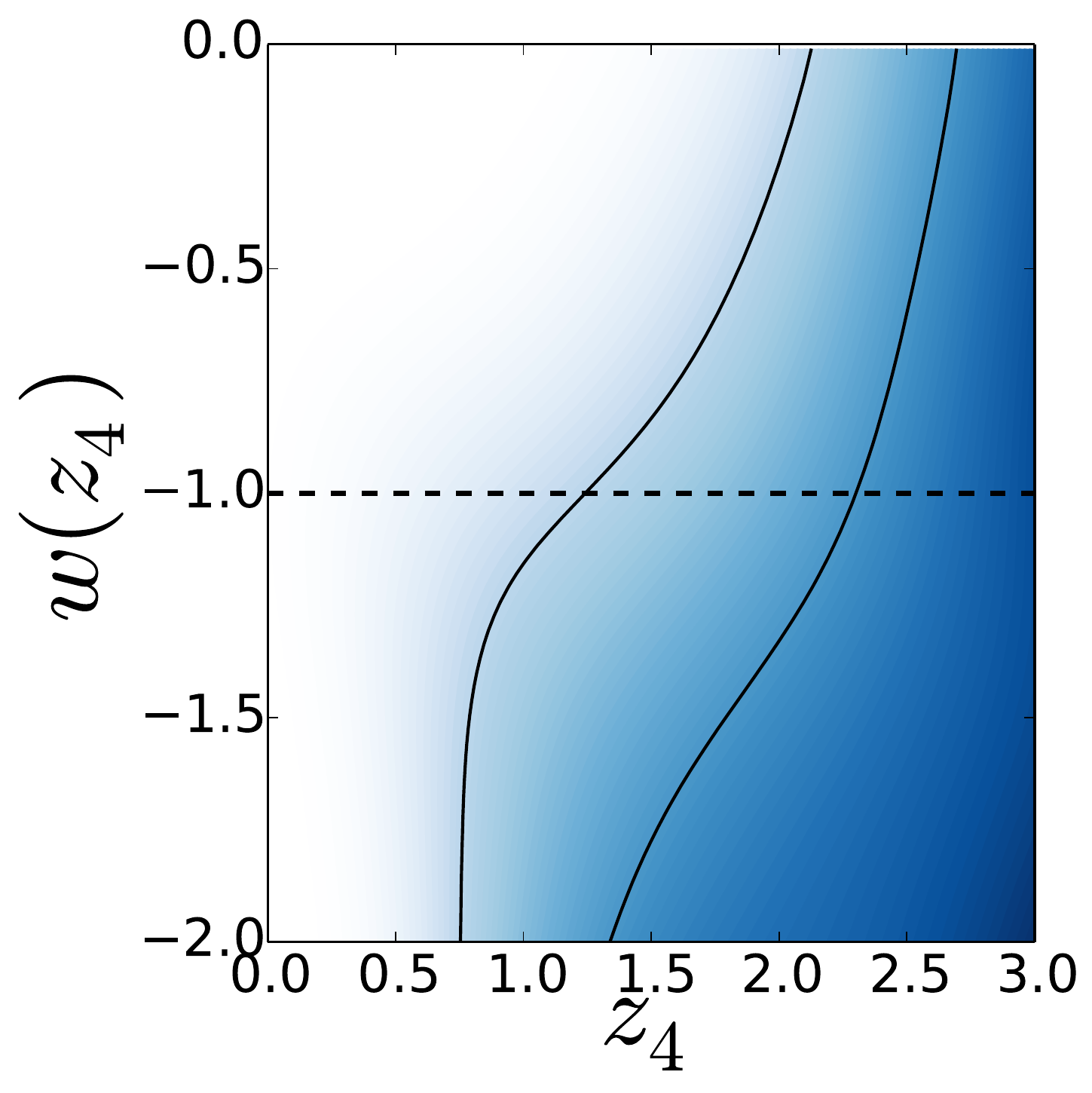}
  \end{minipage}

  \caption{The $w(z)$ priors, $w(z)$ reconstructions and parameter constraints for each of the 5 model extensions beyond $\Lambda$CDM\@. The leftmost plots are the prior space on the function $w(z)$ as a result of our uniform nodal reconstruction parameters and \codeF{CosmoMC}'s sampling, and the central plots show the constraints on $w(z)$ as a result of the data. These plots show the posterior probability $\Prob(w|z)$: the probability of $w$ as normalised in each slice of constant $z$, with colour scale in confidence interval values. The $1 \sigma$ and $2 \sigma$ confidence intervals are plotted as black lines. Note that the sigma-deviations are plotted assuming a central value such that a flat prior would not have a uniform colour, thus interpreting regions of the posterior space that are highly unconstrained is more difficult, such as when interpreting the lower bounds of $w$ at high redshifts. Reviewing priors we see a slight favouring in $w(z)$ of the central values, as expected when calculating priors analytically and given that \codeF{CosmoMC} restricts the permissible parameter space. The posteriors show that the data constrains $w(z)$ strongly compared to our priors. Rightmost are the 1D and 2D marginalised posteriors of the additional model parameters. Marginalised plots were produced using \codeF{GetDist} and $w(z)$ reconstructions were produced in python with the cubehelix colour scheme by~\protect\cite{Green2011} for linearity in grey scale.}
\label{fig:wz_FullData_analysis}
\end{figure*}

\begin{figure*}
  \centering
  \includegraphics[width=0.33\textwidth, height=0.19\textwidth]{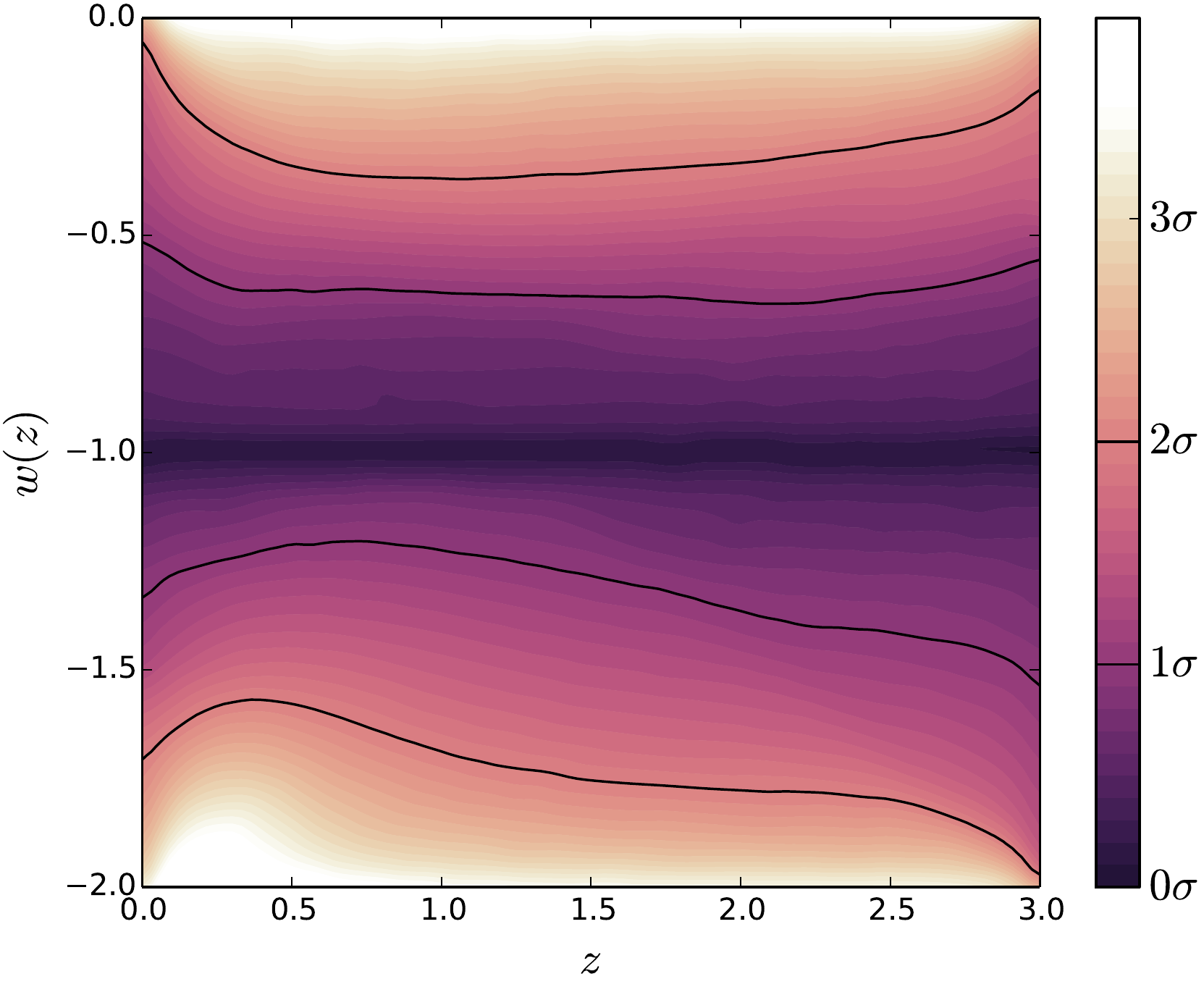}
  \includegraphics[width=0.33\textwidth, height=0.19\textwidth]{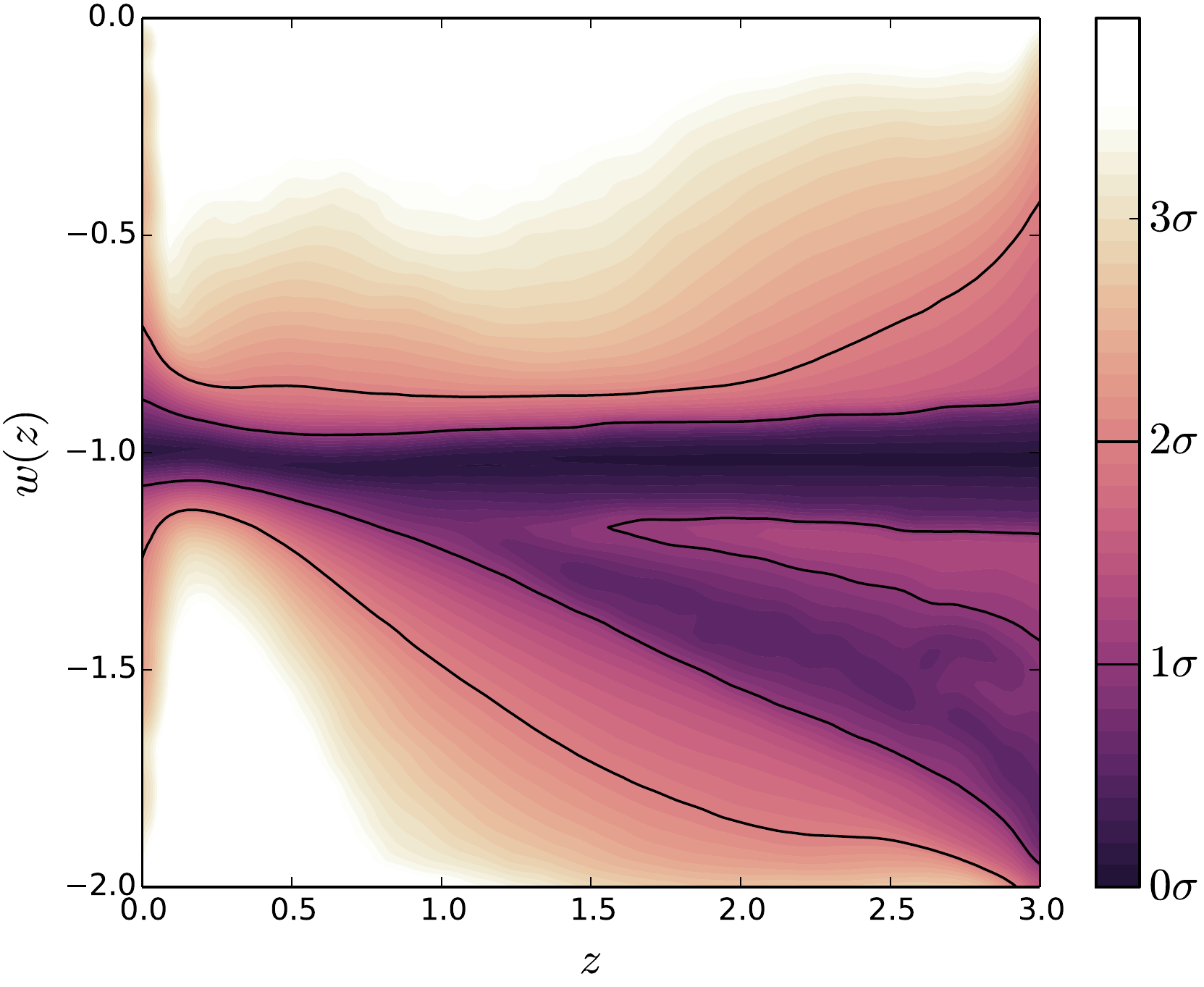}
  \includegraphics[width=0.33\textwidth, height=0.19\textwidth]{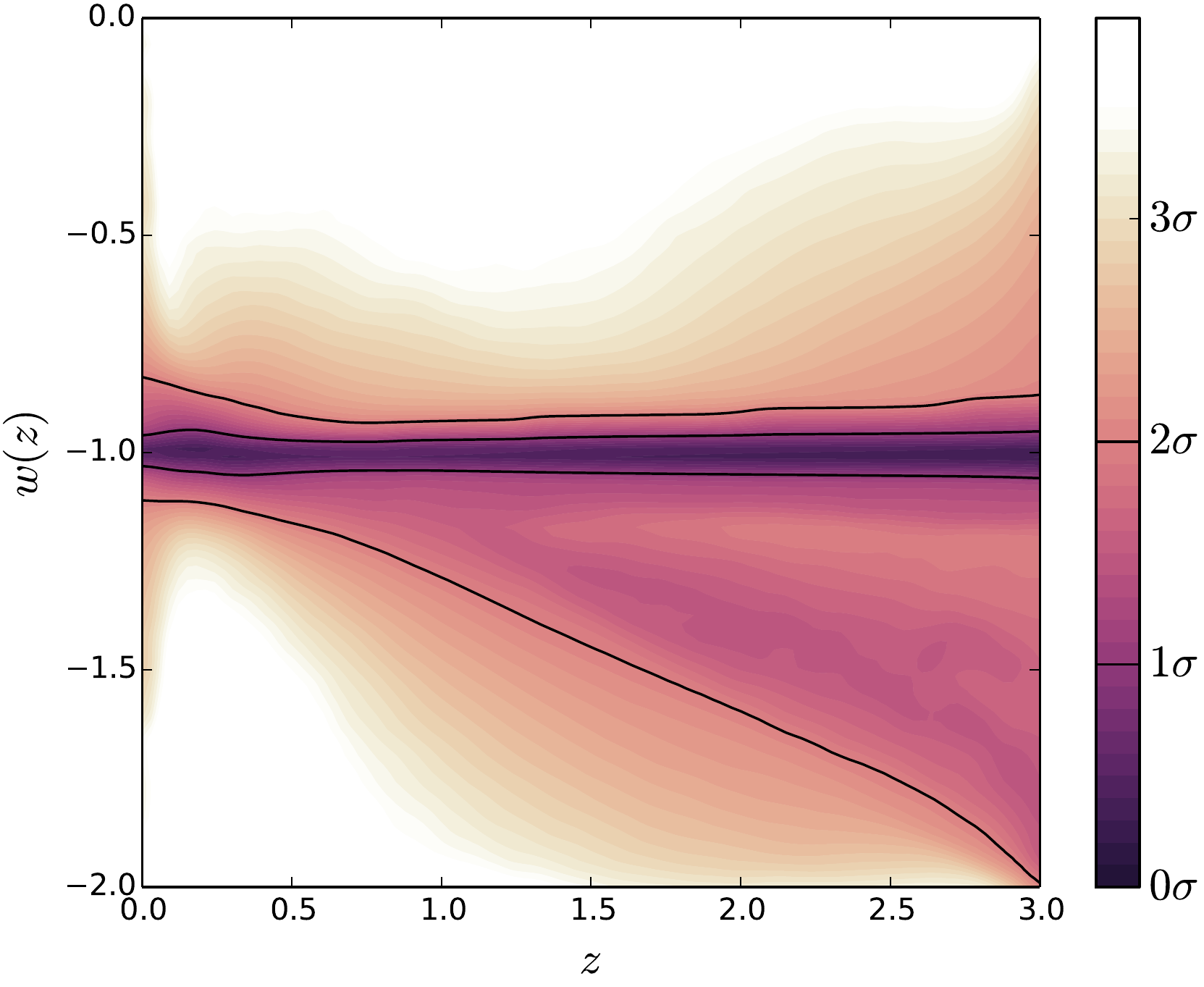}
  \caption{Summarising the DE model extension results for the constraints on the $w(z)$ plane. All models are weighted by their evidences to give a model averaged plane reconstruction~\citep{Parkinson2013,PlanckCollaboration2015_inflation,Hee2015}, and plotted as in Figure~\ref{fig:wz_FullData_analysis}. The three plots show the prior space (left) contracting down to the posterior odds ratio averaged $w(z)$ plane reconstruction for all of the model extensions beyond $\Lambda$CDM (middle) and for all of the models including $\Lambda$CDM (right). For the model extension averaged reconstruction it is clear that there is one solution around $w{=}{-}1$ and another favouring a supernegative equation of state. When including $\Lambda$CDM the significance of the supernegative solution wanes due to the associated large Bayes factor for the $w{=}{-}1$ equation.}
\label{fig:DE_finalPlane}
\end{figure*}

The columns in Figure~\ref{fig:wz_FullData_analysis} show from left to right the prior, posterior and marginalised 1D and 2D posteriors for the $w(z)$ plane reconstructions alongside the Bayes factors for the 5 model extensions compared to $\Lambda$CDM\@. $\Lambda$CDM is the favoured model in the Bayesian model selection analysis. $w$CDM is disfavoured by more than 2 log-units, a slight disfavouring on the Jeffreys scale, whilst all other models are significantly disfavoured at beyond 2.5 log-units. We conclude that the additional flexibility in capturing $w(z)$ features provided by additional parameters does not produce favourable Bayes factors. This is consistent with previous results obtained with Planck~\citep{Hee2015}. The systematic dropping off in Bayes factors for models with increasing numbers of parameters used for defining $w(z)$ suggests that there is not sufficient time dependence in the true equation of state function to overcome the Occam factor associated with the additional parameters~\citep{MacKay2003}. Specifically, one can estimate the evidence integral using a laplace approximation to obtain an Occam factor given by $\sigma_{\theta | \Data} / \sigma_{\theta}$~\citep[page349]{MacKay2003}~\citep{Hee2015}, where $\sigma_{\theta| \Data}$ is the width around the peak of the posterior and $\sigma_{\theta}$ is the same for the prior, and use this to determine the size of the Occam factor between models. When moving from $2$CDM to $3$CDM we obtain an Occam factor of approximately $0.72$, where we assume the posterior on the additional nodal position parameter is equal to the prior, as there is little additional structrual information, and have taken the average full width half max value of the five $3$CDM amplitude parameters to estimate the effect of adding the additional node (the prior is flat so $\sigma_{\theta}$ is the width, $2$). This shows that the observed Bayes factor drop of $0.54$ (with errors on order $0.29$) is comparable to the Occam factor and therefore the information gained from the additional node, which should compensate the effect, is small.

The plane reconstructions show clear constraining power compared to the priors. In all cases that allow for time dependence there is the suggestion that a supernegative equation of state fits the data best at higher redshifts. Specifically, the $t$CDM model deviates from $\Lambda$CDM by $1 \sigma$ already before $z{=}1$ whilst the models with internal nodes, which are able to identify more flexibly where deviations occur, suggest a $1 \sigma$ deviation around $z{=}1.5$. No model deviates at $2 \sigma$ however. It should also be noted that the tightest constraints on the EoS are around redshift $z{=}0.1{-}0.5$, and all models tend to $\Lambda$CDM in this region. This suggests that conclusions are still data limited but that time dependent behaviour of a supernegative EoS is hinted at by the combinations of $Planck+BAO+JLA+\Lya$.

We can look at the marginalised posteriors of nodes and amplitude parameters to gather further insights. Interestingly, the 1D marginalised posteriors on the $w(z{=}0)$ parameters of the models seem to favour $w{>}{-}1$, whilst the $w$CDM model does not specifically as the single amplitude parameter has simultaneously to model the late time behaviour. This suggests that using $w$CDM simplifies the dark energy problem in a way that can obscure underlying dark energy physics. Given that the difference in Bayes factors between $w$CDM and any of the more flexible models is indistinguishable on the Jeffreys scale, using a more flexible model is statistically valid and therefore advisable if wishing to analyse $w(z)$.

Looking at the 2D marginalised node positions in the $w(z)$ plane it is clear that in all cases the lowest redshift node is well defined as agreeing with $\Lambda$CDM\@. In $1$CDM, where there is only 1 internal node, the plane reconstruction takes a very similar form to $t$CDM as a result. For the $2$CDM and $3$CDM models, the additional nodes then have considerable freedom and the plane reconstruction shape at higher redshifts reflects this via a more constant value of $w$ from about redshift 2 onwards. The last node for both the $2$CDM and $3$CDM models is largely consistent with $\Lambda$CDM as the amplitude is poorly constrained beyond $z{=}2$, whilst in the range $1.5{<}z{<}2.0$ it deviates by $1 \sigma$, as consistent with the plane reconstructions. Generally, we conclude that all the amplitude parameters are in good agreement with $\Lambda$CDM, which is why the additional parameters do not add sufficient constraining power to generate Bayes factors that favour the models over $\Lambda$CDM\@.

Reviewing the model averaged plane reconstructions shown in Figure~\ref{fig:DE_finalPlane} we observe the conclusions noted above quite clearly in the bifurcation of probabilities on $w(z)$. In the central plot averaging over all models that allow for deviation from $\Lambda$CDM, a supernegative solution creates a second peak in the posterior of $w$ for $z{>}1.5$. As the reconstruction colour represents posteriors on $w$ in constant slices of $z$ measured by $\sigma$ confidence intervals with respect to the maximum, the dual peak structure defined by the $1 \sigma$ contour suggests that the data is sufficiently powerful to resolve a distinct supernegative solution. This supernegative structure is well within the $1 \sigma$ confidence intervals of the posterior distribution, fitting the data well, whilst $w(z){=}{-}1$ creates the peak probability that defines the $0 \sigma$ confidence interval. When including $\Lambda$CDM in the model averaging, to produce the right-hand plot, again the statistical significance and consistent identification of deviations away from $\Lambda$CDM in the reconstructions identifies the alternative supernegative equation of state structure. However, the significant Bayes factor favouring of $\Lambda$CDM ensures that the functional reconstruction heavily favours $w{=}{-}1$ for all redshifts. When including $\Lambda$CDM in the model averaging, we conclude that a supernegative equation of state fits the observed data at best to within the $1.5 \sigma$ confidence interval. It should be noted that the model averaging has been done over 4 models with very similar features identified, which no doubt adds to the strength of the bifurcation when averaging.

\section{Results: Kullback-Leibler divergence and dataset analysis}
\label{sec:dkl}

\begin{figure*}
  \centering
  \begin{minipage}{0.22\textwidth}
    \centering
    $P - 0.42 \: (0.33)$\\ 
    \includegraphics[width=1\textwidth, height=0.5\textwidth]{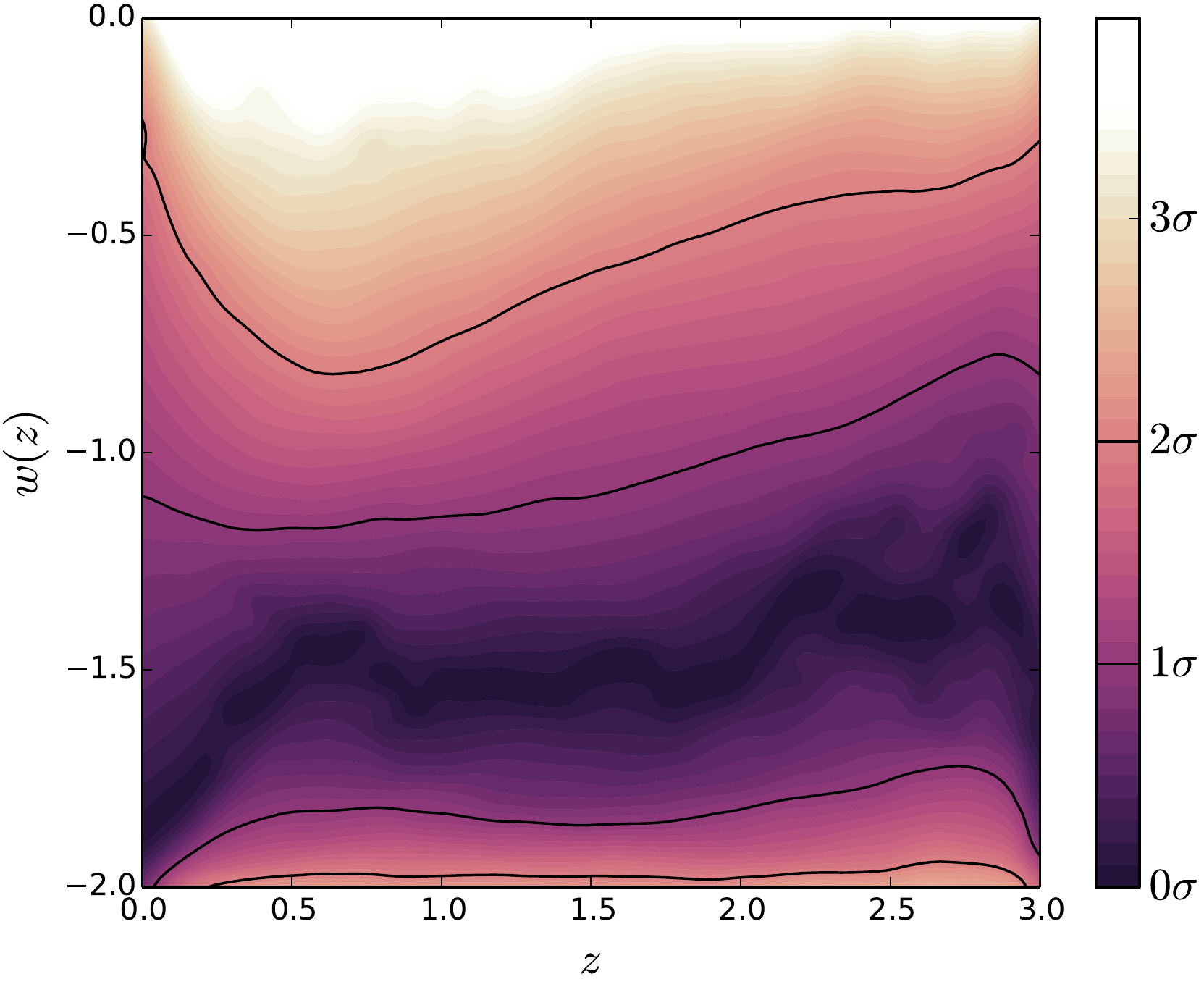}
  \end{minipage}
  \begin{minipage}{0.22\textwidth}
    \centering
    $Pa - 0.60 \: (0.45)$\\ 
    \includegraphics[width=1\textwidth, height=0.5\textwidth]{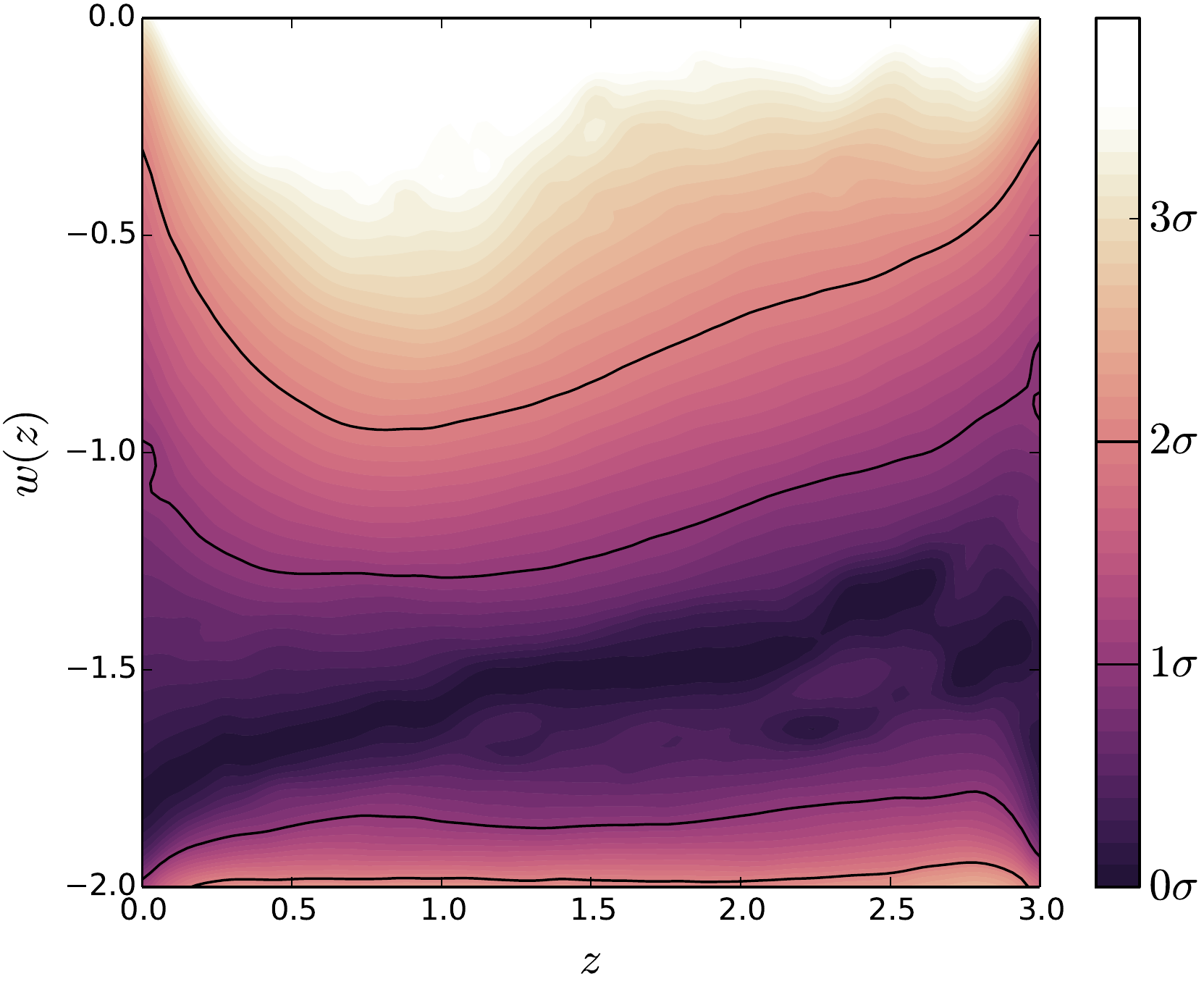}
  \end{minipage}
  \begin{minipage}{0.22\textwidth}
    \centering
    $Pb - 0.59 \: (0.42)$\\ 
    \includegraphics[width=1\textwidth, height=0.5\textwidth]{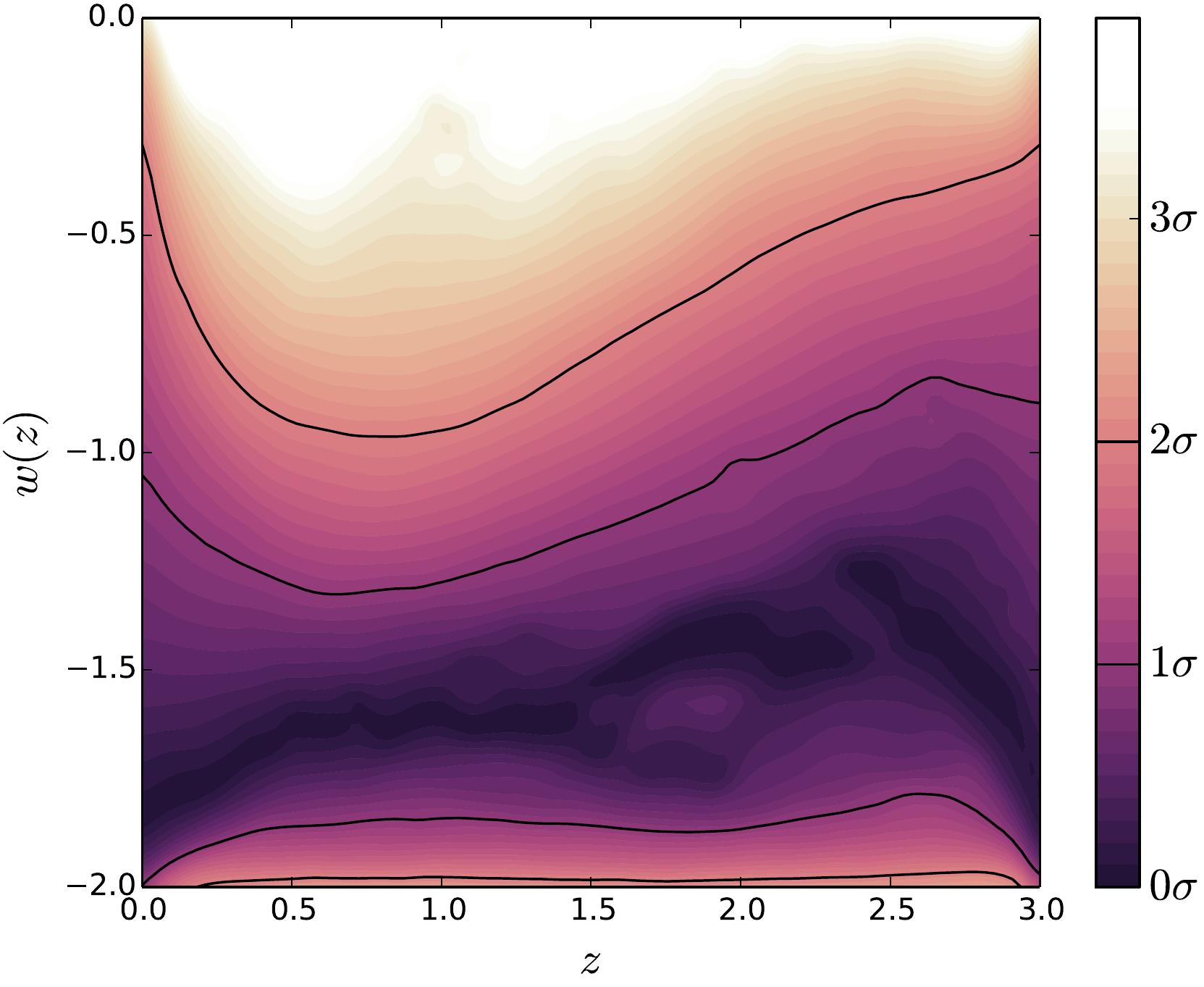}
  \end{minipage}
  \begin{minipage}{0.22\textwidth}
    \centering
    $Pab - 0.70 \: (0.47)$\\ 
    \includegraphics[width=1\textwidth, height=0.5\textwidth]{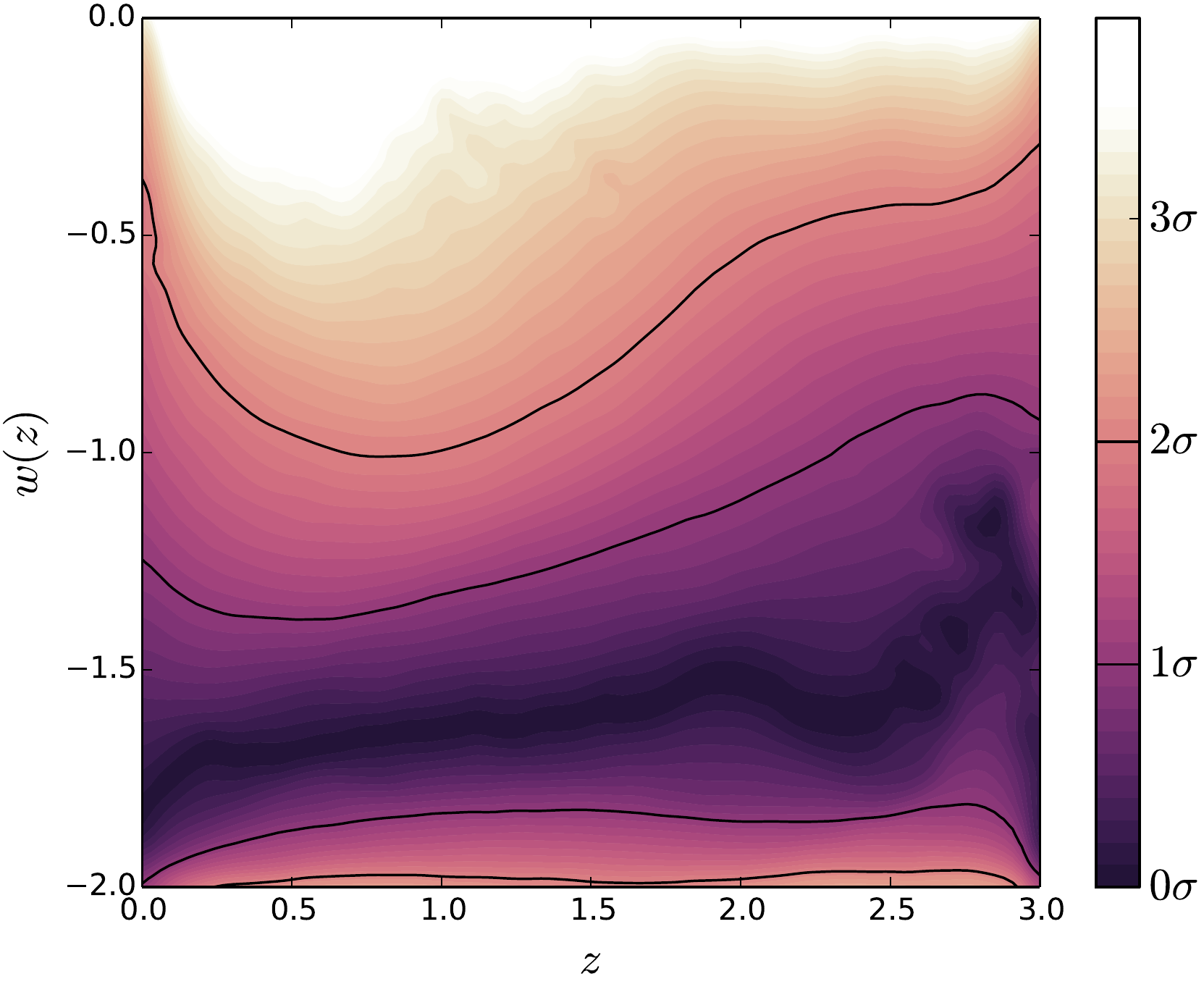}
  \end{minipage}
  \begin{minipage}{0.22\textwidth}
    \centering
    $PB - 0.35 \: (0.64)$\\ 
    \includegraphics[width=1\textwidth, height=0.5\textwidth]{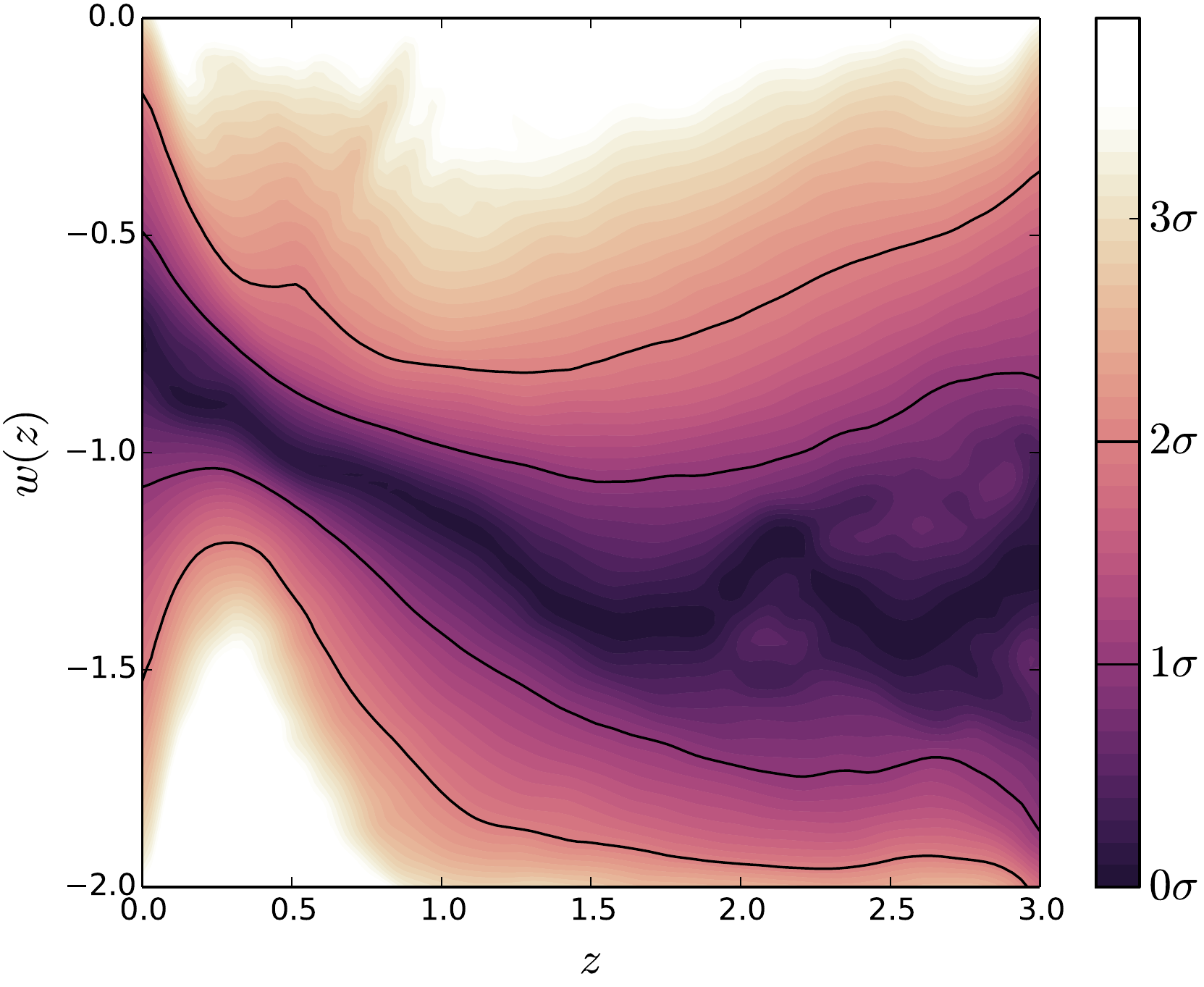}
  \end{minipage}
  \begin{minipage}{0.22\textwidth}
    \centering
    $PaB - 0.39 \: (0.67)$\\ 
    \includegraphics[width=1\textwidth, height=0.5\textwidth]{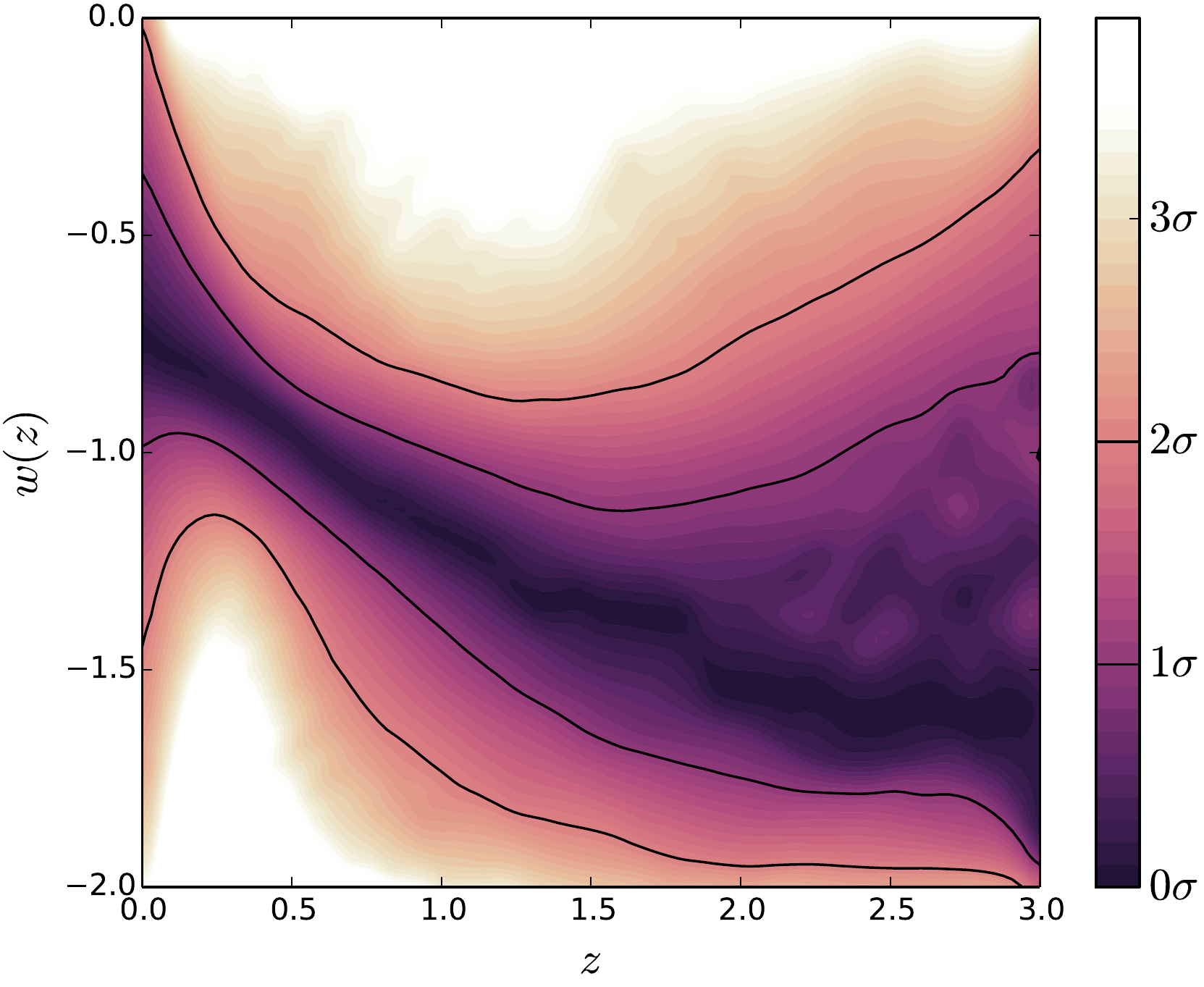}
  \end{minipage}
  \begin{minipage}{0.22\textwidth}
    \centering
    $PbB - 0.42 \: (0.65)$\\ 
    \includegraphics[width=1\textwidth, height=0.5\textwidth]{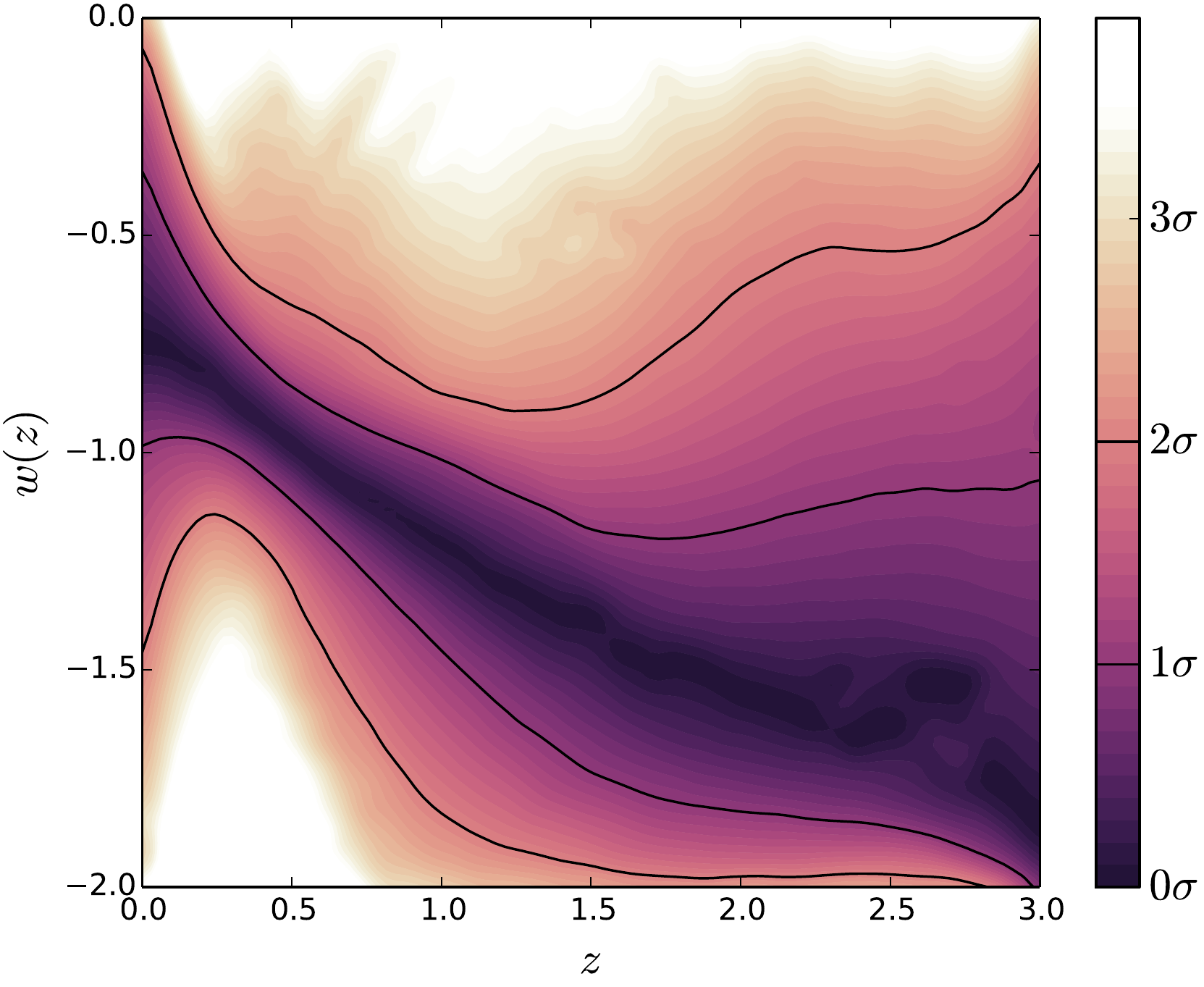}
  \end{minipage}
  \begin{minipage}{0.22\textwidth}
    \centering
    $PabB - 0.47 \: (0.73)$\\ 
    \includegraphics[width=1\textwidth, height=0.5\textwidth]{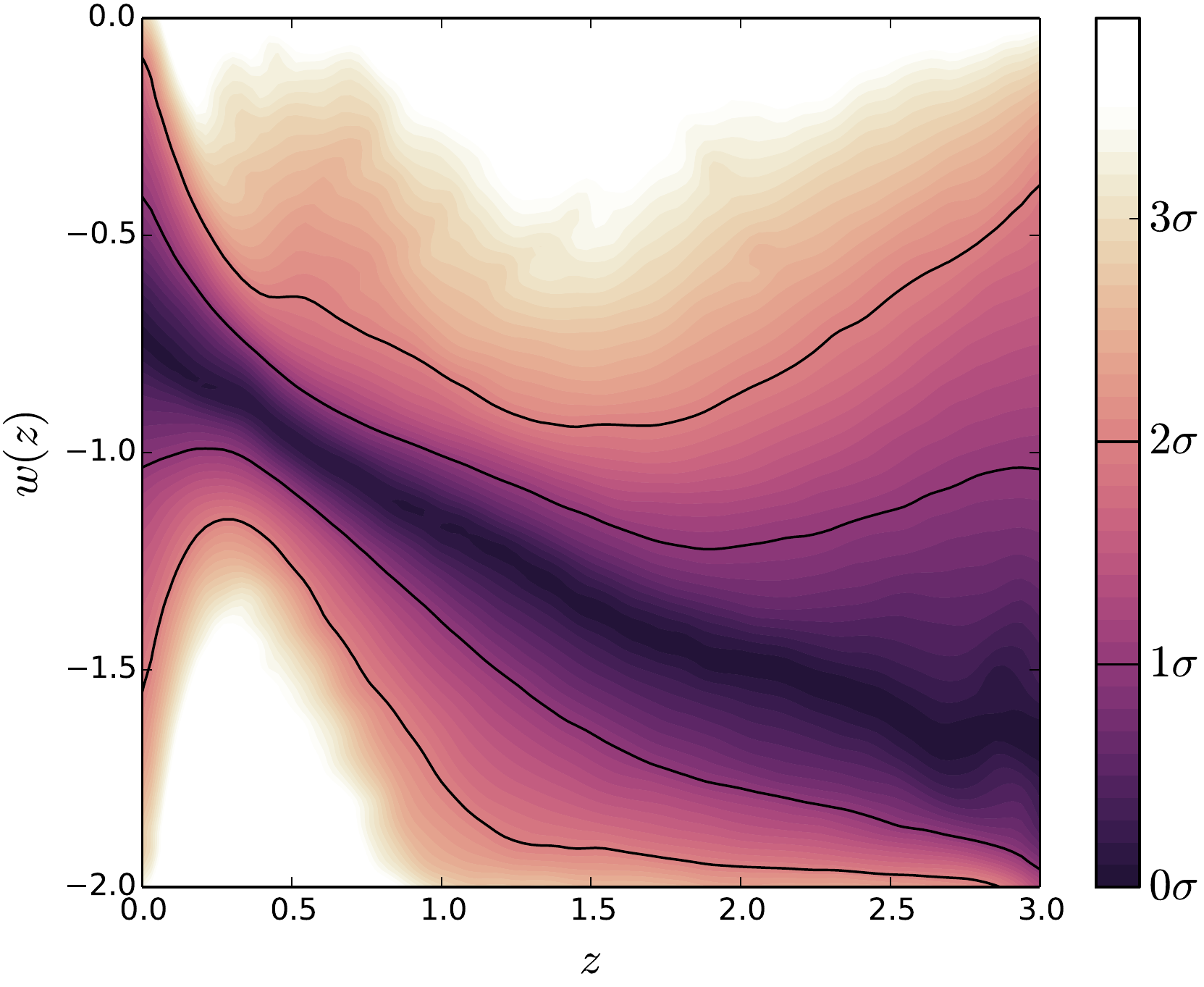}
  \end{minipage}
  \begin{minipage}{0.22\textwidth}
    \centering
    $PS - 0.43 \: (0.63)$\\ 
    \includegraphics[width=1\textwidth, height=0.5\textwidth]{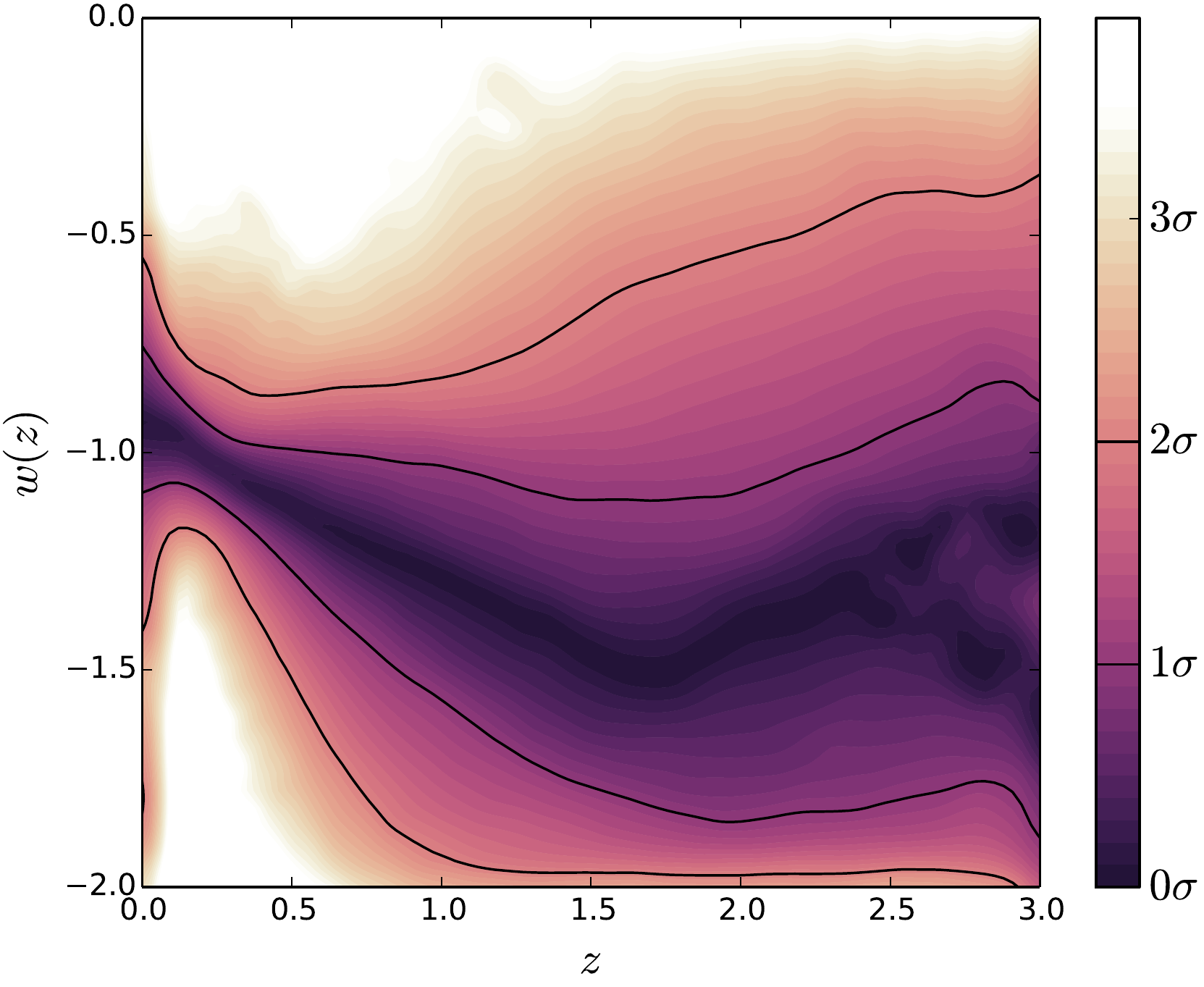}
  \end{minipage}
  \begin{minipage}{0.22\textwidth}
    \centering
    $PaS - 0.45 \: (0.64)$\\ 
    \includegraphics[width=1\textwidth, height=0.5\textwidth]{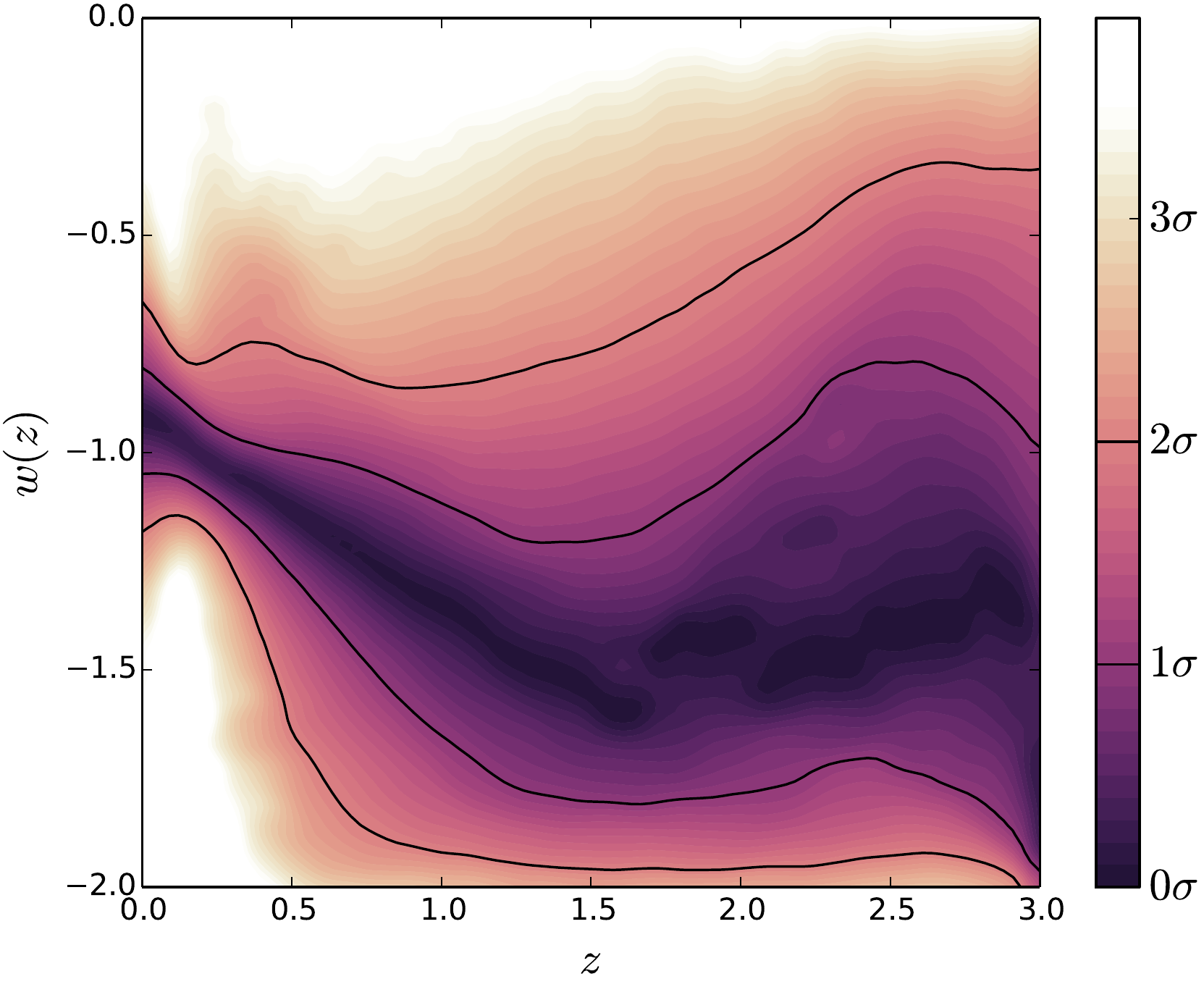}
  \end{minipage}
  \begin{minipage}{0.22\textwidth}
    \centering
    $PbS - 0.61 \: (0.74)$\\ 
    \includegraphics[width=1\textwidth, height=0.5\textwidth]{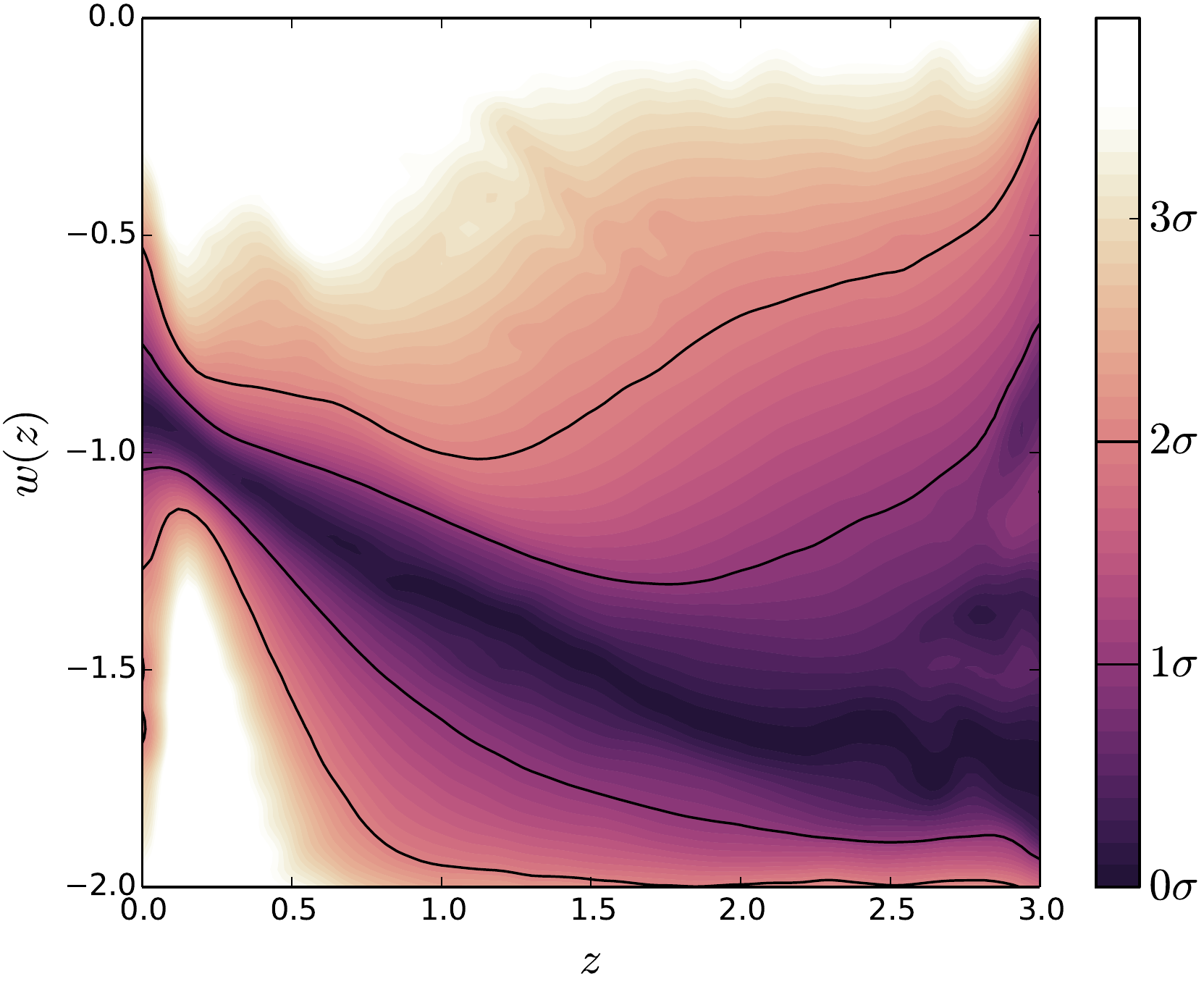}
  \end{minipage}
  \begin{minipage}{0.22\textwidth}
    \centering
    $PabS - 0.59 \: (0.73)$\\ 
    \includegraphics[width=1\textwidth, height=0.5\textwidth]{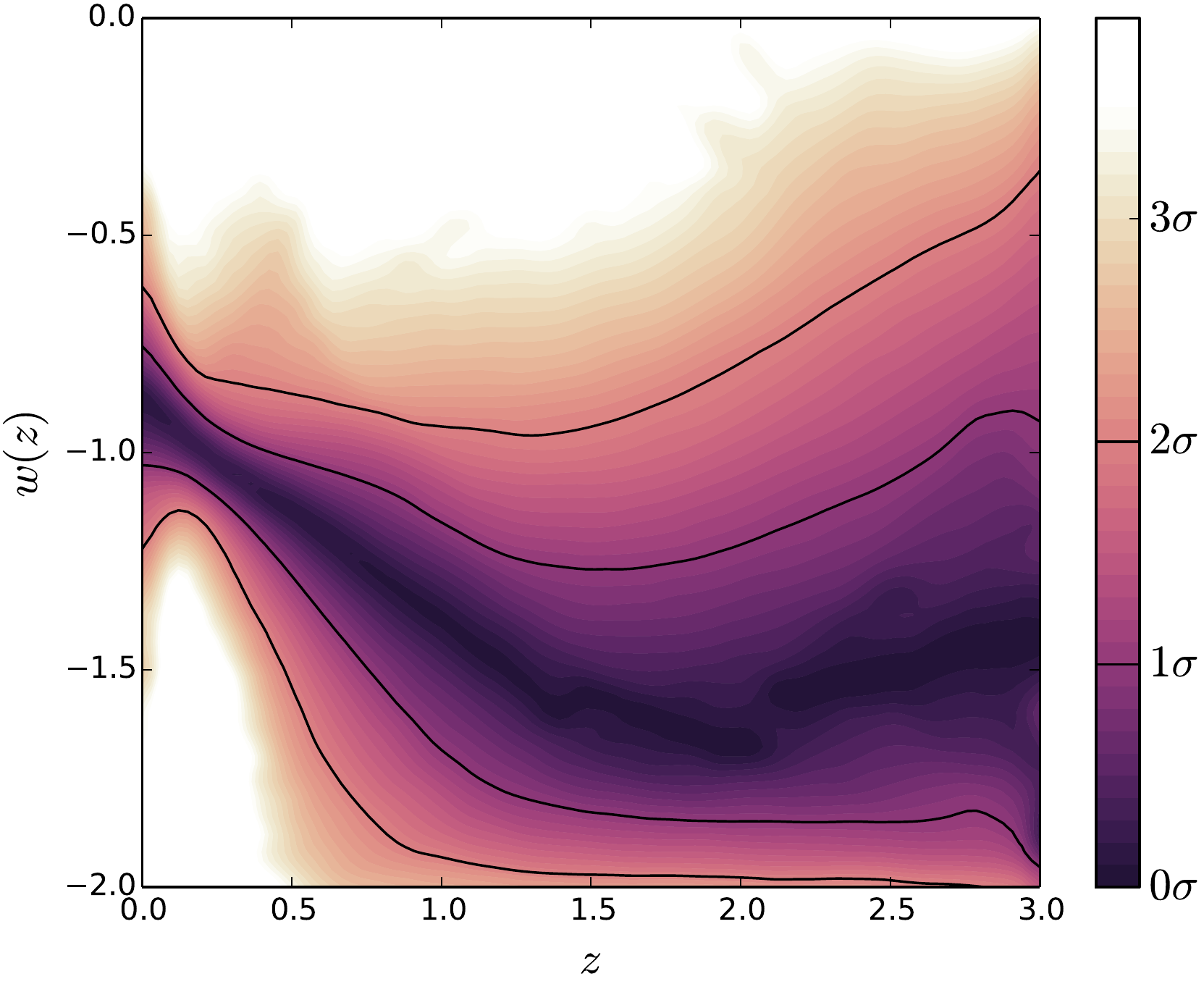}
  \end{minipage}
  \begin{minipage}{0.22\textwidth}
    \centering
    $PBS - 0.46 \: (0.82)$\\ 
    \includegraphics[width=1\textwidth, height=0.5\textwidth]{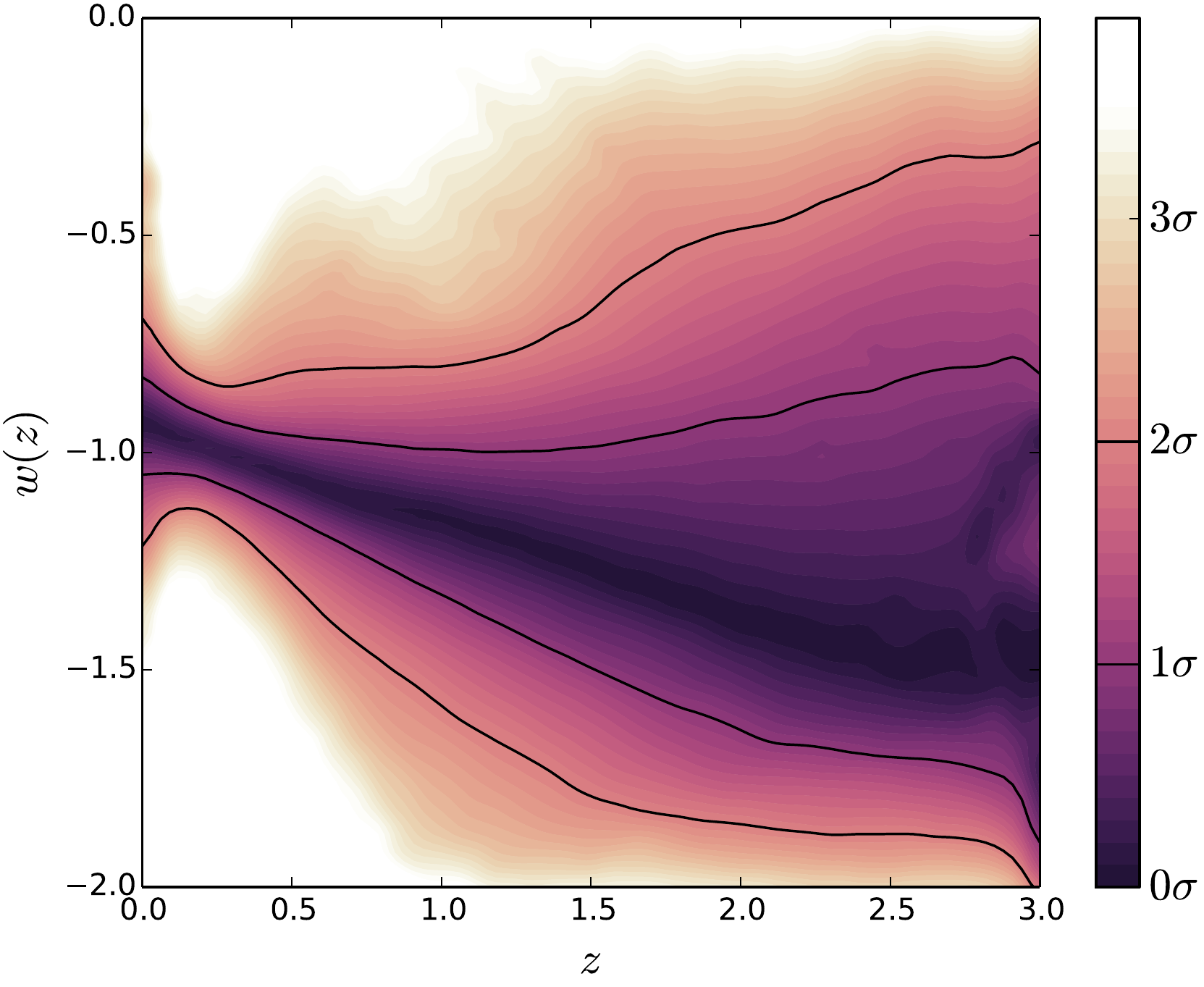}
  \end{minipage}
  \begin{minipage}{0.22\textwidth}
    \centering
    $PaBS - 0.46 \: (0.83)$\\ 
    \includegraphics[width=1\textwidth, height=0.5\textwidth]{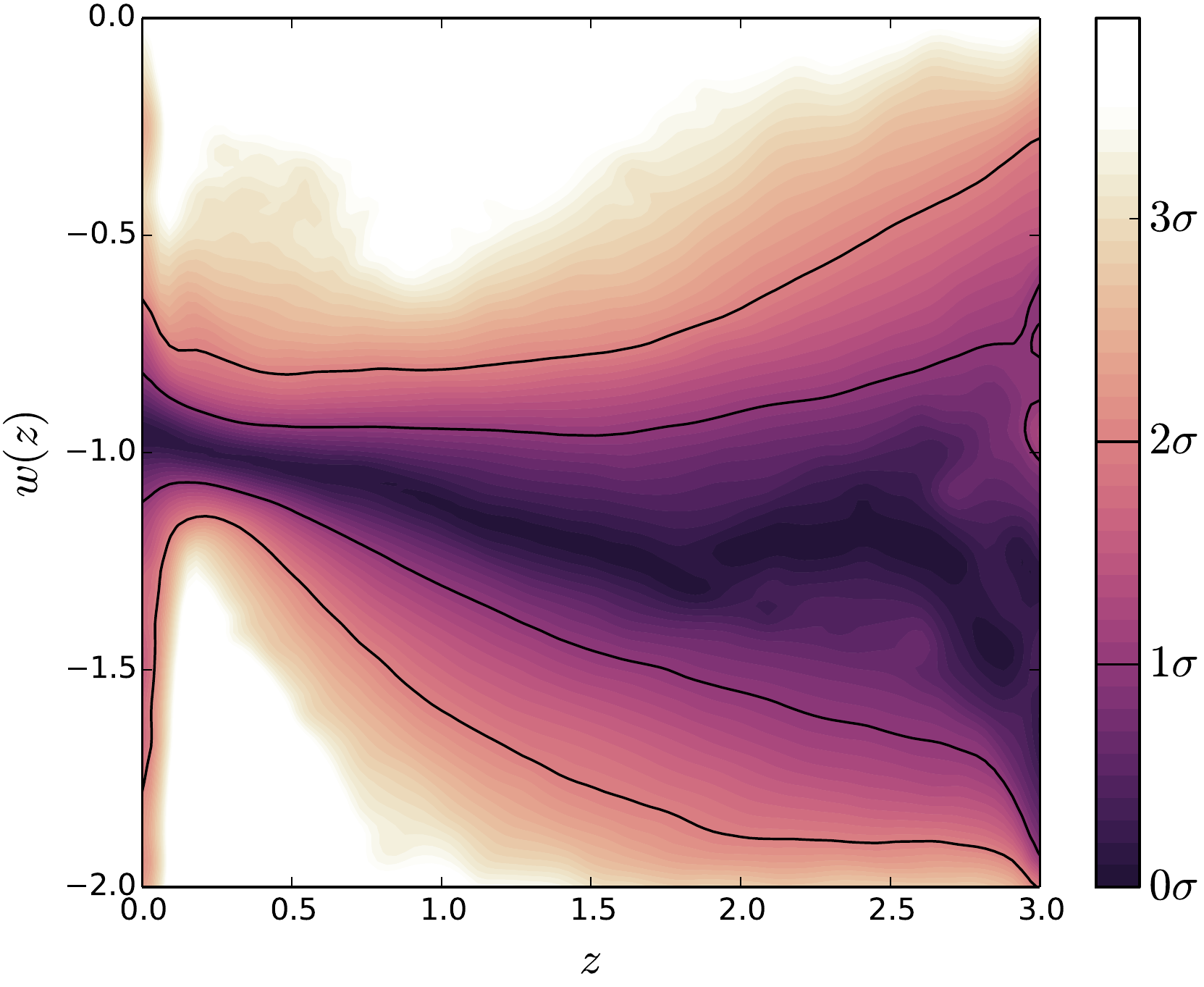}
  \end{minipage}
  \begin{minipage}{0.22\textwidth}
    \centering
    $PbBS - 0.51 \: (0.84)$\\ 
    \includegraphics[width=1\textwidth, height=0.5\textwidth]{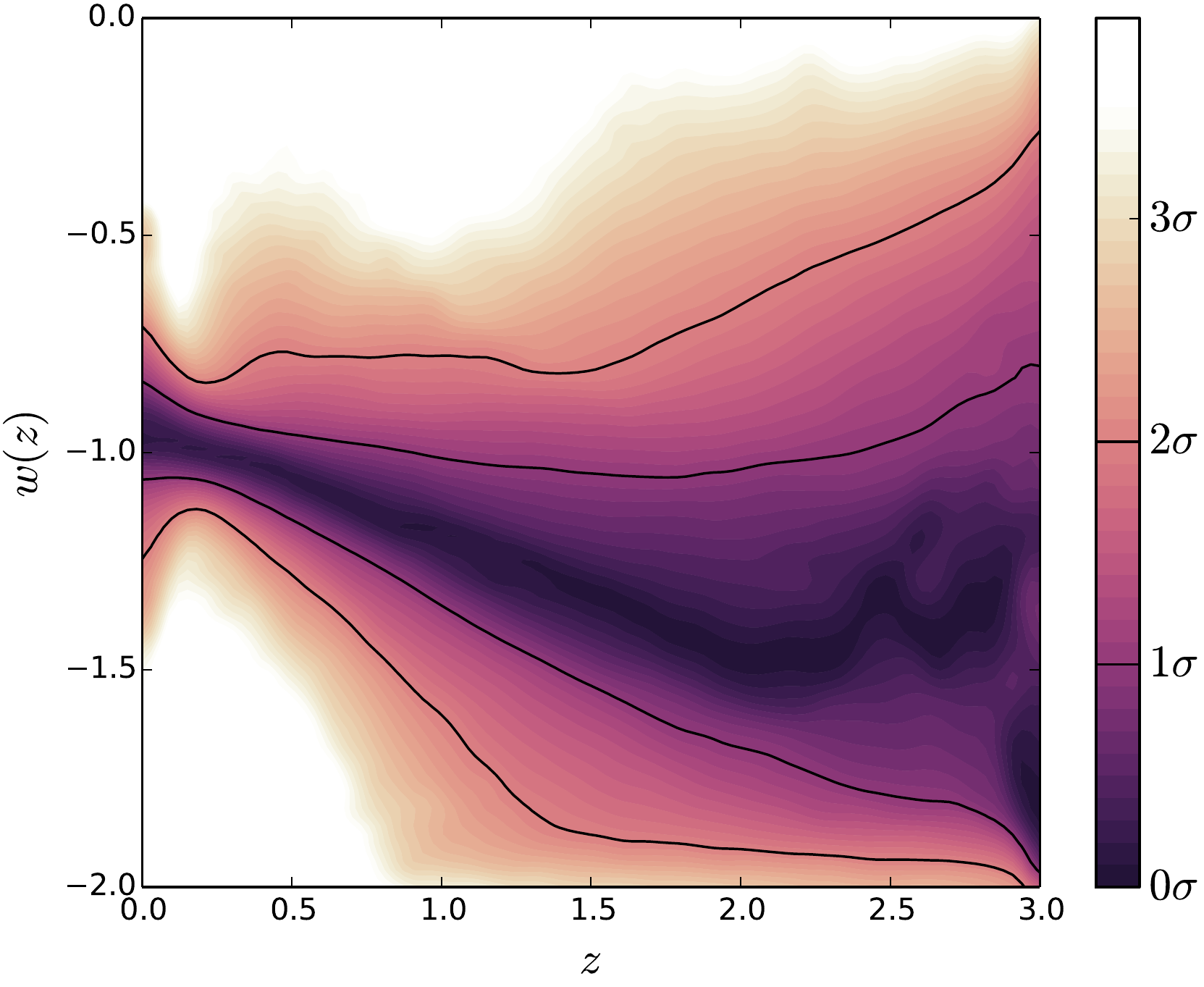}
  \end{minipage}
  \begin{minipage}{0.22\textwidth}
    \centering
    $PabBS - 0.57 \: (0.91)$\\ 
    \includegraphics[width=1\textwidth, height=0.5\textwidth]{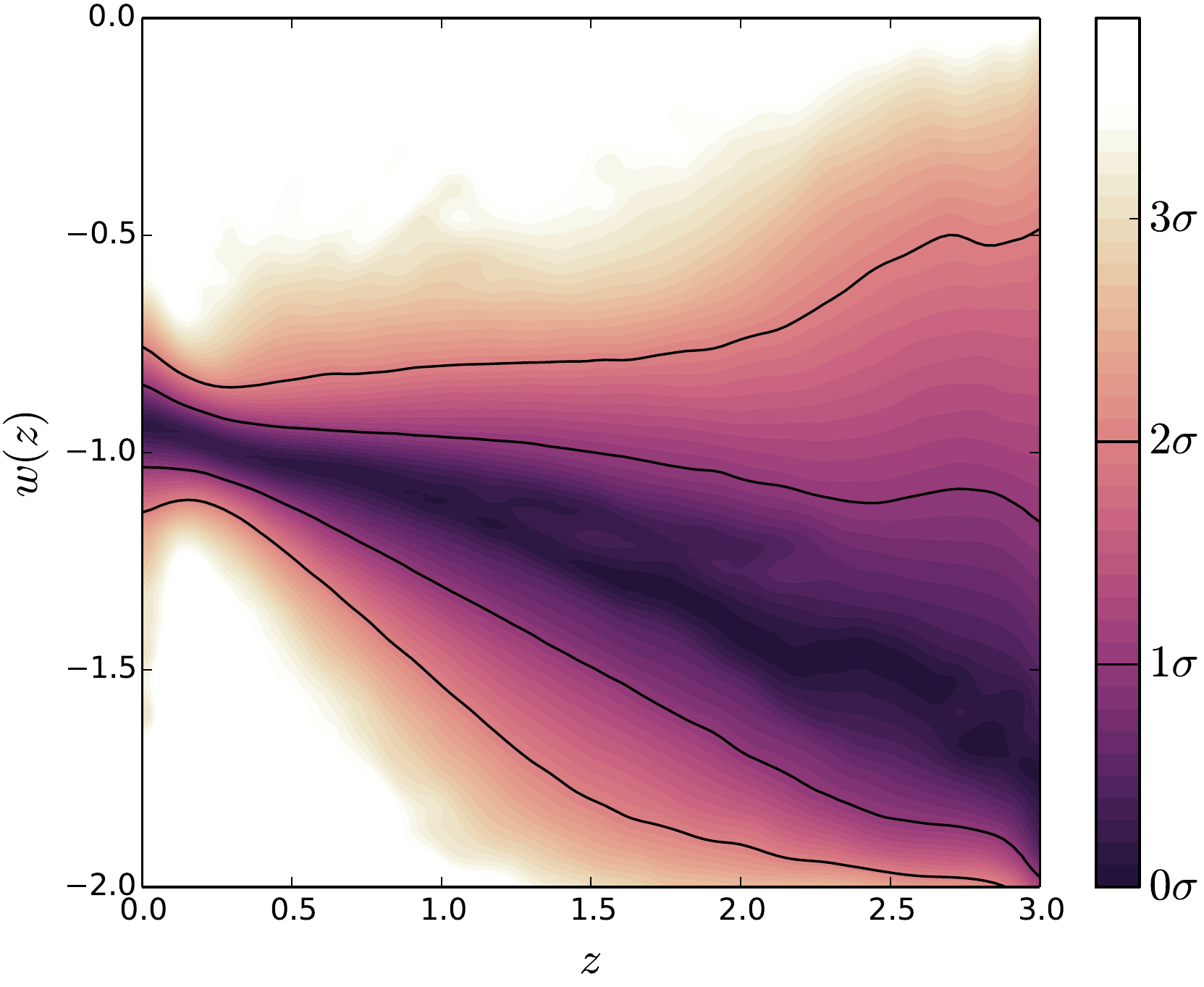}
  \end{minipage}
  \caption{Plane reconstructions of $w(z)$ using the $2$CDM model for Planck data with each possible combination of the $\LyAndreu$, $\LyBusca$, $BAO$ and $JLA$ datasets (abbreviated to $P$, $a$, $b$, $B$ and $S$ respectively). Results are laid out in a grid with columns of $\Lya$ combinations (without any, with $a$, with $b$, and with both) against rows of $BAO$ and $JLA$ combinations (without either, $B$, $S$, and both). $\DKL$ values for the $w(z)$ plane reconstructions, from $2$CDM prior to each given posterior, are stated next to each dataset combination to quantify the information gained when moving from prior to posterior due to the data. In brackets are the $\DKL$ values when moving from a flat reconstruction to the posterior, which capture the overall constraining of the posterior whilst ignore any shifts between prior and posterior peaks. Reviewing each row from left to right shows that the $\Lya$ datasets add only some constraining power, whilst reviewing each column from top to bottom shows that $BAO$ and $JLA$ datasets are both numerically and graphically significant.}
\label{fig:DKL_all}
\end{figure*}

\begin{figure*}
  \centering
  \begin{minipage}{0.22\textwidth}
    \centering
    $P$\\ 
    \includegraphics[width=1\textwidth, height=0.5\textwidth]{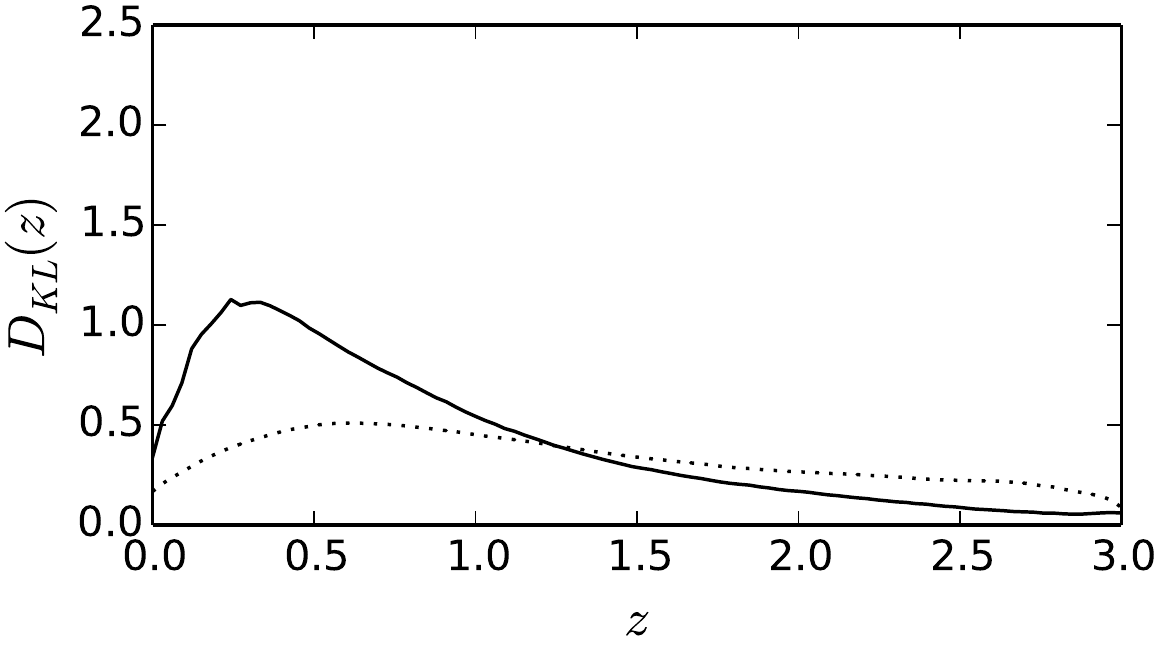}
  \end{minipage}
  \begin{minipage}{0.22\textwidth}
    \centering
    $Pa$\\ 
    \includegraphics[width=1\textwidth, height=0.5\textwidth]{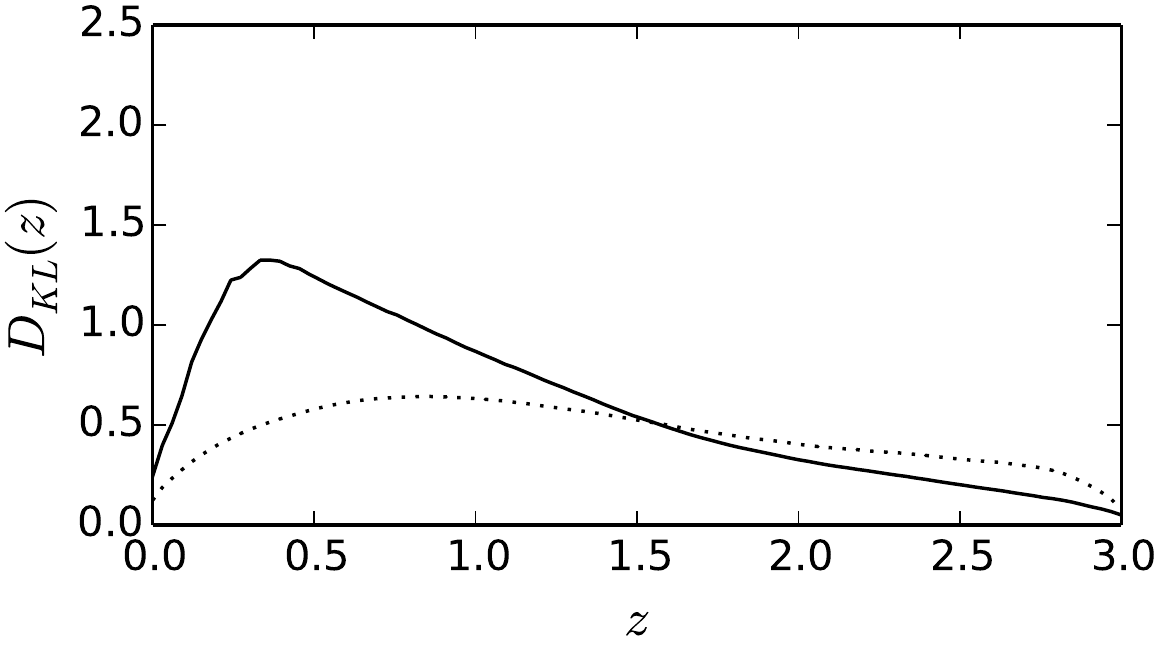}
  \end{minipage}
  \begin{minipage}{0.22\textwidth}
    \centering
    $Pb$\\ 
    \includegraphics[width=1\textwidth, height=0.5\textwidth]{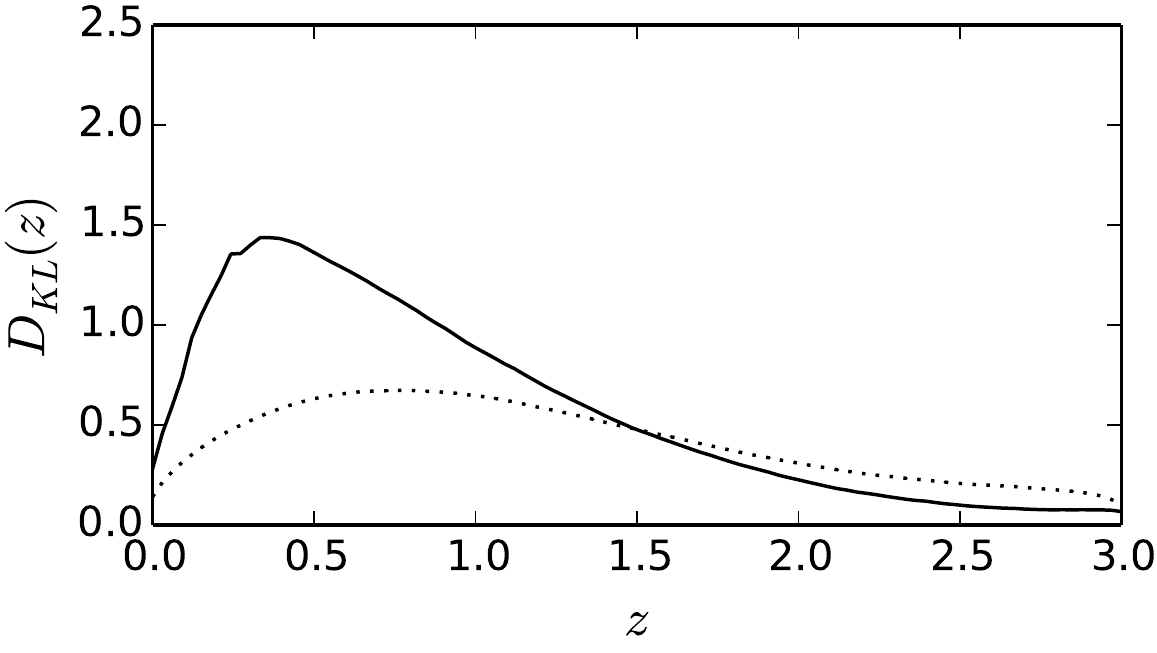}
  \end{minipage}
  \begin{minipage}{0.22\textwidth}
    \centering
    $Pab$\\ 
    \includegraphics[width=1\textwidth, height=0.5\textwidth]{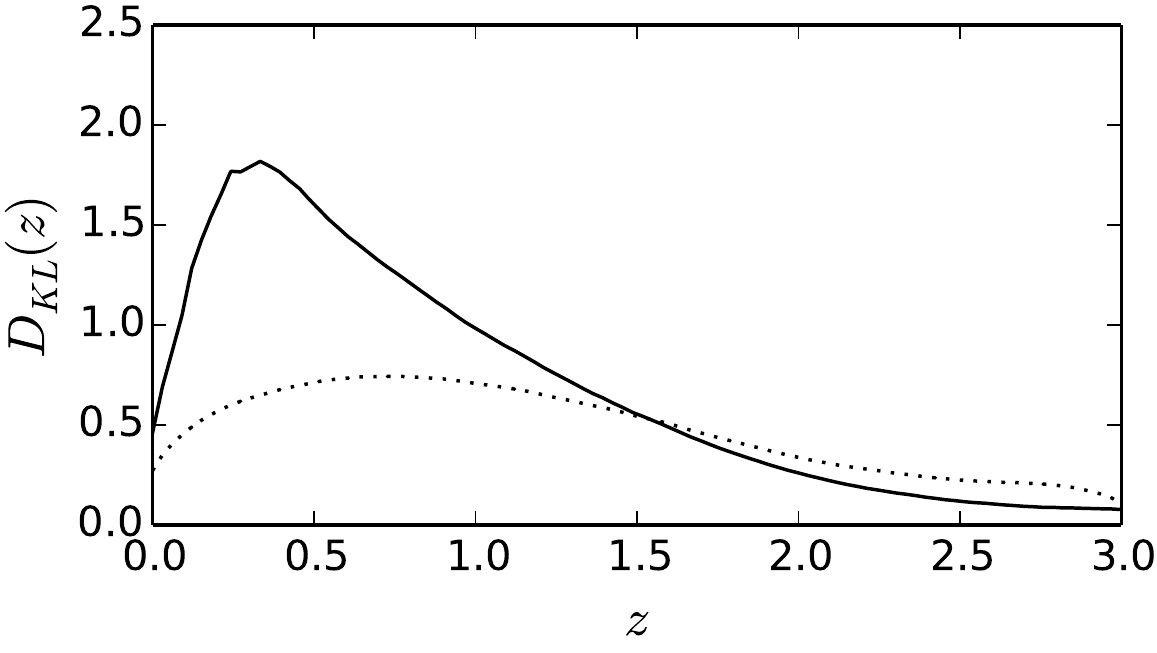}
  \end{minipage}
  \begin{minipage}{0.22\textwidth}
    \centering
    $PB$\\ 
    \includegraphics[width=1\textwidth, height=0.5\textwidth]{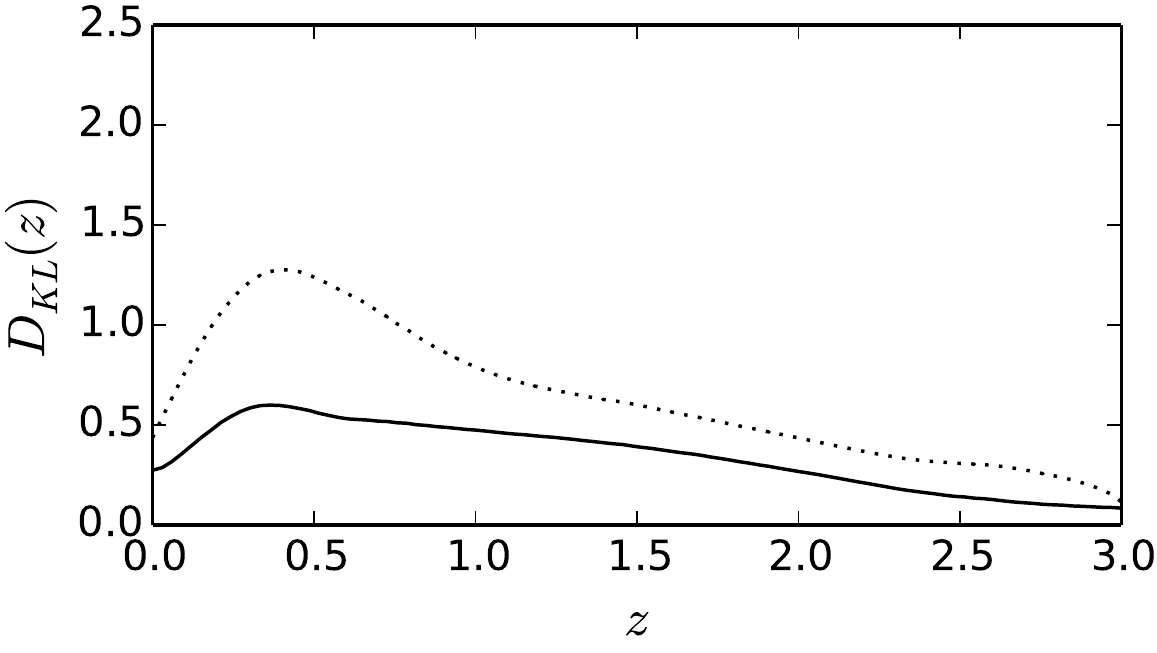}
  \end{minipage}
  \begin{minipage}{0.22\textwidth}
    \centering
    $PaB$\\ 
    \includegraphics[width=1\textwidth, height=0.5\textwidth]{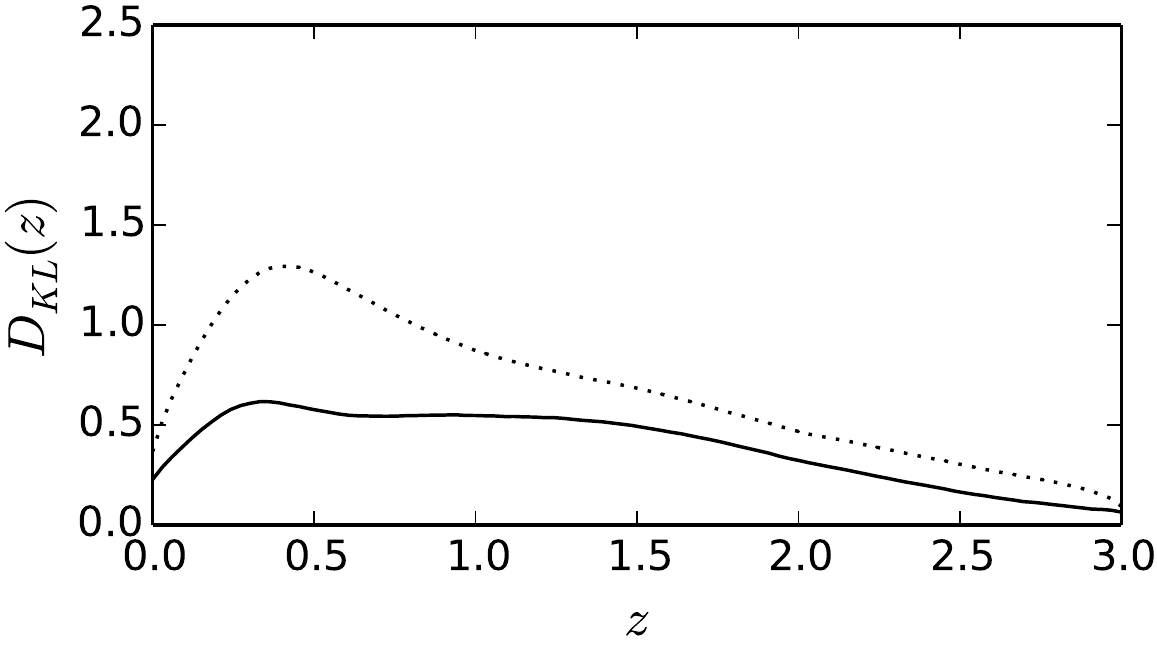}
  \end{minipage}
  \begin{minipage}{0.22\textwidth}
    \centering
    $PbB$\\ 
    \includegraphics[width=1\textwidth, height=0.5\textwidth]{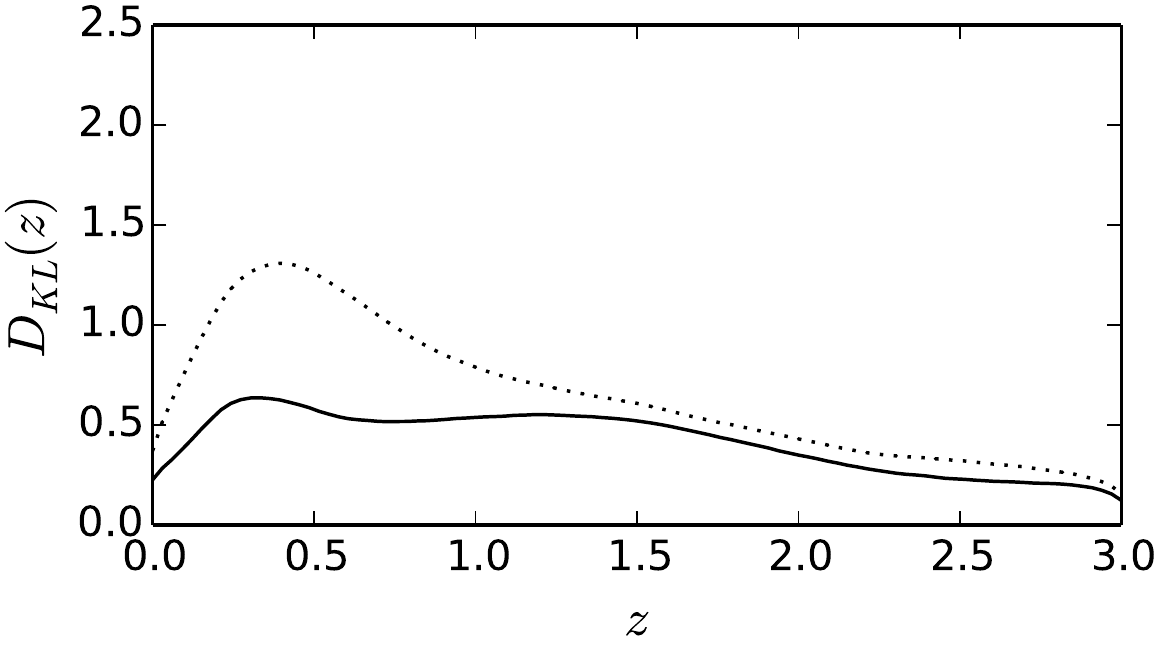}
  \end{minipage}
  \begin{minipage}{0.22\textwidth}
    \centering
    $PabB$\\ 
    \includegraphics[width=1\textwidth, height=0.5\textwidth]{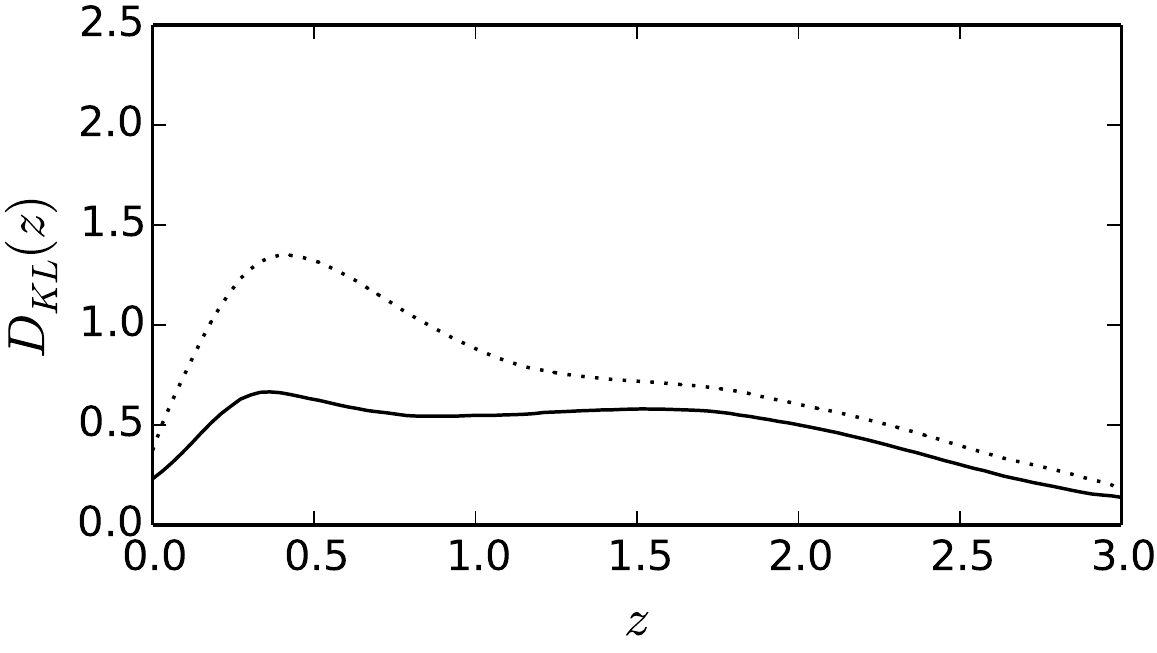}
  \end{minipage}
  \begin{minipage}{0.22\textwidth}
    \centering
    $PS$\\ 
    \includegraphics[width=1\textwidth, height=0.5\textwidth]{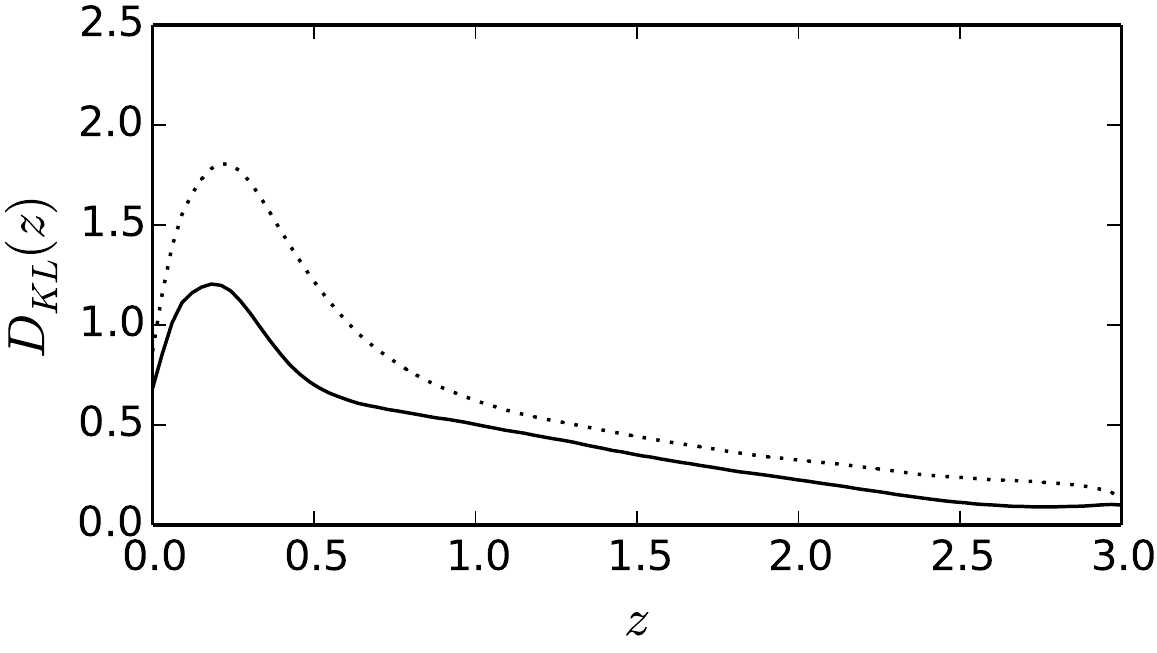}
  \end{minipage}
  \begin{minipage}{0.22\textwidth}
    \centering
    $PaS$\\ 
    \includegraphics[width=1\textwidth, height=0.5\textwidth]{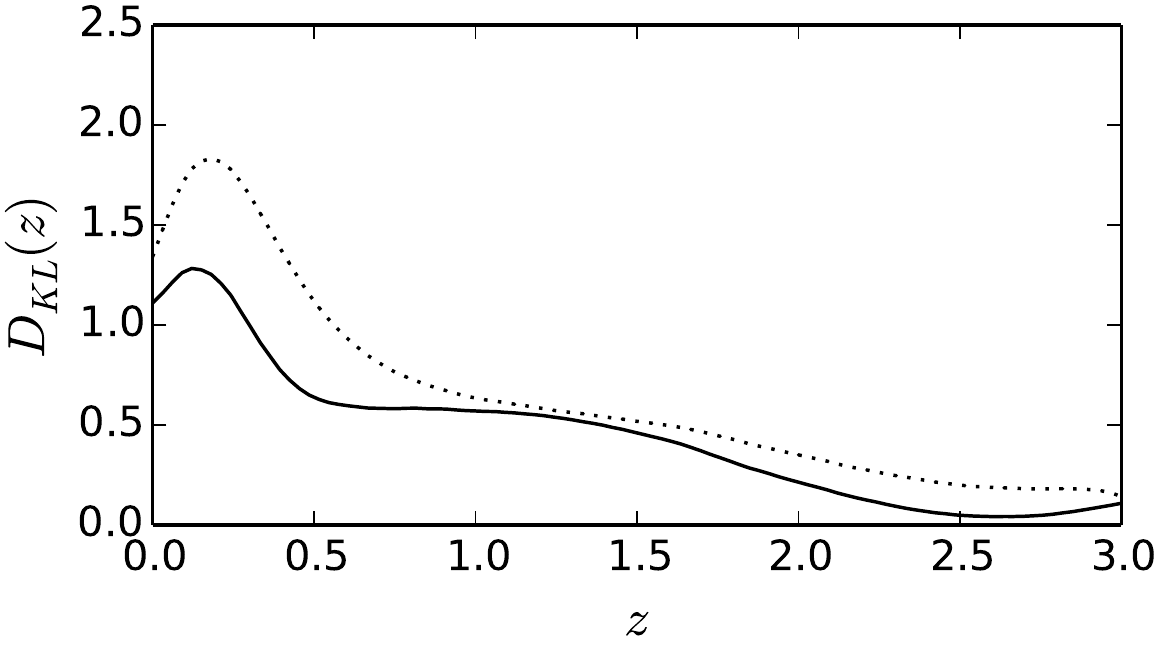}
  \end{minipage}
  \begin{minipage}{0.22\textwidth}
    \centering
    $PbS$\\ 
    \includegraphics[width=1\textwidth, height=0.5\textwidth]{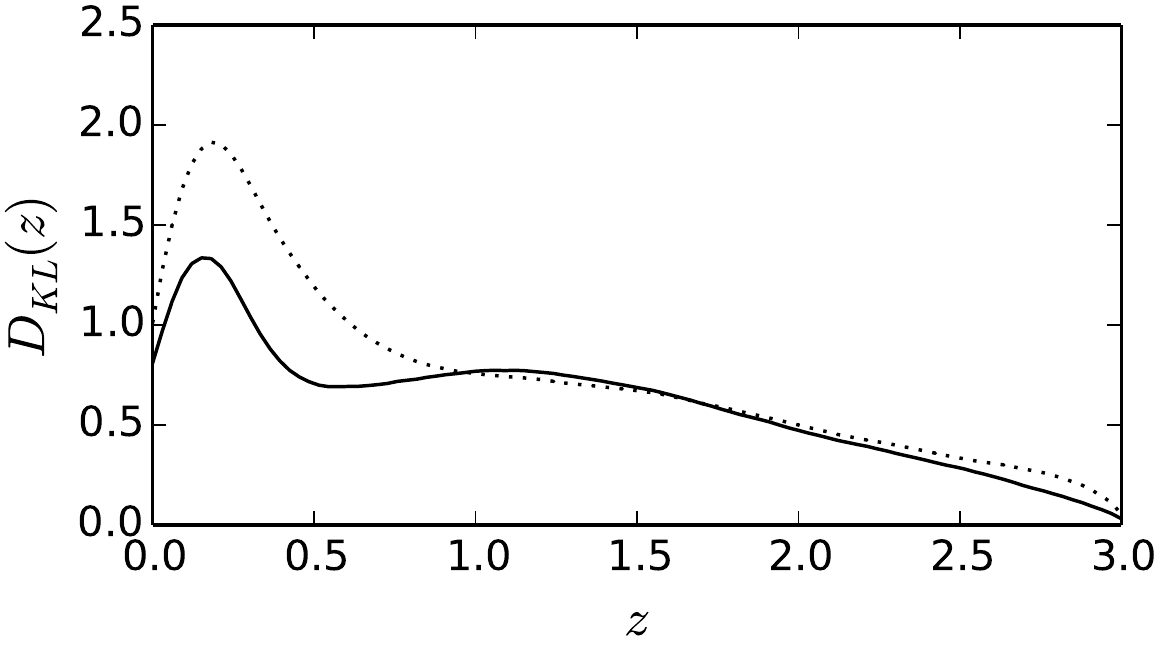}
  \end{minipage}
  \begin{minipage}{0.22\textwidth}
    \centering
    $PabS$\\ 
    \includegraphics[width=1\textwidth, height=0.5\textwidth]{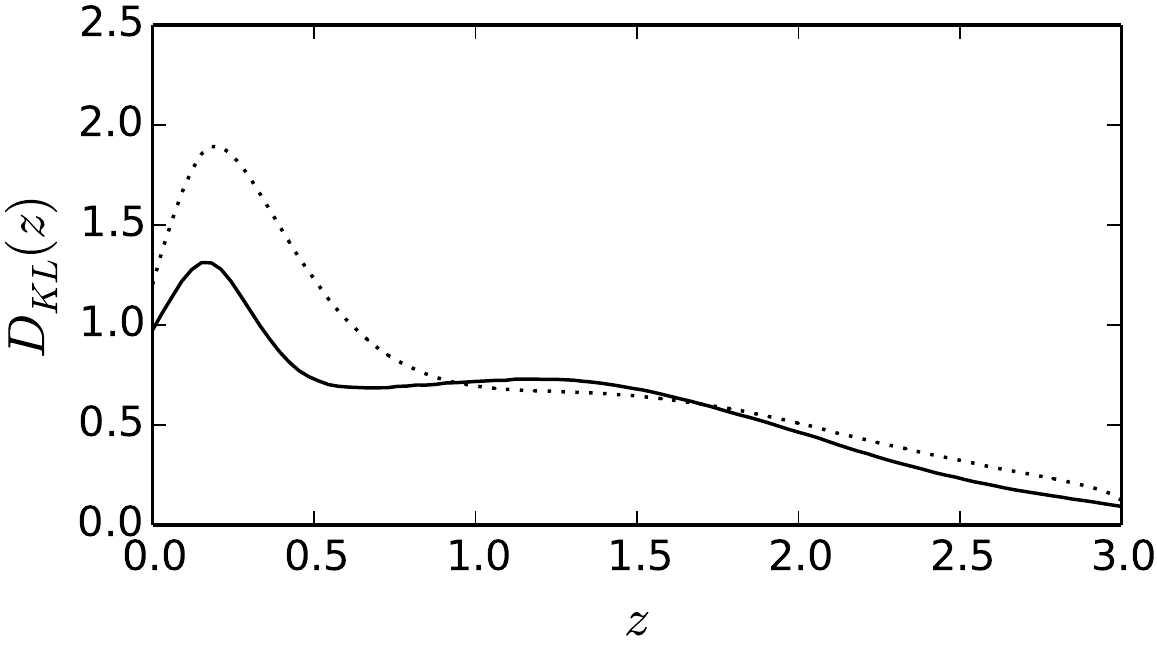}
  \end{minipage}
  \begin{minipage}{0.22\textwidth}
    \centering
    $PBS$\\ 
    \includegraphics[width=1\textwidth, height=0.5\textwidth]{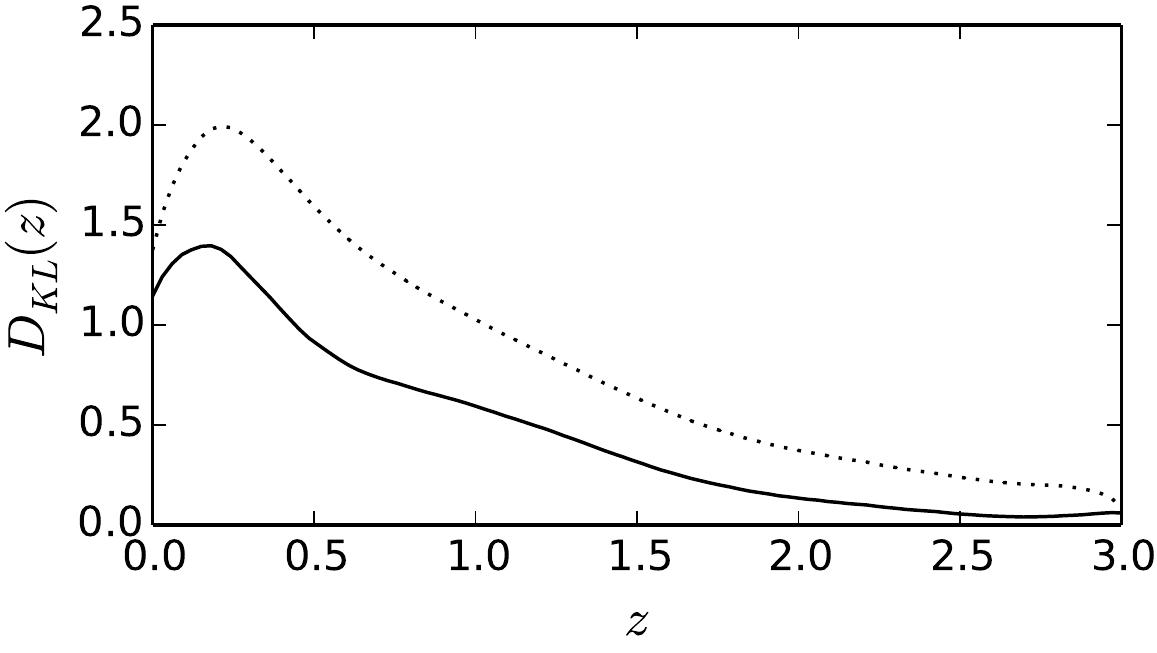}
  \end{minipage}
  \begin{minipage}{0.22\textwidth}
    \centering
    $PaBS$\\ 
    \includegraphics[width=1\textwidth, height=0.5\textwidth]{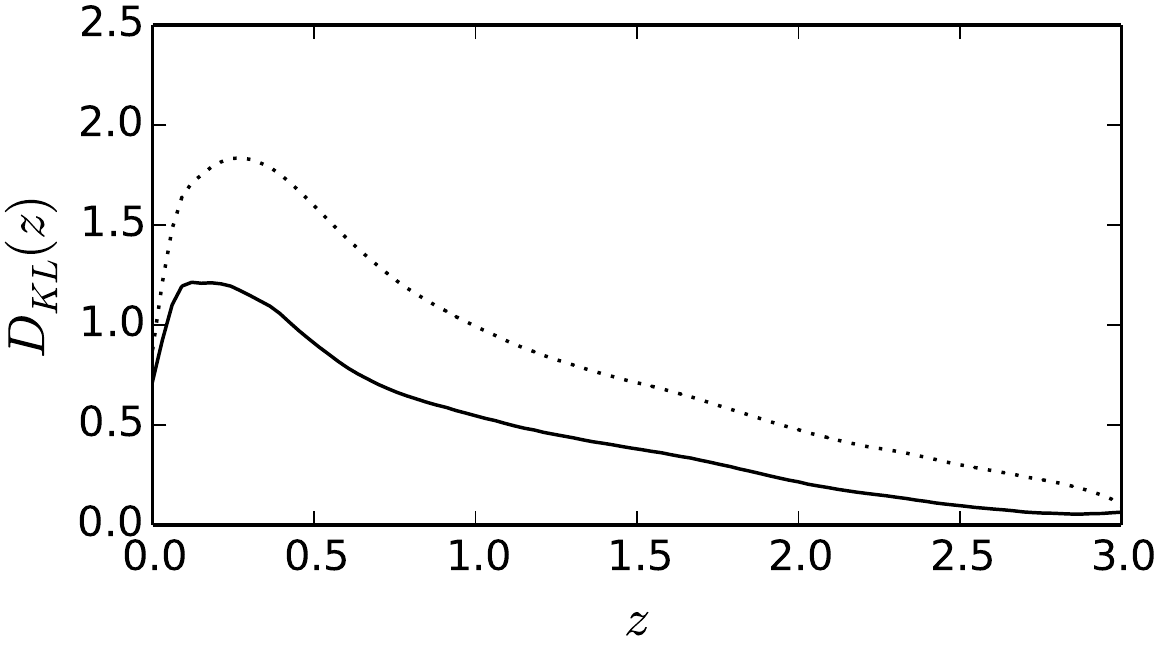}
  \end{minipage}
  \begin{minipage}{0.22\textwidth}
    \centering
    $PbBS$\\ 
    \includegraphics[width=1\textwidth, height=0.5\textwidth]{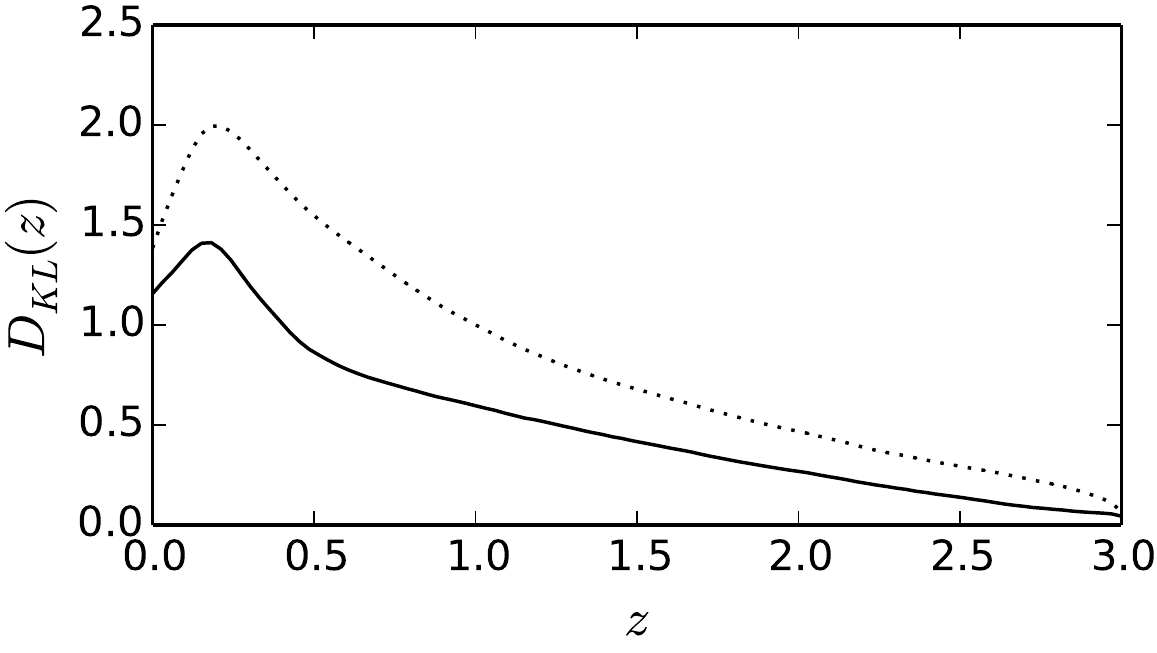}
  \end{minipage}
  \begin{minipage}{0.22\textwidth}
    \centering
    $PabBS$\\ 
    \includegraphics[width=1\textwidth, height=0.5\textwidth]{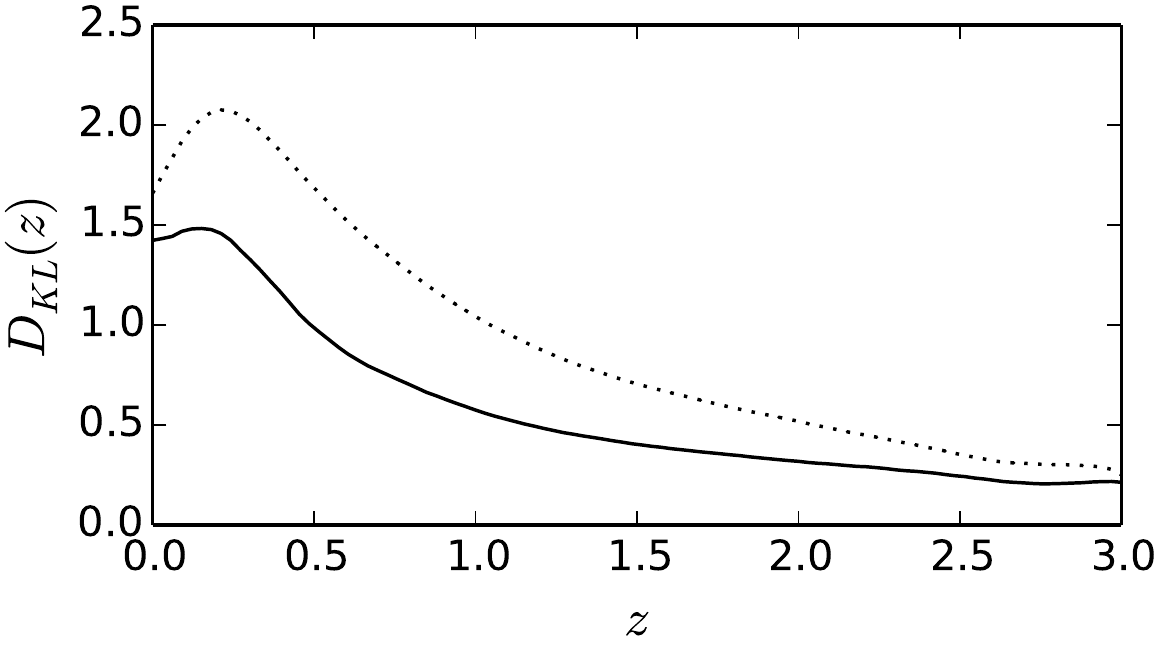}
  \end{minipage}
  \caption{$\DKL(z)$ for all combinations of datasets, laid out as in Figure~\ref{fig:DKL_all}, quantifying the constraining power observed qualitatively in the plane reconstructions. The solid lines use the \codeF{CosmoMC} priors when computing $\DKL(z)$ and demonstrate the additional information gained by using the data in updating our knowledge from the \codeF{CosmoMC} priors to the posteriors. The dashed lines use flat priors across the $w(z)$ plane when calculating $\DKL(z)$ and quantify more intuitively how constrained the plane appears visually, without including the effect of the posterior shifting from the \codeF{CosmoMC} prior peaks. Using the \codeF{CosmoMC} priors shows that the $\Lya$ datasets add much information due to this shift, whilst the posteriors themselves are less tightly constrained than when using $BAO$ and $JLA$ data.}
\label{fig:DKLz_all}
\end{figure*}

To understand how the various datasets constrain the $w(z)$ equation of state we analyse every combination of the datasets using the $2$CDM model and the Kullback-Leibler divergence ($\DKL$). We chose the $2CDM$ model for its flexibility to capture features whilst not being as computationally demanding as $3$CDM\@. For each combination we present the $w(z)$ plane reconstruction to identify features visually, the single value $\DKL$ to understand the total information gained and dataset constraining power, and the distribution $\DKL(z)$ to localise these effects as a function of redshift. As discussed in Section~\ref{sec:method_DKL}, the $\DKL$ values and $\DKL(z)$ functions are presented for each dataset using both the \codeF{CosmoMC} priors to calculate the $\DKL$, which reflect the dataset information content when updating our knowledge from prior to posterior, and also using a flat prior when calculating the $\DKL$ to quantify only the strength of the posterior distribution constraints.

Figure~\ref{fig:DKL_all} shows the plane reconstructions and plane $\DKL$ for each dataset combination in a grid of $\Lya$ versus non-$\Lya$ datasets. The $\DKL$ values in brackets show the constraining power only, whilst the $\DKL$ values not in brackets show the information content of the datasets. Note that for the top row, containing only $Planck$ and $\Lya$ dataset combinations, the information content is larger than the constraining power. As discussed in Section~\ref{sec:method_DKL}, this happens when the posterior peak shifts from the prior peak, and the $\DKL$ analysis therefore is consistent with the observed posterior reconstruction shift to a supernegative equation of state for these four dataset combinations. From reviewing the constraining power in plane reconstructions along each row (where the combinations vary in use of $\Lya$ datasets), it appears that $\Lya$ datasets do not strongly affect the constraining power despite their large information content. Comparing $P$ with $Pab$ we observe an increase in constraining power of $0.14 \nats$ only. When reviewing plots along the columns (where the combinations vary in use of $BAO$ and $JLA$), we visually notice a more pronounced constraint on $w(z)$ and an increase of $0.49 \nats$ when comparing $P$ with $PBS$. In general comparing $PBS$ and $PabBS$, on either measure of information content or constraining power, shows an increase of $0.1 \nats$, which suggests that the $\Lya$ datasets can complement the analysis even if not significantly changing the constraining power.

Figure~\ref{fig:DKLz_all} shows the $\DKL$ as a function of redshift, calculated again using both the \codeF{CosmoMC} priors (solid lines; information content) and flat priors (dashed lines; constraining power). Comparing $PB$ and $PS$ we observe that the peak information content of the $BAO$ dataset is significantly smaller than the $JLA$ peak information content. Specifically, the $PB$ dataset has a peak of $0.6 \nats$ at $z{=}0.3$ which is of lower magnitude but later redshift than the $PS$ peak of $1.2 \nats$ around $z{=}0.2$. The constraining power functions show that this information content is largely due to tightening posterior constraints, and we conclude that the $JLA$ dataset is more powerful in constraining the dark energy equation of state than $BAO$. Reviewing the $\DKL(z)$ information content for the $\Lya$ dataset combination $Pab$ shows a large and broad peak of almost $2 \nats$ at redshift $0.4$, suggesting that the $\Lya$ dataset contains significantly more information than both the $BAO$ or $JLA$ datasets. However this is due to a shift, and the constraining power has a significantly lower peak of only $0.6 \nats$ but over a broad redshift range.

When analysing which datasets may primarily support deviations from $\Lambda$CDM, it is interesting to note that the addition of the $\Lya$ datasets pushes the high redshift constraints away from $w{=}{-}1$ further towards the supernegative. The $PB$ combination plane reconstruction shows that $w{=}{-}1$ is on the $1 \sigma$ contour over the range $1.5{<}z{<}2.0$, whilst the $PabB$ combination disfavours $w{=}{-}1$ at more than $1 \sigma$ for $z{>}1.5$. This is similar for the $PS$ and $PabS$ comparison. Generally though, the plane reconstructions of most combinations either favour or approach a supernegative $w(z)$ for $z{>}1.5$ at a $1 \sigma$ level even without the $\Lya$ datasets and the constraints often broaden out for $z{>}2$ to be consistent with $\Lambda$CDM due to a lack of data (as can be observed by the trailing off in the $\DKL(z)$ plots at higher redshift). Therefore we do not attribute supernegative behaviour strongly to any single dataset when combining them. Another deviation from $w{=}{-}1$ can be observed in the combination $PaB$ at low redshift, where this time $w{>}{-}1$ is favoured. Generally, the $BAO$ dataset seems to favour a less negative equation of state for $z{<}0.5$, whilst $JLA$ is consistent with $w{=}{-}1$ at the same period and the $\Lya$ datasets favour a supernegative $w$-value at all redshifts (which $Planck$ does too). 

Generally, from the dashed $\DKL$ plots, we conclude that for the $\Lya$ datasets a broad but small peak in $\DKL(z)$ at around $z{=}1$ can be observed to complement the $BAO$ and $JLA$ datasets (when comparing $PB$ with $PabB$, $PS$ with $PabS$ and $PBS$ with $PabBS$) by increasing $\DKL(z)$ for $z{>}1.5$\footnote{Note that taking the difference of two $\DKL(z)$ graphs does not represent the information gained or lost between combinations, but the observed change in shape is what we are commenting on. The addition of $ab$ raises $\DKL(1.5{<}z{<}2)$ slightly and tightens the plane reconstruction contours for higher redshifts.}. Comparing the $PabBS$ plane reconstruction figure (or any dataset combination) with the corresponding $\DKL(z)$ plot shows good agreement with the qualitative conclusion that the datasets provide the most constraining power at redshift $0.1 {-} 0.5$, and now provide a clear quantification of this effect together with a more precise conclusion: the constraining power for the $PabBS$ dataset and \codeF{CosmoMC} prior combination peaks at redshift $0.25$ at $2.1 \nats$ whilst the dataset maximises information gain at redshift $0.2$ with $1.5 \nats$.

\section{Conclusions}
\label{sec:conclusions}

We have presented a detailed Bayesian model selection analysis applied to the nodal reconstruction of $w(z)$, concluding that the Bayes factors on the Jeffreys scale `slightly favour' $\Lambda$CDM when compared to $w$CDM and `significantly disfavour' the $t$CDM, $1$CDM, $2$CDM and $3$CDM models, with an error on the Bayes factors of around $0.29$. Despite this favouring, a model averaging approach presents a bifurcation of the $P(w|z)$ plane reconstruction space which shows that, whilst $w{=}{-}1$ for all redshift is strongly favoured, a supernegative $w(z)$ equation of state at redshift $z{>}1.5$ within the $1.5 \sigma$ confidence intervals of the posterior on $w(z)$ is supported by the data.

To understand this possible deviation we analysed the constraining power of the datasets using the Kullback-Leibler divergence ($\DKL$). We calculated a novel function $\DKL(z)$ to analyse the information gained when moving from the prior distribution of $w(z)$ to the posterior distribution, in slices of constant $z$, as well as a single $\DKL$-value for the whole plane. For each we used both \codeF{CosmoMC} priors and flat priors to observe information gain due to the data and the overall constraining power respectively, and we analysed each permutation of datasets using the $2$CDM model. We observed that the $BAO$ and $JLA$ datasets constrained the $w(z)$ plane much more strongly than the $\Lya$ datasets used. These two datasets had a strong peak at redshifts ${<}0.5$ whilst the $\Lya$ datasets peaked more broadly at $z{=}1$. As expected, the combination of all datasets had the greatest constraining power, specifically the $Planck$ dataset alone had $\DKL{=}0.33 \nats$, the combination with $BAO$ and $JLA$ datasets had $\DKL{=}0.82 \nats$ and the combination $Planck+BAO+JLA+\Lya$ had $\DKL{=}0.91 \nats$. The same dataset combination had a maximum information gain at redshift $0.2$ of $1.5 \nats$. Reviewing the plane reconstructions and $\DKL(z)$ functions showed that the $\Lya$ datasets provided additional constraints at $z{>}1.5$ that favours a supernegative equation of state, with $\Lambda$CDM disfavoured at $1 \sigma$ significance. 

Generally, many of the dataset combinations disfavoured $\Lambda$CDM at $1 \sigma$ significance around $1.5{<}z{<}2$, with higher redshifts being too poorly constrained to draw conclusions. For redshifts below $1.5$, the $\Lya$ datasets favoured a supernegative $w(z)$, the $JLA$ dataset typically agrees with $\Lambda$CDM and the $BAO$ dataset tends towards $w{>}{-}1$ values (around $1 \sigma$ significance at $z{=}0.25$). Concluding on the higher redshift deviations, we do not attribute this supernegative favouring to a particular dataset, but note that the inclusion of $\Lya$ data adds prominence as it provides a small amount of much needed constraining power over that range. 

In the future, the conclusions of an analysis with these techniques will strengthen as data quality improves. The nodal reconstruction has again been shown to be useful in constraining cosmological models and developing a model independent data driven analysis~\citep{Vazquez2012c,Vazquez2012,Aslanyan2014,PlanckCollaboration2015_inflation,Hee2015}. In addition, the novel formalism introduced here of the Kullback-Leibler divergence as a function of redshift provides a quantitative analysis of dataset information content applied to specific cosmological problems. Future applications of this method with upcoming mission and survey data or for forecasting with mock-data will provide useful insights into the value of datasets in constraining our cosmological models.

\section*{Acknowledgments}
This work was performed using the Darwin Supercomputer of the University of Cambridge High Performance Computing Service (\href{http://www.hpc.cam.ac.uk}{http://www.hpc.cam.ac.uk}), provided by Dell Inc\@. using Strategic Research Infrastructure Funding from the Higher Education Funding Council for England and funding from the Science and Technology Facilities Council (STFC). Parts of this work were undertaken on the COSMOS Shared Memory system at DAMTP, University of Cambridge operated on behalf of the STFC DiRAC HPC Facility; this equipment is funded by BIS National E-infrastructure capital grant ST/J005673/1 and STFC grants ST/H008586/1, ST/K00333X/1. SH and WH thank STFC for financial support. We wish to thank the reviewer for their insightful additions.

\bibliographystyle{mnras}
\bibliography{library_mnras}

\appendix
\label{lastpage}
\end{document}